\documentclass[12pt]{article}\usepackage[hyperfootnotes=false]{hyperref}
\usepackage{epsfig}
\usepackage{float,dsfont}
\usepackage{empheq}
\usepackage{subcaption}

\usepackage{caption}

\newcommand{\ignore}[1]{}

\usepackage[export]{adjustbox}

\usepackage{braket}
\usepackage{amsmath}
\usepackage{amssymb}
\usepackage{graphicx}
\usepackage{fullpage}
\newlength{\abstractwidth}
\renewcommand{\thefootnote}{\fnsymbol{footnote}}
\renewcommand{\thanks}[1]{\footnote{#1}} % Use this for footnotes
\newcommand{\starttext}{
\setcounter{footnote}{0}
\renewcommand{\thefootnote}{\arabic{footnote}}}

\newcommand{\be}{\begin{equation}}
\newcommand{\bea}{\begin{eqnarray}}
\newcommand{\eea}{\end{eqnarray}}
\newcommand{\beq}{\begin{equation}}
\newcommand{\ee}{\end{equation}}

\newcommand{\TFD}{| \textrm{TFD} \rangle}
\newcommand{\TFDt}{| \textrm{TFD}(t) \rangle}
\newcommand{\EE}{\mathbb E}

\def\lsim{ \lower .75ex \hbox{$\sim$} \llap{\raise .27ex
\hbox{$<$}} }
\def\gsim{ \lower .75ex \hbox{$\sim$} \llap{\raise .27ex
\hbox{$>$}} }

\newcommand*\widefbox[1]{\fbox{\hspace{2em}#1\hspace{2em}}}

\def\eq{&=&}

\def\la{\langle}
\def\ra{\rangle}

\def\simleq{\; \raise0.3ex\hbox{$<$\kern-0.75em
\raise-1.1ex\hbox{$\sim$}}\; }
\def\simgeq{\; \raise0.3ex\hbox{$>$\kern-0.75em
\raise-1.1ex\hbox{$\sim$}}\; }

\def\bi{\begin{itemize}}
\def\ei{\end{itemize}}

\def\sc{\setcounter{equation}{0}}

\def\CC{{\cal{C}}}

\def\bsub{ \begin{subequations}
\begin{
}[box=\widefbox]{align}  }
\def\esub{ \end{empheq}
\end{subequations}}

\def\bn{\bigskip \noindent}

  \def\kl{$k$-local}

    \def\b{\bf{b}}
        \def\r{\bf{r}}

\makeatletter
\g@addto@macro\normalsize{%
  \setlength\abovedisplayskip{10pt}
  \setlength\belowdisplayskip{20pt}
  \setlength\abovedisplayshortskip{10pt}
  \setlength\belowdisplayshortskip{20pt}
}
\makeatother

\usepackage{color}

\begin{document}

\begin{titlepage}

\rightline{}
\bigskip \bigskip \bigskip
\bigskip\bigskip\bigskip\bigskip
\bigskip

\centerline{\Large \bf {  The Python's Lunch: }}
\centerline{\large \bf { geometric obstructions to decoding Hawking radiation }}
\bn

\bigskip
\begin{center}

\bf Adam R. Brown$^{ab}$,   Hrant Gharibyan$^{c}$, Geoff Penington$^{b}$,  \\
 and  Leonard Susskind$^{ab}$  \rm

\bigskip

$^a${\it Google, Mountain View, CA 94043, USA}

$^b${\it Stanford Institute for Theoretical Physics,  \\ Stanford University, Stanford, CA 94305, USA}

$^c${\it Institute for Quantum Information and Matter, Caltech, Pasadena CA 91125, USA}

\end{center}

\bn

\begin{abstract}

 According to Harlow and Hayden [arXiv:1301.4504] the task of distilling information out of Hawking radiation appears to be computationally hard despite the fact that the quantum state of the black hole and its radiation is relatively un-complex. We trace this computational difficulty to a geometric obstruction in the Einstein-Rosen bridge connecting  the black hole and its radiation. Inspired by tensor network models, we conjecture a precise formula relating the computational hardness of distilling information to geometric properties of the wormhole -- specifically to the exponential of the difference in generalized entropies between the two non-minimal quantum extremal surfaces that constitute the obstruction. Due to its shape, we call this  obstruction the `Python's Lunch', in analogy to the reptile's postprandial bulge.

\end{abstract}

\vspace{2cm}

\begin{figure}[H]
\hspace*{-1in}
\begin{raggedleft}
\includegraphics[scale=.6, inner]{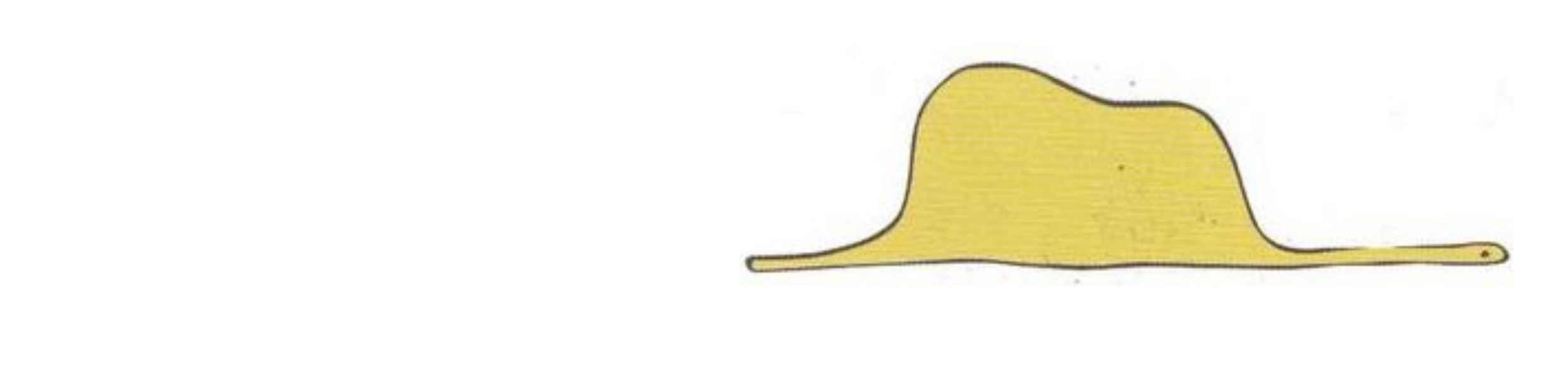}
\end{raggedleft}
\end{figure}

\vspace{3cm} 
\end{titlepage}

\starttext \baselineskip=17.63pt \setcounter{footnote}{0}
\tableofcontents

%\large
\sc

\section{Introduction}
Computational complexity is relevant to the study of quantum gravity in (at least) two ways: in its traditional role as a measure of the difficulty of carrying out tasks \cite{HarlowHayden}; and as a possible holographic dual for properties of the spacetime behind horizons \cite{Susskind2014, Brown2015, Brown2015-1}. 

Harlow and Hayden \cite{HarlowHayden} studied the complexity of the  task  of distilling a single qubit of information from Hawking radiation. They argued that  the complexity of distillation grows exponentially with the entropy $S$ of the black hole. 

Later, in the context of the AdS/CFT duality, one of us proposed a holographic identification between the computational complexity of an entangled pair of boundary states and the size of the Einstein-Rosen bridge in the dual two-sided black hole \cite{Susskind2014}.

At first sight there seems to be some tension between these two roles of complexity. While the complexity of decoding Hawking radiation is exponential in $S$,  the volume of the wormhole connecting the black hole to its radiation is  only polynomial.

The source of the discrepancy is that we are using two different definitions of complexity. The decoding task \cite{HarlowHayden} is only hard because we are restricted to act solely on the radiation outside the black-hole horizon. In Ref.~\cite{Susskind2014} there is no such restriction. The distinction between {\it restricted} and {\it unrestricted} complexity will be a central theme of this paper. In particular we will be interested in the distinction between the holographic dual of unrestricted complexity, which was the subject of \cite{Susskind2014, Brown2015, Brown2015-1}, and the holographic dual of restricted complexity, which will be a subject we will develop in this paper. The main point of this paper is not to prove the Harlow-Hayden conjecture -- like almost  everything else in complexity theory this is too hard -- but to explain how it may be related to the geometry of wormholes. \\

Consider one possible decoding `strategy' for distilling information while acting solely on the Hawking radiation\footnote{We emphasize that this strategy is chosen more for illustrative clarity than engineering practicality.}. The first step in this strategy is to gather the radiation and collapse it into a  second black hole. This new black hole is entangled with the first black hole, and the entanglement can be interpreted, according to ER=EPR \cite{Maldacena2013}, as a wormhole connecting them. At the Page time the wormhole would have a volume of order $S^2,$ far less than the exponential complexity claimed by Harlow and Hayden, but still too large to easily implement the decoding. The next step would be to apply unitary operations to the second black hole in order  to shorten the wormhole and bring it to the thermofield-double state. In that state  the two horizons have no separation between them, and the structure of the entanglement is especially simple. Once this is accomplished the decoding should be easy. The only potentially hard step in this strategy, therefore, is shortening the wormhole. 

Since the only potentially hard step is shortening the wormhole, and since the Harlow-Hayden argument shows that decoding information from the Hawking radiation alone is indeed exponentially hard, we conclude that shortening the wormhole from one side must be exponentially hard. This situation suggests that there must be some kind of obstruction in the wormhole, an obstruction which prevents us from efficiently shortening the wormhole from one side. Moreover  this obstruction  cannot be large volume, since the volume is not large. 

\begin{figure}[t]
\begin{center}
\includegraphics[scale=.4]{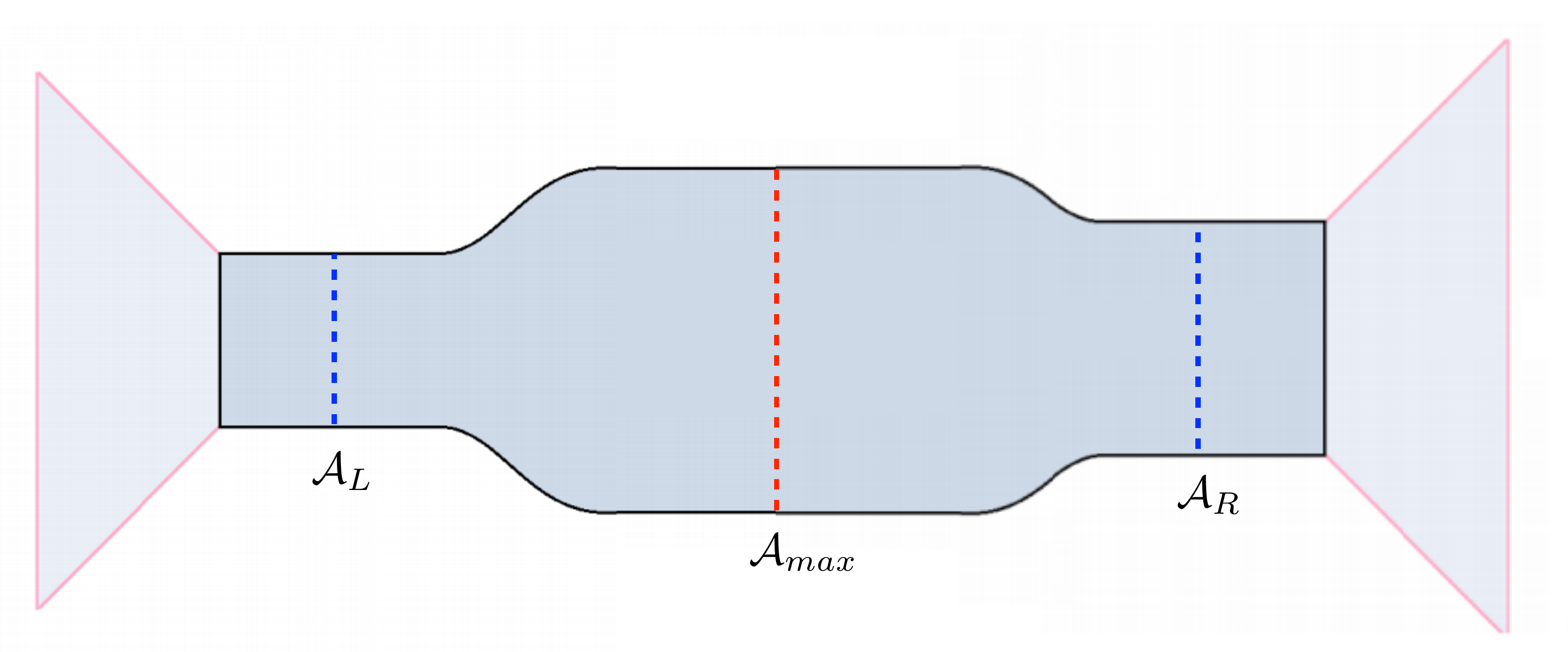}
\caption{A spatial slice through a `Python's Lunch' geometry. On the far left, the wormhole opens up to one asymptotic region with infinite cross-sectional area; on the far right, the wormhole opens up to the other asymptotic region also with infinite cross-sectional area. In AdS-Schwarzschild black holes the cross-sectional area reaches a minimum in the middle of the wormhole, and increases on either side. By contrast, in the Python's Lunch geometry the cross-sectional area first shrinks, then grows, then shrinks, then grows again, giving rise to a bulge in the middle of the wormhole---the eponymous Lunch. $\mathcal A_L$ and $\mathcal A_R$ are the areas of the minimal surfaces on each side and $\mathcal A_\textrm{max}$ is the area of the luncheon bulge. }
\label{lunch}
\end{center}
\end{figure}

In this paper we will conjecture that the geometric obstruction is a bulge in the wormhole, which  because of its shape we call  the ``Python's Lunch'', as depicted in Fig.~\ref{lunch}. We will estimate the complexity of bypassing the Python's Lunch, and find that, consistent with the Harlow-Hayden claim, it is indeed exponential.  In Eq.~\ref{eq:bigconjecture} we will conjecture that the restricted complexity is dual to the size of the Python's Lunch via
\begin{align}
\CC_R[U_{PL}] \sim  \exp  \Big[ \frac{1}{2} \frac{\mathcal{A}_\textrm{max} - \mathcal{A}_R}{4G \hbar} \Big] ,
\end{align}
where $\mathcal{A}_\textrm{max}$ is the maximum cross-section of the wormhole and $\mathcal{A}_R$ is the size of the throat connecting the wormhole to the radiation. In Eq.~\ref{conjCov} we will make a covariant generalization of this conjecture. This proposal for the geometric dual of the \emph{restricted} complexity is complementary to existing conjectures about the geometric duals to \emph{unrestricted} complexity \cite{Susskind2014, Brown2015, Brown2015-1}. 

\bn
 Despite our focus on restricted complexity, in  Sec.~\ref{measureHawking} we find that one-sided Python's Lunches can also teach us about {unrestricted} complexity. We suggest an improvement to the definition of unrestricted holographic complexity conjectured to be dual to volume \& action in Refs.~\cite{Susskind2014, Brown2015, Brown2015-1}. Specifically, we argue that these conjectures should have defined complexity to permit not only unitary gates but also non-unitary projections.

\section{The Shortening of Wormholes} \sc

In much of what follows we will assume that black holes can be modeled as   ``quantum computers" by which  we mean collections of $N$ qubits evolving by means of  \kl \ all-to-all Hamiltonians or discrete gates. The number of qubits is determined by the entropy of the black hole, \footnote{Modeling black holes as a quantum computer has been extremely fruitful in the study of quantum information scrambling \cite{Nahum:2017yvy, Khemani2017}, onset of random matrix behavior \cite{Gharibyan:2018jrp} and derivation of the RT formula \cite{Hayden:2016cfa}.}
\begin{equation}
N\sim S.
\end{equation}
We will encounter both (unitary) operator complexity and relative state complexity.  

The complexity of a unitary operator $U$ may be defined as the minimal number of $2$-qubit gates $g$ needed to prepare it; in other words the smallest $n$ for which 
\be 
U=g_ng_{n-1}....g_1.
\ee
There are other definitions but for our purposes this definition will do. The complexity of $U$ will be denoted by $\CC(U).$ By construction, it satisfies 
\be 
\CC(U) = \CC(U^{\dag}) \ . 
\label{Cu=Cu*}
\ee

The relative complexity of two states $|\psi\ra$ and $|\phi\ra$ is defined as the complexity of the \it least complex \rm unitary that connects them $|\psi\ra = U |\phi\ra.$ In other words it is the minimum number of gates required to transform $|\phi\ra$ to $|\psi\ra,$
$$|\psi\ra = U|\phi\ra =g_ng_{n-1}....g_1 |\phi\ra.$$
Relative complexity is denoted by $\CC(\psi, \phi).$ Due to Eq.~\ref{Cu=Cu*}, it is symmetric in its arguments
\be 
\CC(\psi, \phi)=\CC(\phi, \psi).
\label{CC=sym}
\ee
Unitary matrices represent operators,
\be 
U = \sum_{IJ} U_{IJ}|I\ra\la J |
\label{Uoperator}
\ee
where $I,J$ label a complete basis of $N$ qubit states in the computational basis.
The same matrix  plays a second role in representing a maximally entangled state of $2N$ qubits,
\be 
|\Psi\ra = \frac{1}{2^{N/2}} \sum_{IJ} U_{IJ}|I\ra |\bar{J} \ra,
\label{Ustate}
\ee
where $|\bar{J} \ra$ is the time reversal of $|J \ra$.

A special case is $U_{IJ} = \delta_{IJ}$  which describes the infinite temperature thermofield-double state,
\be 
\TFD \Bigl|_{T = \infty} =  \frac{1}{2^{N/2}} \sum_I |I\ra|\bar{I}\ra \ . 
\label{TFD}
\ee
The infinite temperature $\TFD$ state may also be written as a product of $N$ Bell pairs,
\be 
\TFD \Bigl|_{T = \infty}  = |\text{Bell} \ra^{\otimes N}.
\label{Tbell}
\ee

\bn

The thermofield-double  state of a two-sided black hole is 
a finite temperature state of an infinite number of qubits, but is often modeled as an infinite temperature state of a finite number of qubits. We too will make that approximation, so by $\TFD$ we will mean the state in Eq.~\ref{Tbell}. 
The  state $\TFD$ is the simplest  case of a maximally entangled state.

The two subsystems called $A$ and $B$  are under the control of Alice and Bob respectively.
The natural evolution of the system is governed by an overall Hamiltonian which is the sum of two non-interacting terms,
\be 
H=H_A \otimes  \mathds{1} +  \mathds{1} \otimes H_B \, .
\label{H=HA+HB}
\ee
For simplicity we will assume that the two Hamiltonians are identical and real (so that we don't have to worry about the details of time reversal).\\
The natural  time evolution operator is a product,
\be 
U(t) = U_A(t)  \otimes U_B(t) = e^{-iH_A t} \otimes e^{-iH_B t}.
\label{U=UU}
\ee 
 We will define the time-evolved state $\TFDt$ by 
\be  
\TFDt  \equiv U(t) \TFD
\label{Tt=UT}
\ee 
In the maximally entangled case $|\textrm{TFD}(t)\ra$ can be constructed by evolving on only  one side for a total time $2t$,
\bea
\TFDt &\equiv& U_A(t)  \otimes U_B(t)\,\TFD \cr \cr
\eq  U_A(2t) \otimes \mathds{1} \,\TFD  \cr \cr
\eq  \mathds{1} \otimes   U_B(2t) \, \TFD
\label{T=UUT}
 \eea
 
As $t$ evolves,  the linearly growing  complexity of the state is dual to the linearly growing volume of the wormhole. For sub-exponential times the complexity  is the sum of the complexity of $U_A(t) $ and $U_B(t)$ which is equal to the complexity of $U_A(2t)$ and of $U_B(2t).$

\begin{figure}[H]
\begin{center}
\includegraphics[scale=0.45]{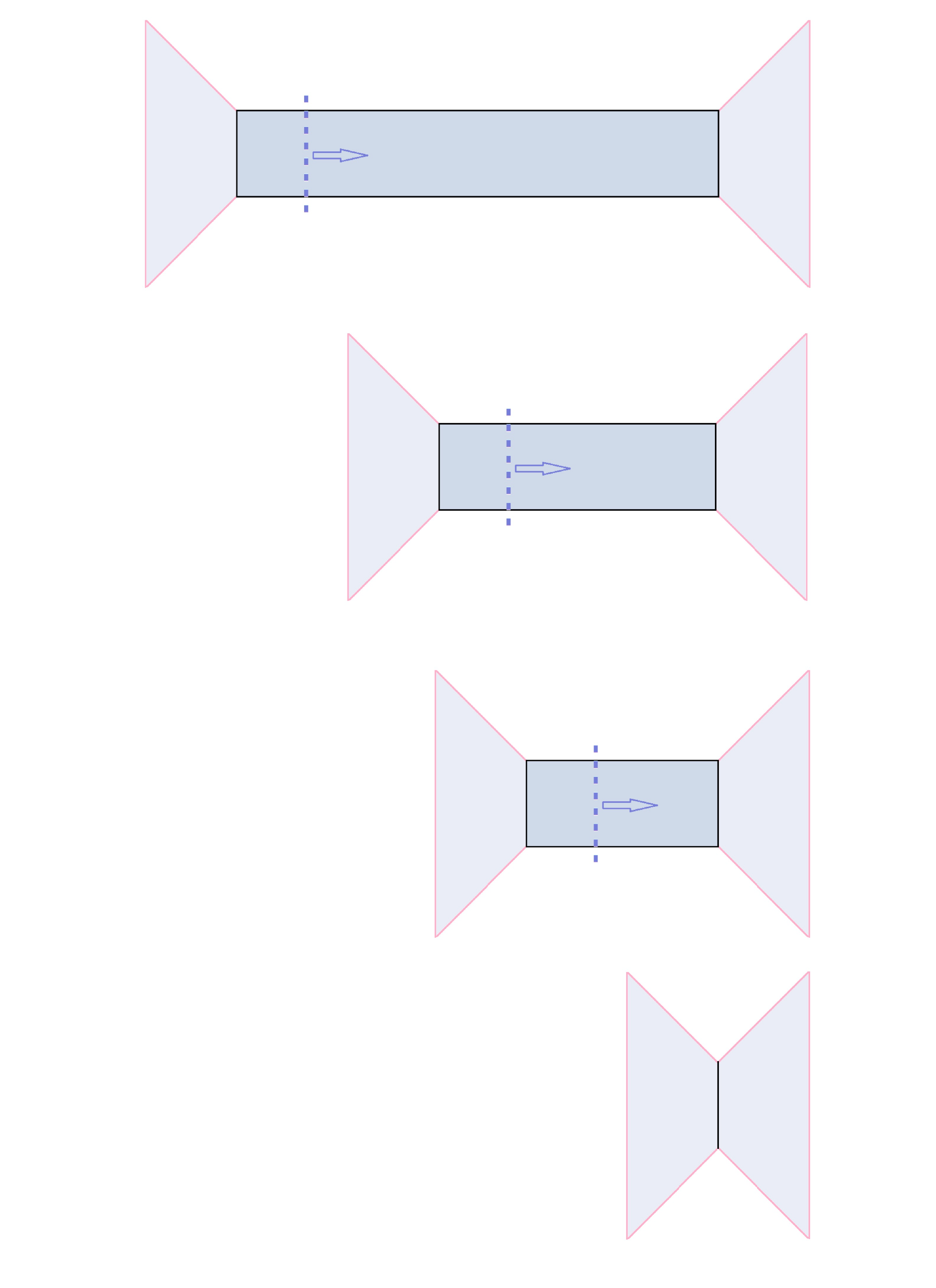}
\caption{Successive spatial slices through the wormhole. Since the two sides are maximally entangled, Alice is  able to shorten the wormhole by unitary operations $U_A \otimes \mathds{1}$ that act only on her side. }
\label{ERB_slide}
\end{center}
\end{figure}

 Given the evolved state $\TFDt$  Alice can return it to the initial {TFD} by applying
 $U^{\dag}_A (2t).$ We will think   of doing this in a series of small steps of low complexity,
 \be 
\TFD =( U^{\dag}_A)^{\epsilon} ( U^{\dag}_A)^{\epsilon}\cdots ( U^{\dag}_A)^{\epsilon} \TFDt .
 \label{steps}
 \ee
 Pictorially  Alice  is shortening the long wormhole  by a series of incremental small steps, as illustrated in Fig.~\ref{ERB_slide}.
 
Bob may also accomplish the shortening by  acting on $\TFDt$  with  $U^{\dag}_B (2t),$ 
or Alice and Bob may act together with $U^{\dag}_B(t_1)  U^{\dag}_A(2t - t_1) $.

\bn

\noindent Here are some questions to consider:

\begin{itemize}
\item Why might one be interested in shortening a wormhole? There are a number of reasons. One that we have already mentioned is that it would be a step in decoding Hawking radiation. 

A second would involve the use of the entanglement as 
 a resource for quantum teleportation. Quantum teleportation requires the use of pre-existing entangled qubits. Not only must these qubits be entangled, they must also be brought to have low complexity in order to successfully teleport. In the language of ER=EPR, if we want to make a wormhole traversable \cite{Gao2016, Maldacena2017}, we first need to make it as short as possible.

\item Why might one want to shorten the wormhole by processes which do not couple the two sides? If the two sides are being used to communicate over a long distance then coupling them quantum-mechanically may be unfeasible. Thus there are practical reasons why one might be interested in the complexity of shortening a wormhole by acting on it from one side.

\item Are there situations in which it is easy to shorten a wormhole by interactions which involve both sides, but in which it is extremely difficult to do so from one side?

\item  Most of all we are interested in whether the answer to the previous question correlates with geometric properties of the wormhole, and if so, what properties?

\end{itemize}

  With regard to this last question, we will argue that there is a particular kind of geometric obstruction which prevents us from efficiently shortening a wormhole by one-sided operations, even though the wormhole has small volume and can be easily shortened by two-sided operations. The shape of the  obstruction suggests the name ``Python's Lunch".
  
  \sc
\section{Restricted and Unrestricted Complexity} \label{sec:restrictedcomplexity}

\begin{figure}[H]
\begin{center}
\includegraphics[scale=.2]{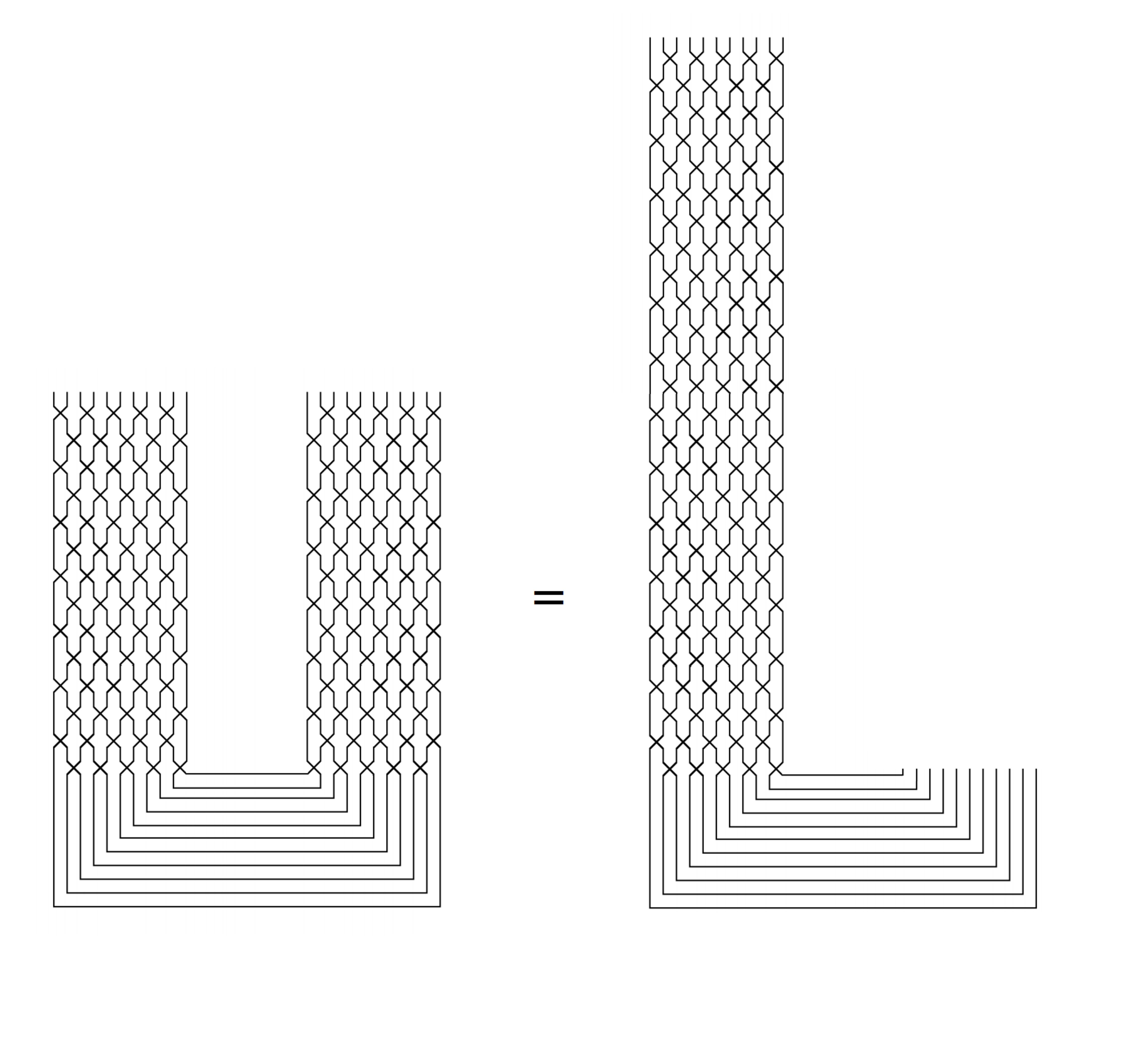}
\caption{ Left: the quantum circuit that prepares $U_A(t)  \otimes U_B(t) \TFD$. Right: the quantum circuit  that prepares  $U_A(2t)  \otimes \mathds{1} \TFD$. The two states are the same. }
\label{TN}
\end{center}
\end{figure}

The restricted complexity $\CC_R$ of a maximally entangled state of $2N$ qubits  is the number of gates needed to construct it from the TFD state under the restriction that all gates act only on one side.\footnote{What we call the `restricted complexity', Aaronson \cite{Aaronson2016} calls the `separable complexity'.} We will sometime use $C_{R, A}$ to indicate that the restriction is on a specific subsystem ($A$ in this case). Without loss of generality we can assume the gates all act on Alice's side or we may distribute them symmetrically between the two sides. A useful picture is provided by the tensor network (TN) description. The state $|\Psi \ra$ is represented as a TN as  in  Fig.~\ref{TN}.
\bn
The restricted relative complexity of $|\Psi\ra$ and $\TFD$ is also the complexity of the unitary operator $U$ corresponding to $| \psi \ra$, as defined in \eqref{Uoperator}.

\begin{figure}[t]
\begin{center}
\includegraphics[scale=.2]{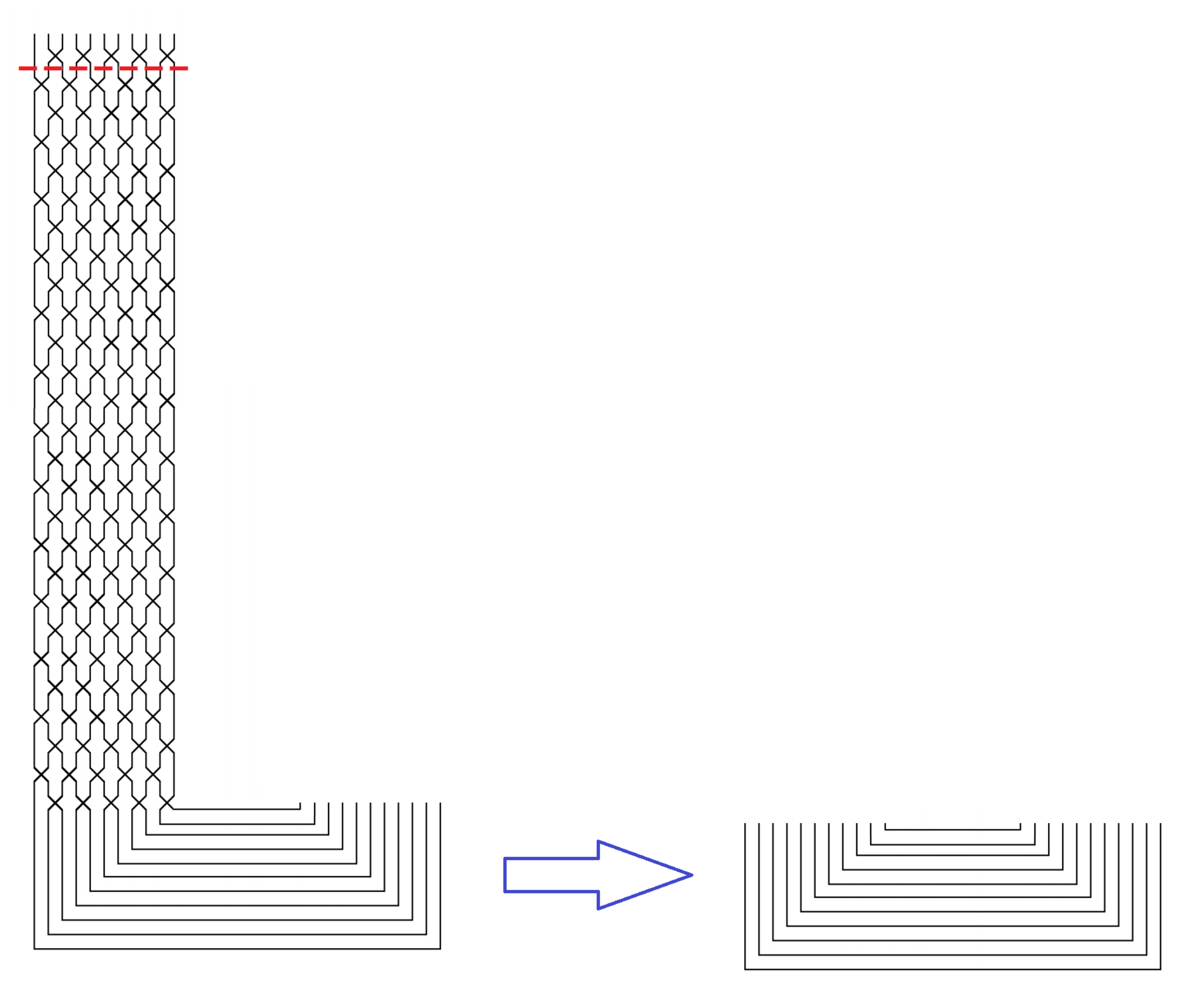}
\caption{ By acting with $U_A(t)^{\dagger} \otimes \mathds{1}$ in a series of incremental $k$-local steps, Alice can undo time evolution, thus mapping  $U_A(t) \otimes \mathds{1} \TFD$ back to $\TFD$.   }
\label{TN2}
\end{center}
\end{figure}

By acting with a layer of gates on Alice's side, a layer can be removed from the TN. By repeating this operation enough times, as in  Fig.~\ref{TN2}, the state can be brought to the simple state $\TFD.$ The minimal number of gates needed to carry out the shortening operation defines\footnote{Note that this definition of complexity is in terms of processes which bring the state back to a simple state rather than processes which prepare the state starting with the simple state. For circuits built from unitary gates the two are the same, but for more general concepts of complexity that may make use of non-unitary elements the two may differ significantly. Later we will consider tensor networks that include non-unitary elements such as projections where this distinction is relevant.} the restricted complexity of $|\Psi\ra.$  

The restricted complexity would be an appropriate measure of the difficulty of the task of shortening the wormhole  if the two computers were too far apart to directly couple.\\

\begin{figure}[t]
\begin{center}
\includegraphics[scale=.2]{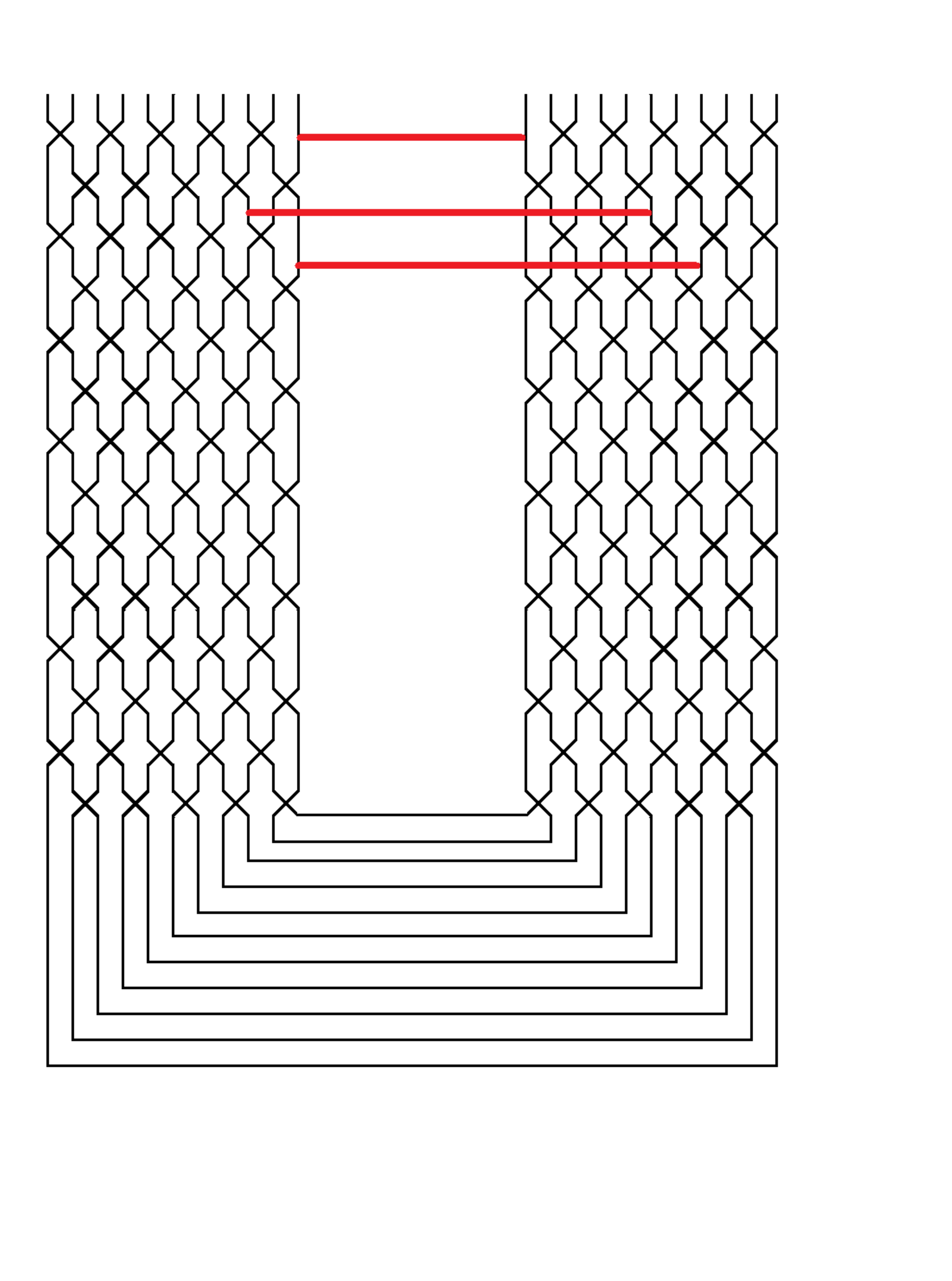}
\caption{A unitary $U_{AB}$ cannot in general be decomposed as $U_A \otimes U_B$. In the example in this figure, the horizontal red links between the left and right sides represent gates which couple qubits on the two sides.    }
\label{TN3}
\end{center}
\end{figure}

The unrestricted complexity $\CC_U$ is the number of $2$-qubit gates needed to complete the shortening  task, allowing  for  gates which couple the two  computers. Fig.~\ref{TN3} shows such an unrestricted circuit.

It is obvious that $\CC_R \geq \CC_U.$ In this paper we will be interested in the conditions under which the  restricted complexity may be exponentially larger than the unrestricted complexity. This subject is not new; it was introduced by Harlow and Hayden \cite{HarlowHayden} in the context of black hole physics, and elaborated on by Aaronson  \cite{Aaronson2016}. Our particular interest is to understand this large gap between restricted and unrestricted complexity
through the geometry of the wormholes connecting entangled systems. The question is: can we identify the situations in which $\CC_R \gg \CC_U$ from the shape of the wormhole? To put it another way, can we identify a geometric obstruction to shortening the wormhole from one side?

\bn

If the state $|\Psi\ra$ was prepared by acting with restricted gates on $\TFD,$ and if the number of such gates is not exponentially large, we expect $  \CC_U = \CC_R.$ 
On the other hand if $|\Psi \ra$ was prepared from $\TFD$ by a circuit that allows interaction between the two computers, then we expect $\CC_{R,A} \gg\CC_U$ (assuming the circuit is longer than the scrambling length). In particular if the number of unrestricted gates used in preparing $|\Psi\ra$ is enough to scramble the system then Harlow and Hayden have argued  that the restricted complexity $\CC_{R,A}$ will be exponential in $N,$ and the same for $\CC_{R,B}.$ At the same time the unrestricted complexity may be no bigger than polynomial in $N$.

\section{The Python's Lunch} \label{sec:PL} \sc

\begin{figure}[H]
\begin{center}
\includegraphics[scale=.40]{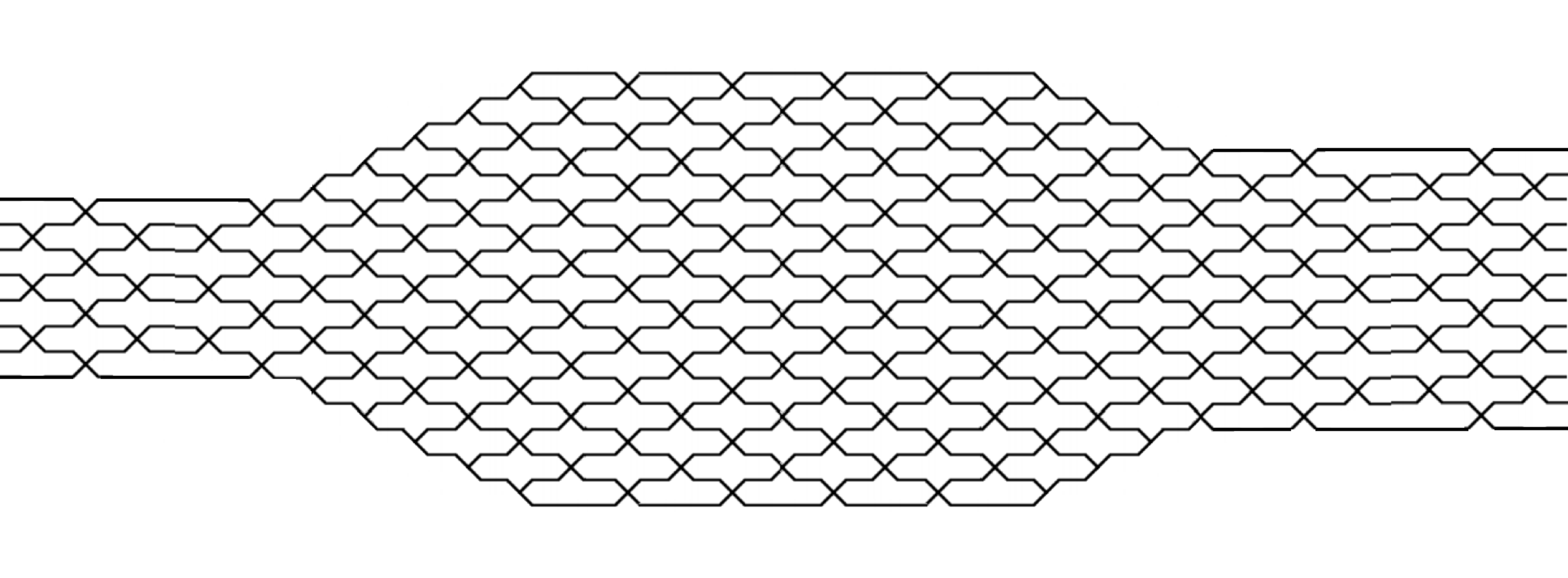}
\caption{The tensor network that corresponds to the Python's Lunch geometry. The throats and bulge (where the girth is constant) are composed of unitary gates, whereas the shoulders (where the girth changes) involve projections like those shown in Fig.~\ref{Iso}.}
\label{Tlunch}
\end{center}
\end{figure}

We now come to the Python's Lunch geometry: a wormhole with a bulge, as illustrated in Figs.~\ref{lunch} and \ref{Tlunch}. For simplicity we will assume that it consists of three regions, all of length polynomial  in $N$, where $N$ denotes the entropy (number of qubits). The two outer regions have area  $A _L \approx N \cdot (4G\hbar)$  and  $A _R \approx (1+\gamma)N \cdot (4G\hbar)$, where $\gamma>1$ is a numerical constant. The bulge between the two outer regions has larger area, 
\be  
A_\textrm{max} = (1+\alpha)N \cdot (4G\hbar), 
\label{A-lunch}
\ee 
where $\alpha>\gamma$ is a constant independent of $N$. 
In order to count as a Python's Lunch, we will mostly assume that the length must  be larger than the scrambling time $t_* \sim \log{N}$. 

An alternative way to look at this geometry is as the tensor network (TN) in Fig.~\ref{Tlunch} which prepares a two-sided state.

If all the vertices in the tensor network were unitary gates, the number of qubits would be the same for every vertical cross-section, but tensor networks (unlike standard quantum circuits) allow certain non-unitary vertices called isometries. Inspection of Fig.~\ref{Tlunch} shows that some of the vertices involve three edges; those are the isometries. They occur in the transition regions where the area varies.

\begin{figure}[H]
\begin{center}
\includegraphics[scale=.6]{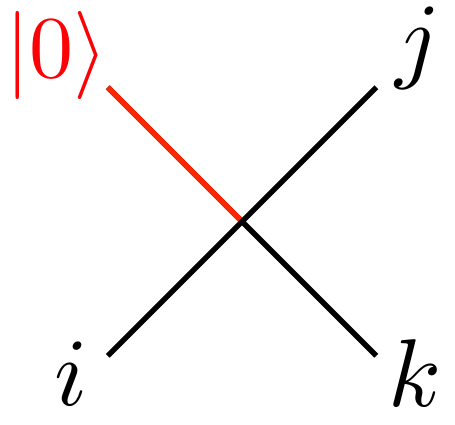}
\caption{An isometry of $| i \rangle \rightarrow |j \rangle | k \rangle$, represented as the projection of a unitary matrix. Such components make up the `shoulders' on the Python's lunch shown in Fig.~\ref{Tlunch}, where the $|0 \rangle$ `leg' is omitted. }
\label{Iso}
\end{center}
\end{figure}

\begin{figure}[t]
\begin{center}
\includegraphics[scale=.5]{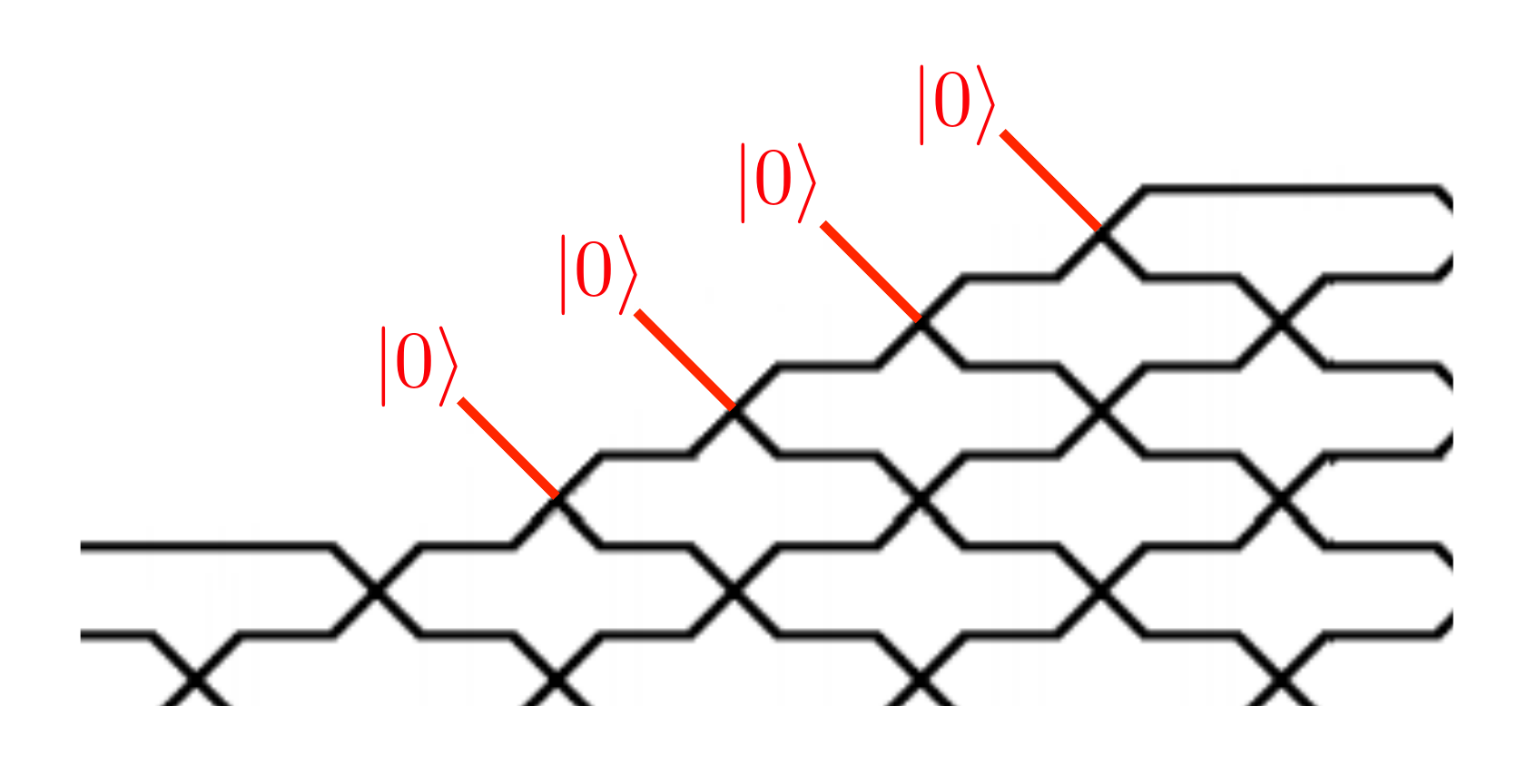}
\caption{Zooming in on the `shoulder' of the Python's Lunch in Fig.~\ref{Tlunch}.}
\label{isometry}
\end{center}
\end{figure}

An isometry can be thought of as a unitary gate in which one of the legs has been projected onto the state $|0\ra$ as shown in Fig.~\ref{Iso}. 
This allows us to draw the TN as a collection of edges connecting unitary gates, but with a subset of the edges being projected. A portion of the TN with isometries is shown in Fig.~\ref{isometry}.
Reading the tensor network from left to right, the tensor network expands when we input an ancilla qubit, and contracts when we post-select on a qubit. In general, the number of input ancilla qubits, $m_L  = \alpha N$, can be different from the number of post-selected qubits, $m_R = (\alpha-\gamma) N = \beta N$. %, assuming we are sweeping from left-to-right.  

One question is whether the operator defined by the TN in Fig.~\ref{Tlunch} is approximately unitary, or equivalently, is the two-sided state that it defines approximately maximally entangled? So long as the right end of the tensor network is larger than the left end (at leading order), the TN will generically be an almost perfect isometry from the left to the right. The state on the left will therefore be almost exactly maximally entangled with the state on the right. We shall assume that this is indeed the case.

With this assumption the TN can be shortened from the right by one-sided unitary operations. But the question is how many one-sided \kl \ operations are required? If the TN is small -- say of polynomial size -- one might conclude that the number of gates should also be small, but that is not the case.

\bn

\subsection{Using Ancilla Qubits}

Let's consider an initial state $|I\ra$ in the computational basis and act on it with the TN, inserted from the left side. The output state is 
\be 
|\psi\ra \propto   \bra{0}^{\otimes m_R} U_{TN} |I\ra \ket{0}^{\otimes m_L} \ , 
\label{U-PL}
\ee
where $U_{TN}$ is the original map from left to right defined by the tensor network, $\ket I$ is the input state on the left,  and $\ket \psi$ is the normalized output state on the right side after post-selection.  
We will be interested in the relative complexity of $|I\ra$ and $|\psi\ra$, when we allow Alice to prepare $m_L$ ancilla qubits that start at $\ket{0}$ and in the end $m_R$ qubits finish in $\ket{0}$ state.

\iffalse 
Here $m$ is the difference between the area of the bulge and the area of the horizons at the ends of the wormhole. From the definition of $\alpha$ in Eq.~\ref{A-lunch},
\be 
m = \alpha N.
\ee
We consider the relative complexity, not of $|I\ra$ and $|\psi\ra,$ but of 
\be 
|I\ra \otimes |0\ra^{\otimes m} \quad \text{and} \quad |\psi\ra \otimes |0\ra^{\otimes m}. 
\ee
We'll call this relative complexity $\CC^{(m)}(\psi, I)$, where the superscript indicates the number of ancilla. For $m=0,$
\be 
\CC(\psi, I)    =  \CC^{(0)}(\psi, I)\, .
\ee
 More generally,
\be 
 \CC(\psi, I)    \geq   \CC^{(m)} (\psi, I) \, .
\ee
Thus $ \CC^{(m)}(\psi, I)$ lower-bounds  $\CC(\psi, I).$

\bn

We will first consider a protocol for applying the TN to the state $|I\ra \otimes |0\ra^{\otimes m}$ in order to prepare the state $|\psi\ra \otimes |0\ra^{\otimes m}.$ 
\fi

We begin on the left side of Fig.~\ref{Tlunch}, and working from left to right, apply the unitary gates. The isometries on the left `shoulder' of the Python's Lunch are straightforward; we simply couple in the ancilla and treat the isometries as unitary circuit elements.

But when we arrive at the right `shoulder' the isometries correspond to final state projections (post-selections). Final state projections are not physically implementable processes, so we must do something new. That new thing is measurement: as Alice arrives at each isometry she simply measures the dangling qubit in Fig.~\ref{isometry} in the $Z$ basis.\footnote{Of course, measurements are still not unitary and so are not traditionally allowed when calculating the complexity of a state. In contrast, the faster Grover-search protocol that we discuss below is genuinely unitary.} If she gets $0$ she moves on to the next isometry and repeats the process. If at the end of all the isometries all measurements give $0$ she moves on to the right of the TN and at the end she has prepared $|\psi\ra$.  

However, if she measures $1$ at some point she starts over and repeats the entire process, until she succeeds in obtaining $0$ for all post-selected qubits. On  average she will have to repeat the procedure $2^{m_R}$ times, for $m_R$ post-selected qubits. The total number of gates she will have to apply is of order  $$2^{m_R}  \cdot \CC_{\text{TN}} $$ where $\CC_{\text{TN}}$ is the number of nodes in the TN.

If this were the minimal protocol we would say that the complexity $\CC(\psi, I)$  is of order $2^{m_R}  \CC_{\text{TN}}$.\footnote{As before, this assumes that we allow measurements in our definition of complexity.}  However, in the appendix we show that there is a more efficient quantum procedure using  a version of 
Grover's algorithm (which was applied to a similar problem by Kitaev and Yoshida in Ref.~\cite{Yoshida:2017non}).

%\bn

\subsection{The Complexity of the Python's Lunch} \label{compOfLunch}

In Appendix \ref{subsec:projectonqudit}, we describe a protocol that uses Grover search to prepare the $|\psi\ra \otimes |0\ra^{\otimes m_{R}}$ from an initial state $\ket{I} \otimes \ket{0}^{m_L}$ with a unitary circuit using $2^{\frac{m_R}{2}} \CC_{\text{TN}}$ gates.  Assuming that the length of the lunch is greater than the scrambling time, there is no reason to think that the task can be performed with fewer gates, thus implying that 
\be 
 \CC (\psi, I) \approx 2^{\frac{m_R}{2}} \cdot \CC_{\text{TN}} \ . 
\ee 
Our initial version of this protocol works only for a single fixed input state $\ket{I}$. 

In contrast, we are really interested in finding a unitary operation $U_{PL}$ that satisfies
\begin{align} \label{eq:UPLG}
\ket{\psi} \ket{0}^{m_R} = U_{PL} \ket{I} \ket{0}^{m_L}
\end{align}
for any input state $\ket{I}$.\footnote{This requirement is not sufficient to specify $U_{PL}$ completely. Instead, our conjecture is that the minimal complexity of \emph{any} unitary operator, satisfying \eqref{eq:UPLG}, is given by \eqref{eq:ComplexUPL}.} In Appendix \ref{app:stateindep}, we such a `state-independent' protocol by a variant on the usual Grover-search algorithm. 

Our state-independent protocol succeeds with high probability for any input state $\ket{I}$, given either the assumption that there exists an exact isometry from left-to-right, or simply the assumption that the right system is parametrically larger than the left system and that the tensor network is scrambling (and so can be modelled using 2-designs). The complexity of this protocol is again given by $2^{\frac{m_R}{2}} \CC_{\text{TN}}$.

Since we have no good reason to think that a faster algorithm exists, we conjecture that 
\begin{equation} \label{eq:ComplexUPL}
\CC [U_{PL}] \approx 2^{\frac{m_R}{2}} \cdot \CC_\textrm{TN} \ . 
\end{equation}
Since we have assumed that $m_R$ is a finite fraction of $N,$ i.e., $m_R  = \beta N$, the complexity of $U_{PL} $ is exponential in $N.$

That is our main technical result: that the complexity of a TN is expected to be $O(2^{\frac{m_R}{2}} \mathcal{C}_{\text{TN}})$ where $m_R$ is the difference between the maximal area of the lunch and the area of the right side (or, more generally, the larger side). In particular when $m_R \sim N$ the complexity is exponential in the number of qubits $N$ at either end. 
The surprising point about this result is that the TN that prepares $\CC(U_{PL}) $ can be as small as $ \CC_{\text{TN}} \sim N\log{N}.$  We can now use the analogy between tensor networks and bulk geometry to conjecture a relationship between the restricted complexity and the geometry of a Python's Lunch. \\
\\
%\noindent
{\bf Restricted Complexity Conjecture:} In a Python's Lunch geometry with min-max-min areas $\mathcal{A}_L, \mathcal A_\textrm{max}$ and $\mathcal A_R$, and with the assumption $\mathcal A_L < \mathcal A_R$, we conjecture that the restricted complexity on the right system is 
\begin{align}
\textrm{\bf {conjecture}}: \ \ \  \CC_R[U_{PL}]= (const) \times \CC_\textrm{TN} \cdot \exp  \Big[ \frac{1}{2} \frac{(\mathcal{A}_\textrm{max} - \mathcal{A}_{R})}{4G \hbar} \Big] \ , \label{eq:bigconjecture}
\end{align}
where $\CC_\textrm{TN}$ denotes the size of the tensor network and is related to the volume/action of the wormhole ($\CC_\textrm{TN} = V/G\hbar l_{AdS}$) from the CV/CA conjectures \cite{Susskind2014, Brown2015, Brown2015-1}. 
\\
\\
In particular, if $A_L \approx N \cdot (4G\hbar) $,  $A_\textrm{max} \approx (1+\alpha)N \cdot (4G\hbar) $, and $A_{R} \approx (1+\gamma) N \cdot (4G\hbar) $ the complexity of decoding from the right system alone is
\begin{equation}
\CC_R[U_{PL}] \approx (const) \times \CC_\textrm{TN} \cdot e^{(\alpha - \gamma) N/2},
\end{equation}
where $\CC_\textrm{TN} \geq  N\log N$. 

\subsection{Post-selection is Superpolynomial}  \label{sec:superpolynomial}

In the previous subsection we argued that we can decode Hawking radiation by projecting out $m$ qubits, and provided a version of a Grover search \cite{Grover-1, Grover-2} that allows to do this projection with a unitary that has complexity $\sqrt{2^{m}}$. Can we rule out the possibility that there is an even faster algorithm that can perform this projection? 

On the one hand, we can almost certainly rule out the possibility that there could be an \emph{exponentially} faster algorithm. There cannot be an algorithm that projects onto $m$ qubits in a time that scales polynomially with $m$. Or -- more precisely --  if there \emph{were} such an algorithm then it would contradict widely held conjectures about computational complexity theory. The complexity class of decision problems you can solve on a quantum computer if you were allowed post-selections (including post-selections onto states with exponentially small amplitude) is called PostBQP. It has been shown \cite{PostBQP} that PostBQP is a fantastically powerful class -- it is equal to the class PP. Conversely, if you could use a normal quantum computer to implement exponentially rare projections in polynomial time, this would imply BQP=PostBQP. Taking these results together would imply BQP=PP. But PP is a very large class that contains all sorts of problems not believed to be efficiently soluble on a quantum computer, including NP.\footnote{Let us see how the ability to post-select would allow us to solve problems in NP. The classic NP-complete problem is 3SAT. An instance of 3SAT is a map $f(\vec{x})$ from $m$ Boolean variables to one Boolean variable (1=true or 0=false). Supposing there is exactly one assignment $\vec{x}$ that makes $f(\vec{x})$ come up true, our job is to find it. This problem is in NP because given a candidate $\vec{x}_\textrm{answer}$ it is easy to check whether $f(\vec{x}_\textrm{answer}) = 1$;  on the other hand it may be hard to find  $\vec{x}_\textrm{answer}$. On a quantum computer, we could test all the possibilities at once by building a circuit that implements $U_\textrm{circuit}|\vec{x} \rangle |0 \rangle = |\vec{x} \rangle |f(\vec{x}) \rangle$ and then plugging in an superposition over all possible inputs,
\begin{equation}
U_\textrm{circuit}\left( \frac{1}{\sqrt{d}} \sum_{\vec{x}} | \vec{x} \rangle |0 \rangle \right) = \frac{1}{\sqrt{d}}   | \vec{x}_\textrm{answer} \rangle |1 \rangle   +  \frac{1}{\sqrt{d}} \hspace{-8mm}  \sum_{\ \ \ \ \ \ \vec{x} \neq \vec{x}_\textrm{answer}}  \hspace{-5mm} | \vec{x} \rangle |0 \rangle ,
\end{equation}
but this generally doesn't help, because even though the wavefunction `knows' the answer, linearity means there is no measurement we can do that has more than an O($1/d$) chance of success that can induce the wavefunction to tell us what it knows. If we were able to project on the (exponentially in $m = \log d$) unlikely outcome that a measurement of the final qubit is $|1 \rangle$, then we could find $| \vec{x}_\textrm{answer} \rangle$ in one step.} It would therefore be in gross violation of widely held complexity assumptions if we could post-select in a time polynomial in $m$. 

On the other hand, it is not obvious how to rule out the existence of a \emph{polynomially} faster algorithm, that would still take a time exponential in $m$. It is not obvious that there can't be a protocol that would improve (say) the square root to a cube root.   (The effect of such a speed up would be to change the coefficient in the exponent of the conjecture Eq.~\ref{eq:bigconjecture} from 1/2 to 1/3.) It has been proved that Grover search amongst $d$ items cannot be implemented faster than $\frac{\pi}{4} \sqrt{d}$ \cite{Bennett:1996iu-1, Bennett:1996iu-2, Bennett:1996iu-3},\footnote{There are also other known lower bounds for more general classes of algorithms that perform amplitude amplification \cite{BCC15, Gilyen2017, Gilyen2019} and which are closely related to our state-independent protocol from Appendix \ref{app:stateindep}.}  but we have an advantage not available in the Grover task, which is that we know in advance which final state we wish to post-select on.

\subsection{Covariant Lunches} \label{sec:covariant}
So far in this paper, to determine  whether the spacetime contains a Python's Lunch we have implicitly assumed the existence of some preferred choice of bulk Cauchy slice. This is, in large part, a limitation of the tensor network toy models that we have been using to guide us and which resemble a bulk Cauchy slice rather than a full bulk spacetime. However, for the non-static spacetimes that we will be considering in future sections, it is not obvious how the correct slice should be chosen. 

In earlier work, the complexity was conjectured to be dual to the volume of the maximal volume slice. An obvious possibility is to work entirely within this slice. 

However, the covariant surface that is analogous to the minimal cut through a tensor network is  the HRT surface \cite{Ryu:2006bv, Hubeny:2007xt, Lewkowycz:2013nqa}, the minimal area extremal surface homologous to one end of the wormhole.  For spacetimes where quantum effects are important, such as evaporating black holes, it is, more precisely, the minimal generalized entropy quantum extremal surface,\footnote{The generalized entropy of a surface $\Sigma_B$ is $S^{\textrm{(gen)}}(\Sigma_B) = A(\Sigma_B)/4G\hbar +S_{\text{bulk}}(\Sigma_B)$, where the first term is the area of the surface $\Sigma_B$, which should be homologous to boundary region $B$, and the second is von Neumann entropy of the bulk fields contained in
the corresponding entanglement wedge (the region between $B$ and $\Sigma_B$). A quantum extremal surface is a surface that is an extremum of the generalized entropy.} also known as the Engelhardt-Wall (EW) surface \cite{Engelhardt:2014gca, Dong:2017xht}.

The existence of a second, locally minimal cut is analogous to the existence of a second extremal surface, satisfying the same homology constraint. In all the cases that we will consider, there will also be a third extremal surface, that lies in between the first two surfaces, and has a larger generalized entropy than either. This third surface has an important qualitative difference compared to the other two surfaces: we cannot choose a Cauchy slice within which any small (but not necessarily local) deformation of this third extremal surface will increase its area (or generalized entropy). In particular, this means that it cannot ever be the HRT (or EW) surface, which is always globally minimal within some Cauchy slice \cite{Wall:2012uf, QuantumMaximin}. Instead of corresponding to one of the narrow constrictions at the ends of the python, this third surface is a covariant definition of the maximum size of the bulge in the middle of the lunch.

In general, none of these surfaces will lie in the maximal volume slice. We therefore should not expect the correct covariant definition of a Python's Lunch to involve the maximal volume slice (although, in many examples, such as evaporating black holes, the maximal volume slice will also look like a Python's Lunch). Instead, we should think of a Python's Lunch as being \emph{defined} by this set of three extremal surfaces, the two `end surfaces' and the `bulge surface' in the middle. With this new covariant definition of Python's Lunch we can modify our conjecture.  \\

\noindent {\bf Restricted Complexity Conjecture (covariant version):} In a covariant Python's Lunch geometry with min-max-min generalized entropies $S^{\textrm{(gen)}}_L, S^{\textrm{(gen)}}_\textrm{max}$ and $S^{\textrm{(gen)}}_R$, and with the assumption $S^{\textrm{(gen)}}_L < S^{\textrm{(gen)}}_R$, the restricted complexity on the right system is, 
\begin{align} \label{conjCov}
\textrm{\bf {conjecture}}: \ \ \ \ \ \CC_R[U_{PL}]= (const) \times \CC_\textrm{TN}   \cdot \exp  \Big[ \frac{1}{2} \Big(S^{\textrm{(gen)}}_\textrm{max} - S^{\textrm{(gen)}}_R\Big) \Big],
\end{align}
where again $\CC_\textrm{TN}$ denotes the size of tensor network. 
% that is the action in the WdW patch from CA conjecture \HG{CITE}.
\\
\\
Naively, a covariant Python's Lunch seems a very specific and unusual feature of a spacetime. It needs to feature three extremal surfaces. Moreover the bulge surface needs to have greater area (or generalized entropy) than either end surface, and, unlike the end surfaces, within any Cauchy slice there should exist small deformations of the bulge surface that decrease its area (or generalized entropy). Nevertheless, every example that we consider of a spacetime with more than one extremal surface will turn out to have a Python's Lunch. 

In Appendix \ref{app:maximin}, we explain this phenomenon. We use `maximin' techniques \cite{Wall:2012uf, QuantumMaximin} to sketch an argument that almost all spacetimes with more than one extremal surface will contain a Python's Lunch. Specifically, we argue that one can generically find a third extremal surface by considering `foliations' of a Cauchy slice from one extremal surface to the other, taking the maximal area (or generalized entropy) surface within that foliation, minimizing the maximum over all foliations, and then maximizing the resulting `minimax' surface over all Cauchy slices. We call this a `maximinimax' prescription for finding the bulge surface.

\section{Evaporating Black Holes}
\sc 

In this section we will see how a Python's Lunch explains the exponential difficulty of decoding Hawking radiation. 

 After the Page time an evaporating black hole is maximally entangled with its own Hawking radiation \cite{Page:1993df}. Harlow and Hayden \cite{HarlowHayden} asked how hard it is to isolate a degree of freedom  $\r$ in the radiation  that is entangled with a particular quantum $\b$ of Hawking radiation that is about to be emitted by the black hole (the AMPS task \cite{Almheiri:2012rt}). A highly related task is to decode the state of a small unknown diary thrown into the (known) black hole, just from the state of the Hawking radiation\footnote{If the diary is maximally entangled with a reference system then the Hayden-Preskill task is the same as the Harlow-Hayden task except the degree of freedom $\bf{r}$ now needs to purify the reference system instead of a late-time Hawking quantum.}. This is the Hayden-Preskill decoding task \cite{Hayden:2007cs} and is also expected to be exponentially hard. If we can get the black hole and Hawking radiation into a simple state, both tasks are simple. The difficulty in doing either task comes from the exponentially large restricted complexity of the combined state of the black hole and Hawking radiation.

Building on earlier ideas in \cite{Almheiri:2018xdw, Hayden:2018khn}, it was shown in \cite{Penington:2019npb, Almheiri:2019psf, Almheiri:2019hni, PSSY, AHMST} that the information-theoretic \emph{achievability} (or otherwise) of the Hayden-Preskill and Harlow-Hayden tasks could be understood holographically using entanglement wedge reconstruction.\footnote{The idea of entanglement wedge reconstruction was originally conjectured in \cite{Headrick:2014cta,Czech:2012bh,Wall:2012uf}, and established in \cite{Jafferis:2015del,Dong:2016eik,Cotler:2017erl} using the ideas of \cite{Faulkner:2013ana, Almheiri:2014lwa}.} After the Page time, a large part of the interior of the black hole is in the entanglement wedge of the early Hawking radiation, and so is encoded in the radiation.\footnote{In this context, the entanglement wedge of the Hawking radiation is defined as the bulk domain of dependence bounded by the EW surface.} This is essentially a holographic derivation of black hole complementarity and ER=EPR \cite{Susskind:1993if,Maldacena:2013xja}.
We shall now see that the exponential computational \emph{difficulty} of the Harlow-Hayden and Hayden-Preskill tasks can likewise be understood holographically as coming from the existence of a Python's Lunch.

\subsection{Preliminary Example} 
\begin{figure}[t]
\begin{center}
\includegraphics[scale=.3]{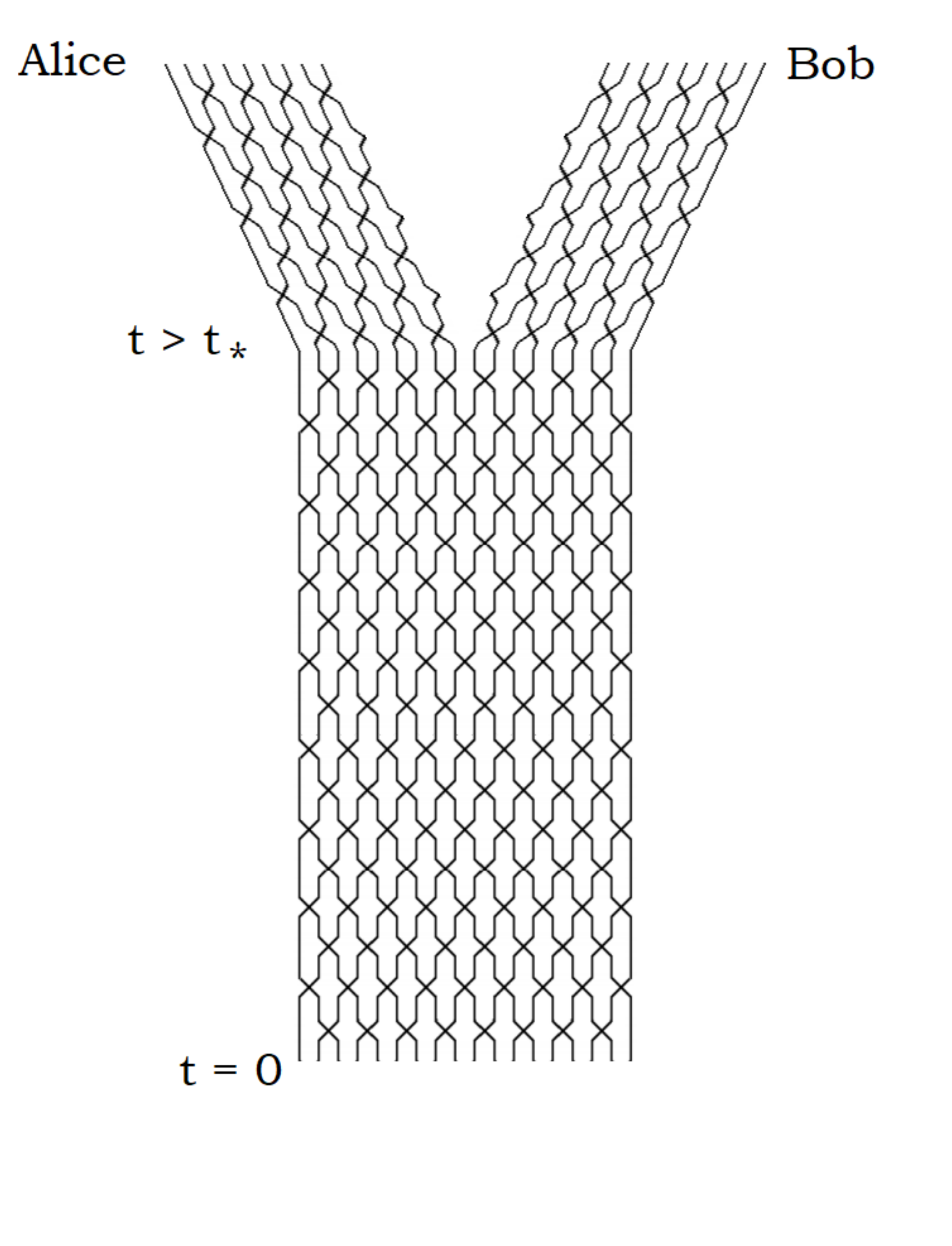}
\caption{A scrambled system of $2N$ qubits can be separated into Alice's and Bob's subsystems which due to the scrambling are close to a maximally entangled state. Since they are maximally entangled, it must be possible to shorten the wormhole by restricted unitary operations. }
\label{split}
\end{center}
\end{figure}

\begin{figure}[t]
\begin{center}
\includegraphics[scale=.3]{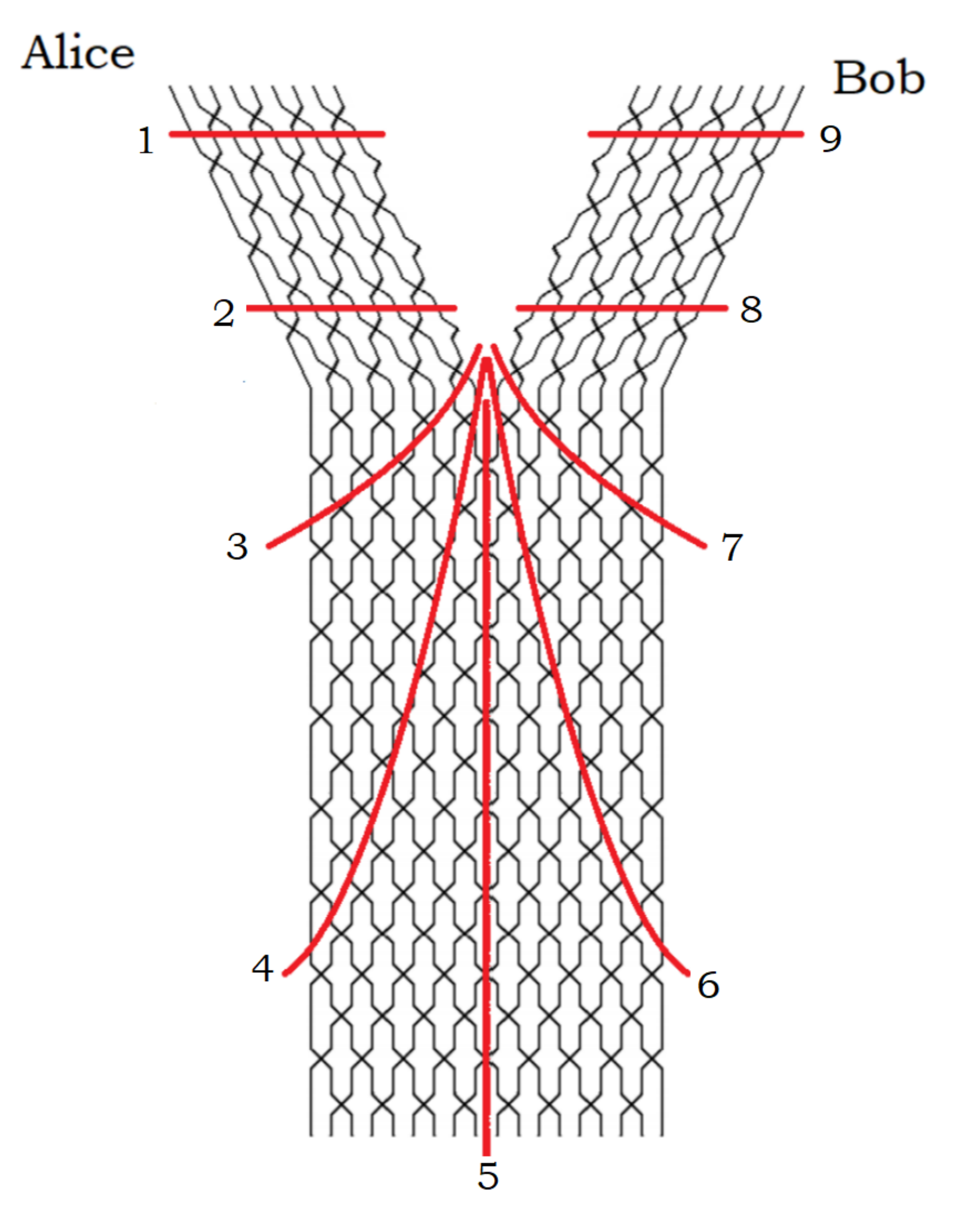}
\caption{A foliation of the tensor network of Fig.~\ref{split} which interpolates between Alice's side and Bob's side. It is clear from the figure that the number of qubits cut by the slices increases and then decreases as the foliation sweeps round. }
\label{split2}
\end{center}
\end{figure}

We will consider a preliminary example.
Consider a quantum computer initialized at $t=0$ in some simple state $|I\ra$ which then evolves for a time greater than the scrambling time $t_*.$  The qubits are then split into two subsystems, Alice's and Bob's shares $A$ and $B.$ The two subsystems continue to evolve for a short time  but with no coupling between them. The process is illustrated\footnote{The black-hole dynamics is better characterized by a 2-local circuit \emph{without} geometric locality, but due to the authors' artistic limitations we will present the illustrations in a geometrically local 1d lattice.} in Fig.~\ref{split}. 
Let us imagine sweeping across the TN by a series of cuts which foliate it as in Fig.~\ref{split2}.
It is obvious that the number of qubits crossed by the cuts first increases and then decreases. At its maximum the number of qubits in at least $N\log{N}.$ Therefore the geometry of the associated wormhole has a Python's Lunch.

Because the system is scrambled at time $t$ the subsystems $A$ and $B$ are approximately maximally entangled. It follows that Alice can act unitarily on her side in order to bring the system to a state close to the TFD. The arguments of the previous section show that the restricted complexity is exponential in $N$ although the number of vertices in the TN is much smaller.

\subsection{Hawking Radiation} \label{sec:evaporation}
\begin{figure}[t]
\begin{center}
\includegraphics[scale=.35]{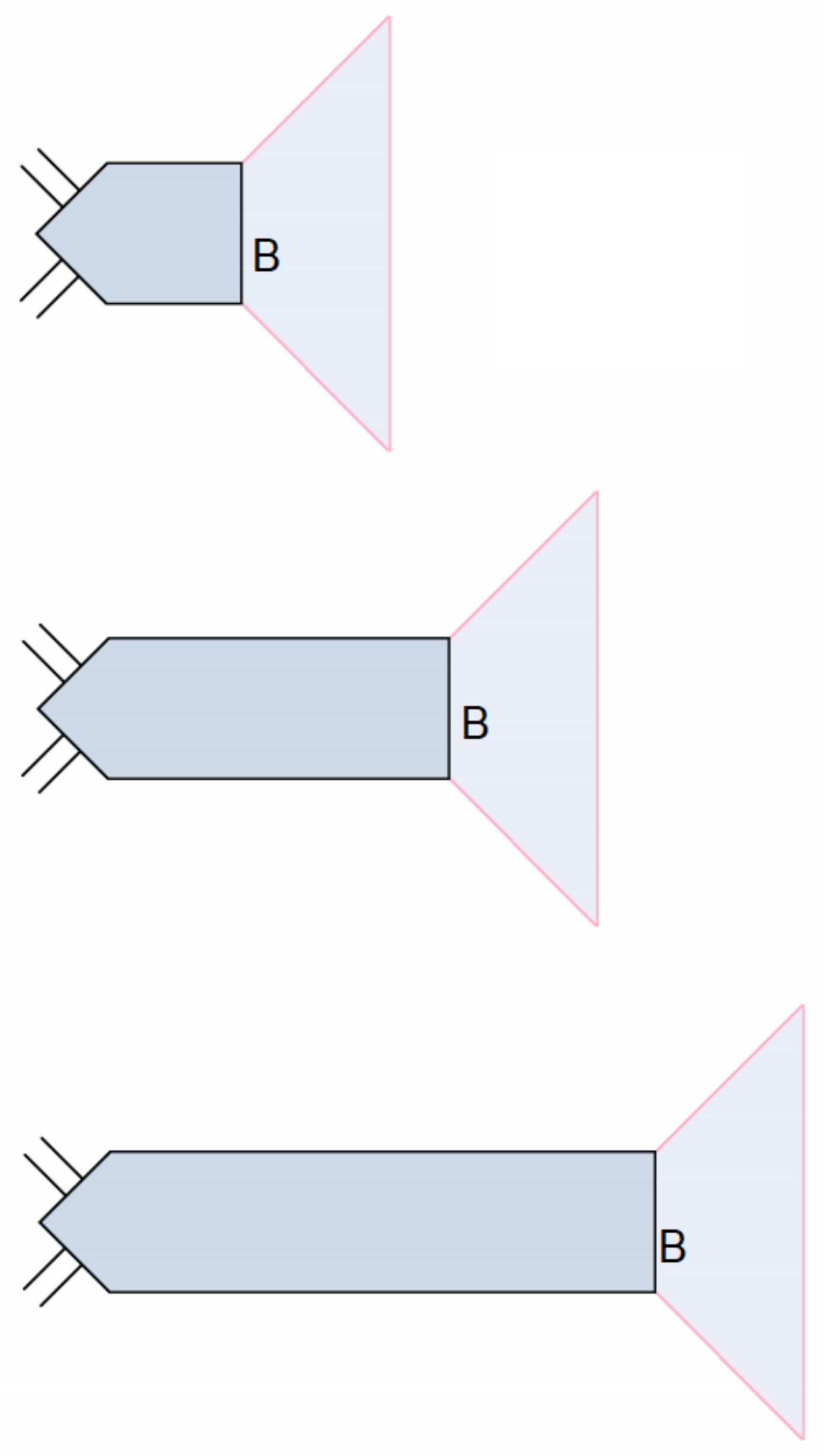}
\caption{Successive spatial slices through a one-sided non-evaporating black hole that formed from collapse. The ``whiskers" on the left side depict the infalling matter. The `bridge to nowhere' grows and becomes elongated as the complexity increases.     }
\label{B1}
\end{center}
\end{figure}

In this section we will explain how the Python's lunch geometry appears during the evaporation of a black hole. Note that, for the moment, we are restricting our attention to a single Cauchy slice through this black hole---a generic `nice' Cauchy slice that stays close to the black-hole horizon. For concreteness we can take it to be the maximal volume slice. In Sec.~\ref{sec:covariantevaporatinglunch}, we will discuss the full covariant description of this lunch.

A classical  one-sided black hole (Bob's black hole) in a pure state is not connected to any purifying system by a wormhole. But it is connected to a growing ``bridge to nowhere" (BTN) whose volume represents the complexity of the state. This is shown schematically  in Fig.~\ref{B1}.

Starting at the horizon and moving into the BTN the area remains constant for most of its length until it quickly shrinks at the far end. The whisker-like lines at the end represent the infalling matter which originally created the black hole. As time increases the BTN grows.

\begin{figure}[t]
\begin{center}
\includegraphics[scale=.3]{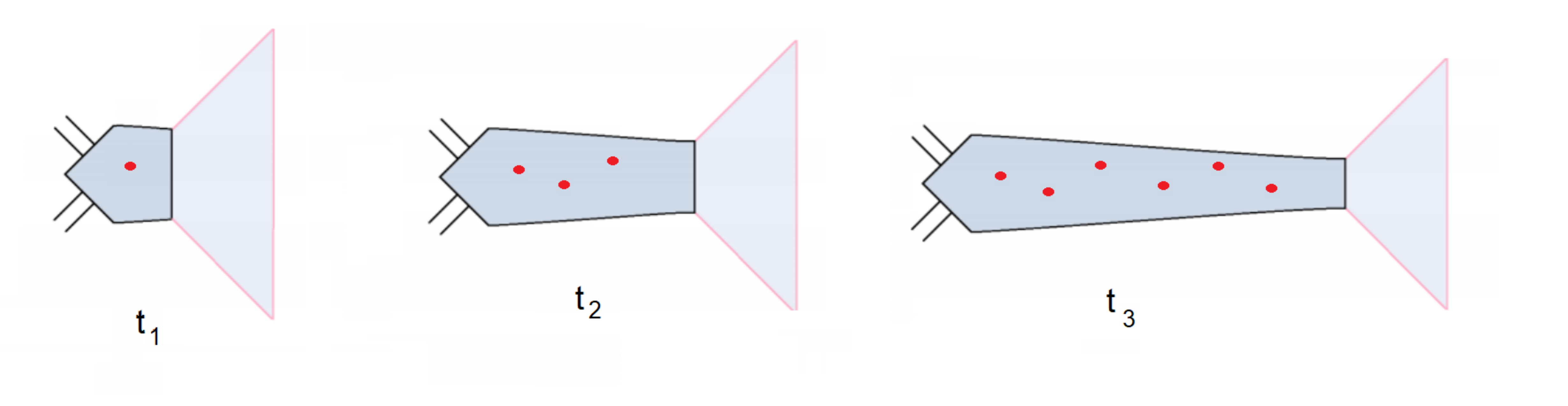}
\caption{ Successive spatial slices through a one-sided evaporating black hole that formed from collapse. The red dots mark where Hawking particles were emitted. }
\label{B2}
\end{center}
\end{figure}

Evaporation modifies this picture and effectively turns the one-sided system into a two-sided system. The black hole becomes entangled with its own Hawking radiation which in effect becomes a second side.  The process which was explained in \cite{Maldacena2013} is depicted in figure 
 \ref{B2} as a time-sequence at times $t_1<t_2<t_3$.
 As the black hole radiates the area of its horizon decreases. In the figure this is shown as a decrease in the thickness of the BTN as one  moves from left to right. The interior modes that purify Hawking radiation are shown as red dots. The partners of Hawking radiation emitted later in the evaporation are at the furthest right of the diagram. These interior modes are entangled with the Hawking radiation, and so they have an entropy that contributes to the generalized entropy of any region containing them, but not containing the Hawking radiation, or vice versa. Alternatively, in the language of ER=EPR, they can be thought as being connected by a Planck-area ``micro-wormhole'' to the Hawking radiation. The homology constraint forces us to cut these micro-wormholes, which increases the generalized entropy.
 
\begin{figure}[t]
\begin{center}
\includegraphics[scale=.3]{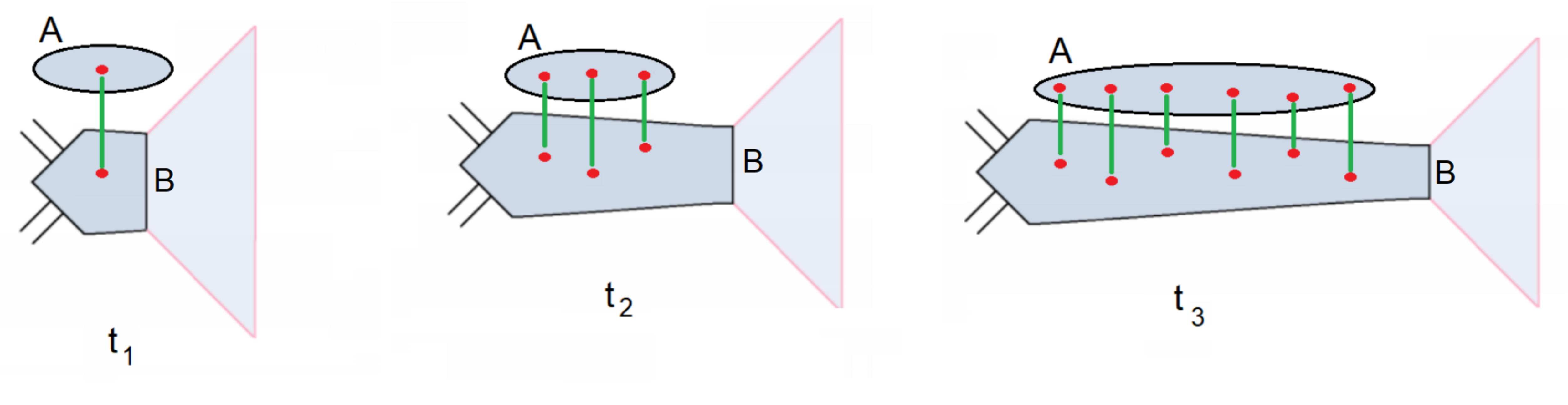}
\caption{The `expandable space blimp'. The one-sided wormhole from Eq.~\ref{B2}, now showing the bulk entanglement connecting  the distant Hawking particles to their anchor points on the red dots.  }
\label{B25}
\end{center}
\end{figure}

Now suppose that Alice collects the radiation in a second system $A$  shown as an elongated ellipse in  Fig.~\ref{B25}. System $A$ will now be connected to the bridge-to-nowhere via either bulk entanglement/micro-wormholes.

Let us assume that, like a tensor network, the entropy of the Hawking radiation is given by the `minimal cut' through this Cauchy slice, where the size of a cut is given by its generalized entropy. Of course, in general, this will only be true if we have chosen our Cauchy slice appropriately. However, we will be able to derive the correct qualitative conclusions by just studying the maximal volume slice.

\begin{figure}[t]

\begin{center}
\includegraphics[scale=.52]{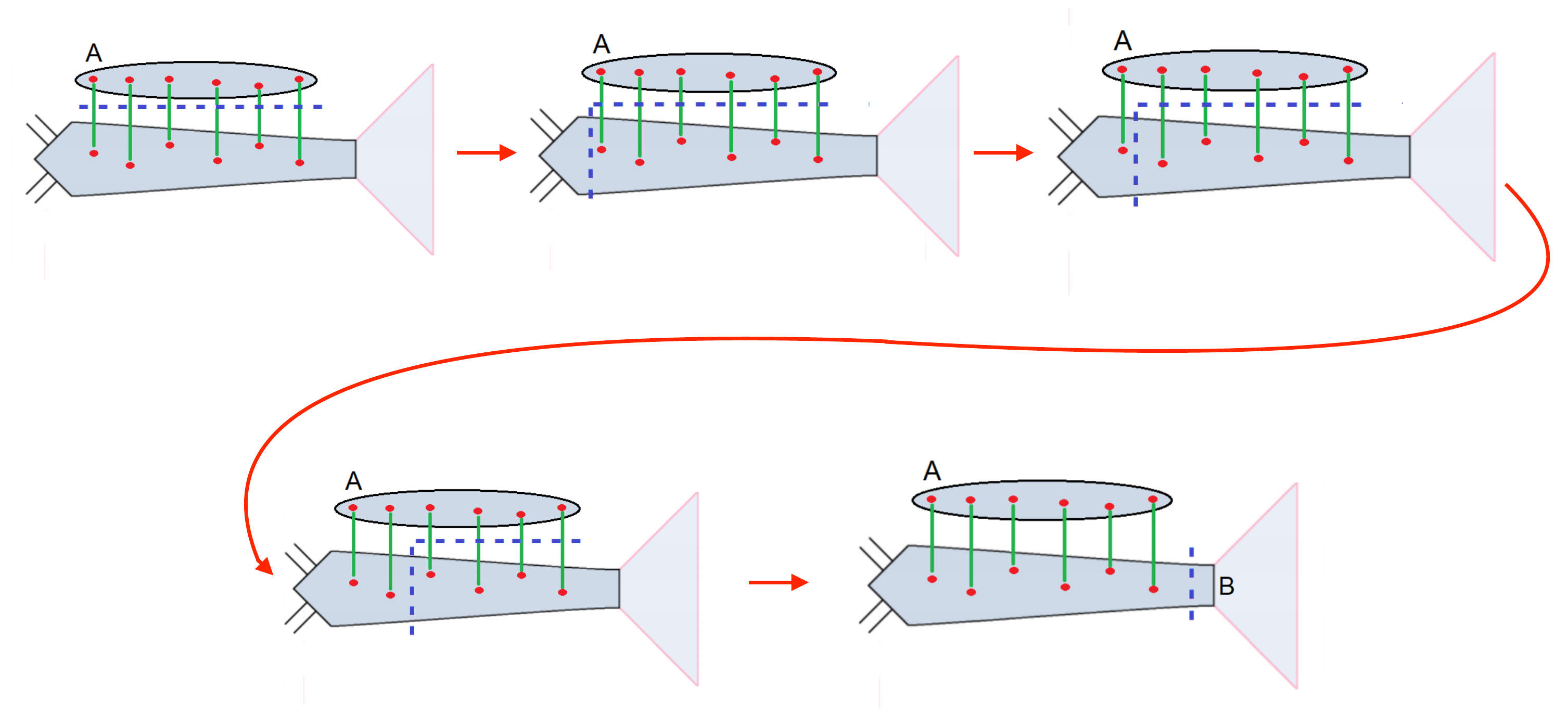}
\caption{A non-temporal sweep of spatial cuts through the expandable space blimp geometry of Fig.~\ref{B25}, analogous to the sweep of the spatial slices in Fig.~\ref{split2}.   }
\label{c5}
\end{center}
\end{figure}

We construct ``cuts" separating $A$  from the boundary at the right end of the ``fish tail." The cuts can be characterized by an area. In Fig.~\ref{c5},  we see  a series of such cuts as we sweep from $A$  to the fishtail. 

%%%%

The generalized entropy of each cut consists of  two contribution. One is the portion that cuts through the bulk entanglement between the interior and the Hawking radiation. The second contribution comes from the classical area required to cut the bridge-to-nowhere. Let us track the generalized entropy as the cut proceeds:

\begin{itemize} 
\item
In the first cut in Fig.~\ref{c5} the only contribution to the generalized entropy comes from the bulk entanglement. That contribution is proportional to the entropy in the radiation.

\item
The next cut also cuts the entanglement between the interior and the radiation. However, it also cuts across the largest part of the BTN. We see that there is a quick increase in the generalized entropy of the cut. 

\item In the third and fourth cuts in Fig.~\ref{c5}, the cut moves to the right. As it does so the both contributions to the generalized entropy decrease.

\item
The final cut in Fig.~\ref{c5} only cuts across the bridge-to-nowhere near the horizon of the black hole.

\end{itemize} 

\begin{figure}[t]
\begin{center}
\includegraphics[scale=.45]{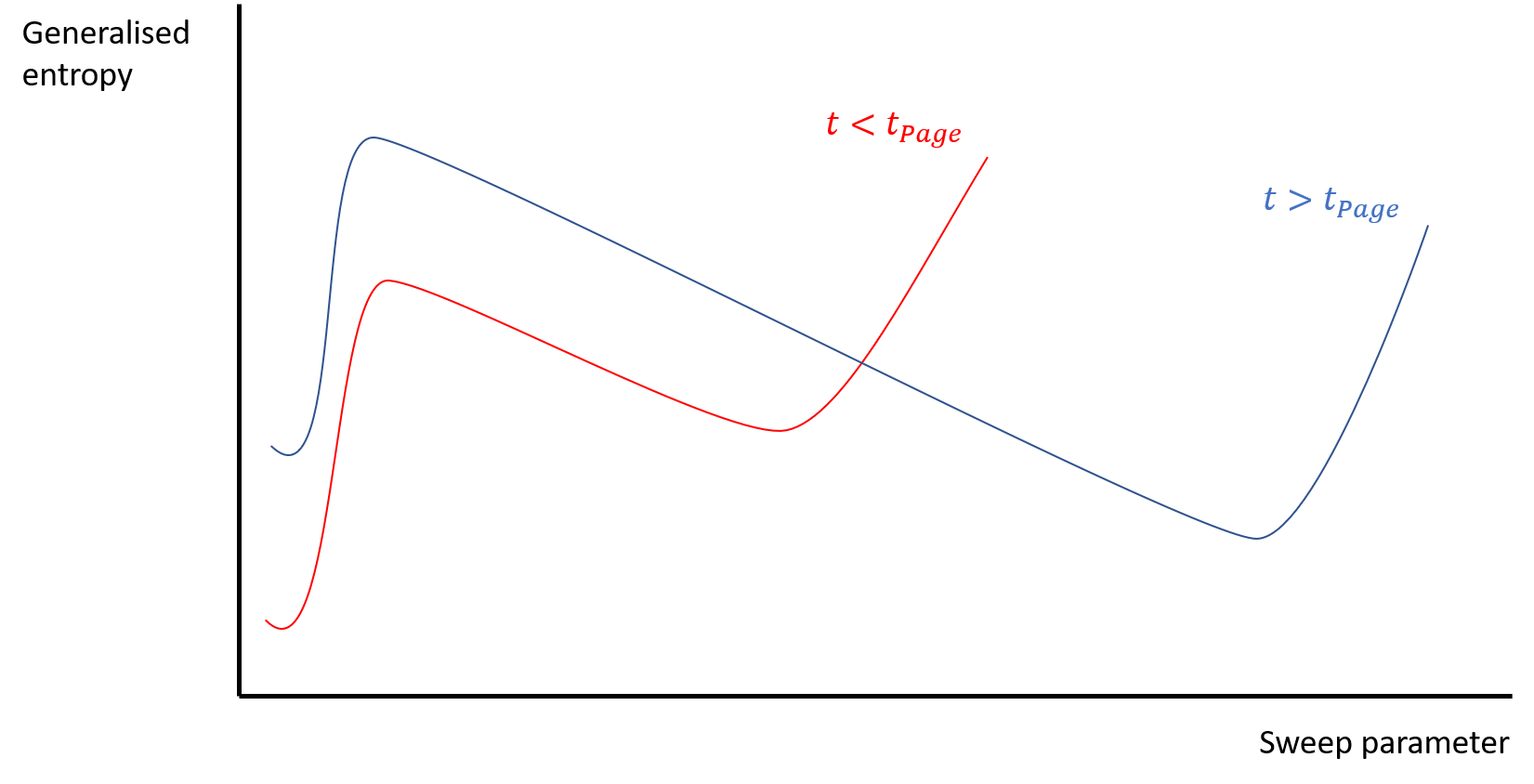}
\caption{ The size of the cut for the slices of Fig.~\ref{c5} as a function of the sweep parameter. For $t < t_\textrm{page}$, the minimum generalized entropy cut is at the beginning, when the bulk entanglement between the black hole and the Hawking radiation is being cut. For $t < t_\textrm{page}$, the minimum generalized entropy cut is near the end, when the cut is near the black hole horizon. In both cases, the largest cut comes near the beginning, when the generalized entropy is the sum of the semiclassical entropy of the radiation, plus the initial Bekenstein-Hawking entropy of the black hole.}
\label{sweep}
\end{center}
\end{figure}

The evolution of the generalized entropy of the cut is shown in Fig.~\ref{sweep}. 

There are two minima to the generalized entropy. One is the cut through the bulk entanglement represented by the green lines in the top picture. As we sweep across the generalized entropy makes a fairly sudden increase, and then a slow gradual decrease to a second minimum -- the horizon -- at the fishtail.  Up to subtleties involving the choice of Cauchy slice, these two minima correspond to the two quantum extremal surfaces found in \cite{Penington:2019npb, Almheiri:2019psf}.

Early on, the horizon area $A_\text{hor}$ (or, more precisely, the Bekenstein-Hawking entropy $A_\mathrm{hor}/4G_N$) is much larger than the bulk entanglement between the interior and the Hawking radiation. At a very late time the horizon shrinks to zero while the bulk entanglement becomes very large. 

At some point there is a crossover where the two minima are degenerate. This defines the Page time. Because the evaporation is irreversible, this happens when the horizon area $A_\text{hor}$ is slightly larger than half its initial area $A_0$ (for Schwarzschild black holes in our universe it happens when $A_\text{hor} \sim 0.6 A_0$ \cite{Page:2013dx}).

The important point is that  the geometry has a Python's Lunch separating the two minima. The generalized entropy at the maximum of the bulge is 
\begin{align}
\frac{A_0}{4 G_N} + S_\text{rad}
\end{align}

Now, suppose that, just after the Page time, Alice, who controls the Hawking radiation, wants to apply gates or a Hamiltonian to shrink the wormhole to the TFD associated with the black hole of area $A_\text{hor}.$  Assuming that the analogy between a tensor network and the Cauchy slice holds, the protocol in the appendix she can do so in a time that is $O(S)e^{A_0/8 G_N}$.  This is consistent with the restricted complexity being exponentially large.
 
Of course, so far we have only considered one way of sweeping through the Cauchy slice, from one minimal cut to the other. If we could find another way of sweeping through the slice that had a smaller maximal generalized entropy, it would suggest that a more efficient protocol exists, since in a tensor network less post-selection would be required.
 
\begin{figure}[t]
\begin{center}
\includegraphics[scale=.2]{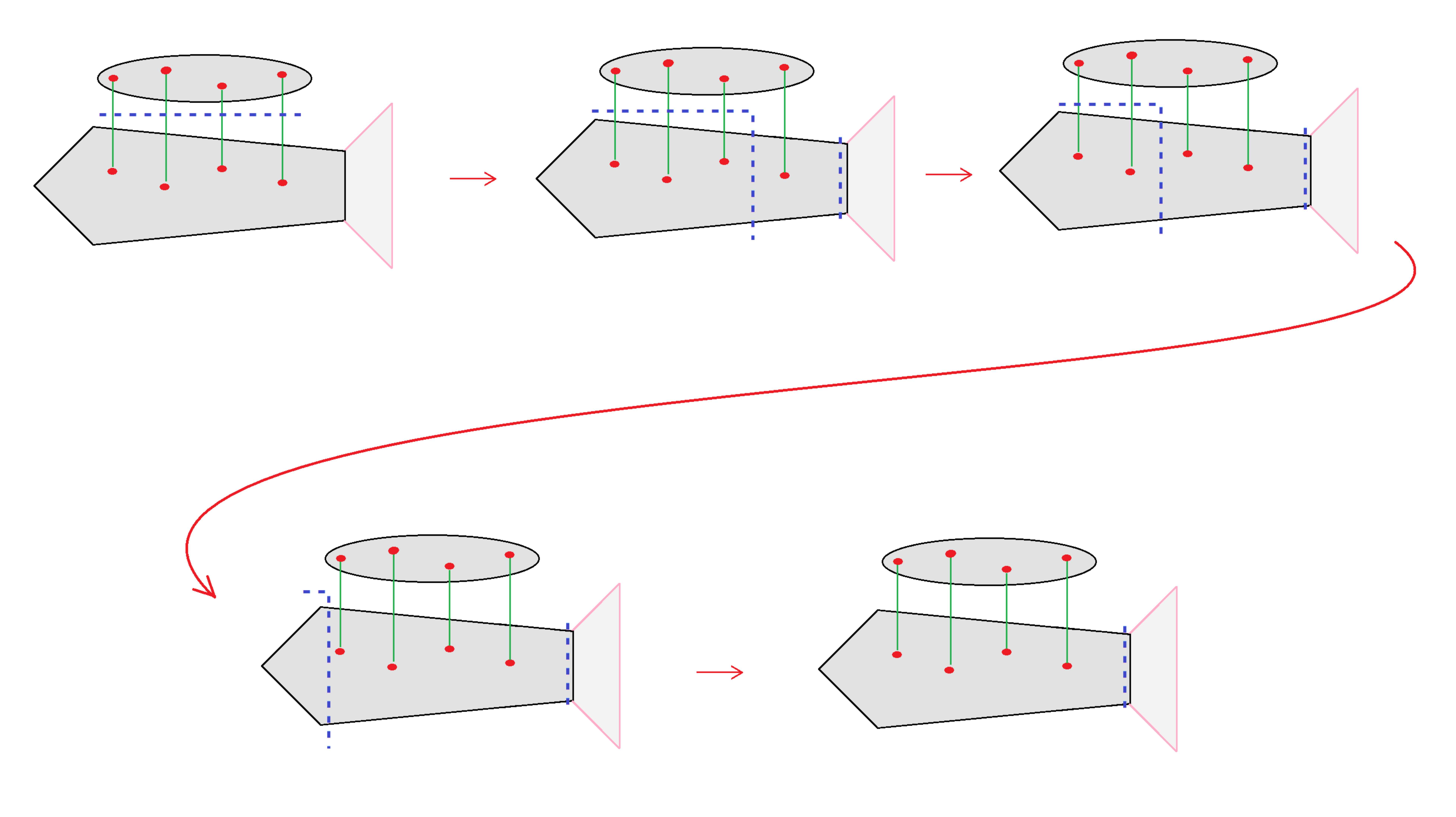}
\caption{At late times, a more efficient way of sweeping between the two minimal cuts is to create two cuts near the horizon, and then to sweep one of these cuts ``backwards'' along the wormhole.}
\label{fig:reversesweep}
\end{center}
\end{figure}

In fact, as far as we can tell, the way of sweeping through the slice just described should be optimal both before and shortly after the Page time. However, at late times, an alternative way of sweeping through the slice becomes preferable. We now analyze this second way of sweeping through the slice, which is shown in Fig.~\ref{fig:reversesweep}.

\begin{figure}[t]
\begin{center}
\includegraphics[scale=.45]{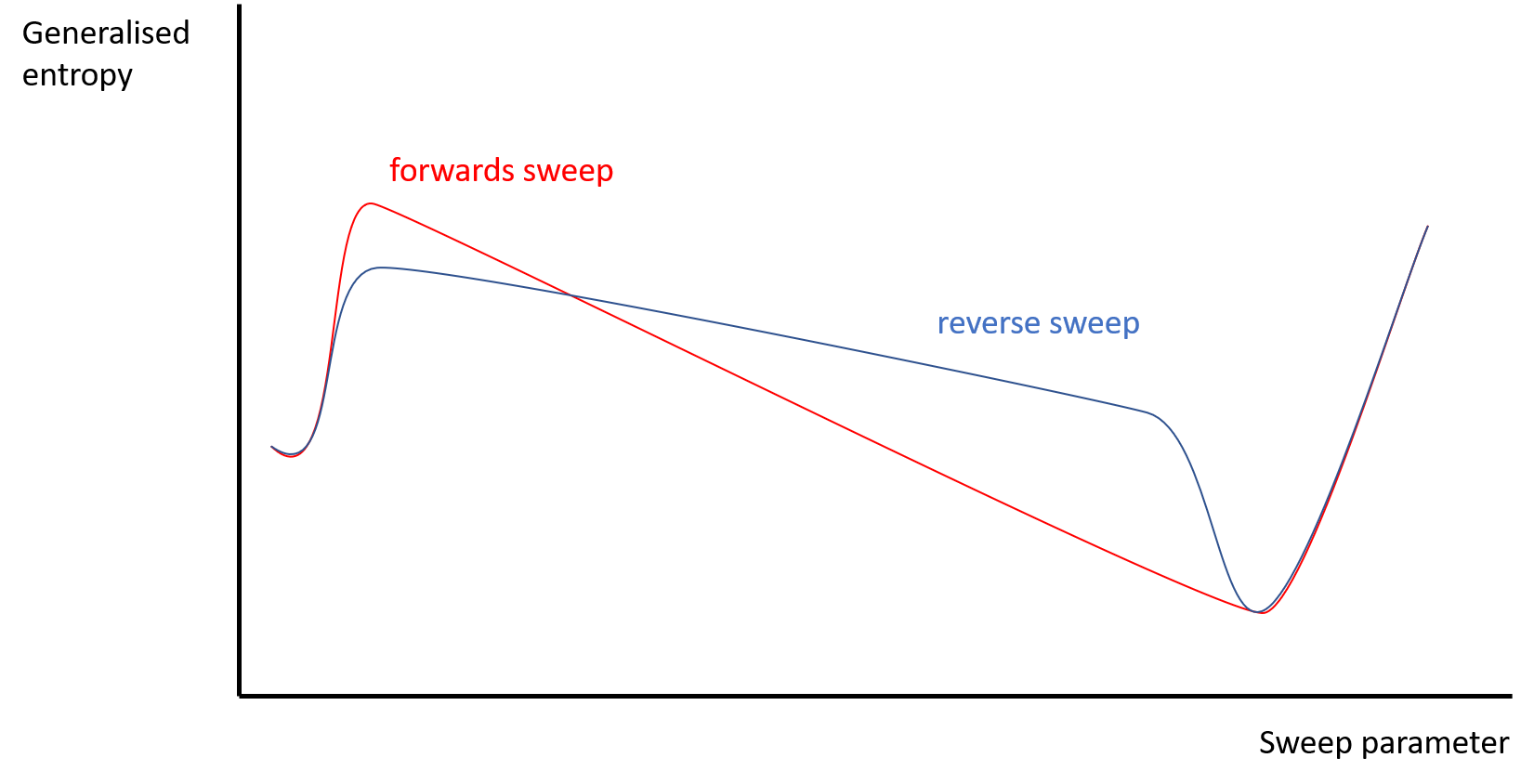}
\caption{A comparison of the generalized entropy as a function of sweep parameter for the `forwards' (Fig.~\ref{c5}) and `reverse' (Fig.~\ref{fig:reversesweep}) ways of sweeping through the bridge to nowhere. Initially, the generalized entropy of both is given by $S_\text{rad}$. The forwards sweep quickly increases by $A_0 / 4 G_N$ and then steadily decreases as the sweep moves along the bridge-to-nowhere. The reverse sweep quickly increases by $2 A_\text{hor} / 4G_N$, slowly decreases as the cut moves backwards along the bridge-to-nowhere, and then finally quickly decreases by $A_0/4 G_N$. The reverse sweep has a smaller maximum size, and hence is more efficient, when $A_\text{hor} < A_0 /2$.}
\label{sweep2}
\end{center}
\end{figure}
\begin{itemize} 
\item
In the first cut in Fig.~\ref{fig:reversesweep} the only contribution to the generalized entropy comes from the bulk entanglement and is proportional to the entropy in the radiation, as before.

\item
The next cut also cuts the bulk entanglement between the interior and the radiation. However, it also includes an additional ``double cut'' near the horizon. This gives an additional area contribution equal to $2 A_\text{hor}/4G_N$.

\item In the third and fourth cuts in Fig.~\ref{fig:reversesweep}, one half of the double cut moves to the left. As it does so, its area increases, but the bulk entanglement decreases. Because black hole evaporation is irreversible, the second effect is slightly larger than the first, and so the generalized entropy slowly decreases.

\item
Finally, the left-moving cut reaches the end of the bridge-to-nowhere and disappears. The generalized entropy therefore has a sharp decrease by $A_0/4G_N$.
\end{itemize}

The evolution of the generalized entropy of the cut is shown in Fig.~\ref{sweep2}. The generalized entropy quickly increases as the double cut is added, reaching its maximum size of
\begin{equation}
S_\text{rad} + 2 \frac{A_\text{hor}}{4 G_N}.
\end{equation} 
It then slowly decreases, before a final sudden drop by $A_0/4 G_N$.

As shown in Fig. \ref{sweep2}, this method of sweeping through the bridge-to-nowhere is therefore more efficient than the forwards sweep when $A_\text{hor} < A_0/2$. Note that this transition happens strictly after the Page time, which, although commonly described as happening at the halfway point in the evaporation, happens when $A_\text{hor} > A_0 /2$.

\subsection{The Covariant Lunch} \label{sec:covariantevaporatinglunch}
Of course, as discussed in Sec. \ref{sec:covariant}, the correct covariant definition of a Python's Lunch is not given by the shape of a single Cauchy slice, but by a set of three quantum extremal surfaces, the two end surfaces and the bulge surface in the middle.  However, as discussed in Appendix \ref{app:maximin}, there exist Cauchy slices within which the most efficient `sweep' has locally minimal generalized entropy  at the end surfaces, and locally maximal generalized entropy at the bulge surface. We can think of these as the Cauchy slices where the tensor network analogy is valid.

For an evaporating black hole, the location of the end surfaces were calculated in \cite{Penington:2019npb, Almheiri:2019psf}. The first end surface is the empty surface, containing no points. Its generalized entropy is simply the semiclassical entropy $S_\text{rad}$ of the Hawking radiation. The second end surface, which becomes the EW surface after the Page time, lies at a radius that is $O(G_N)$ inside the horizon of the black hole, and at an infalling, or retarded, time that is one scrambling time in the past of the current boundary time. Its generalized entropy is given at leading order by the Bekenstein-Hawking entropy $A_\text{hor}/4 G_N$.
\begin{figure}[t]
\begin{center}
\includegraphics[scale=.6]{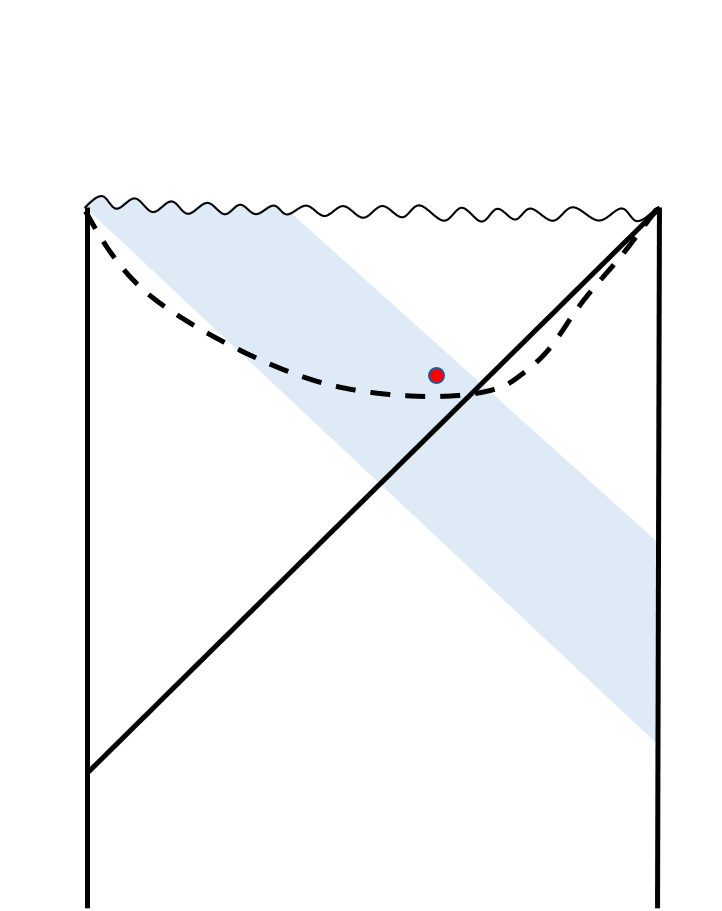}
\caption{The covariant bulge surface at early times (i.e.~for the `forwards' sweep) lies inside the infalling matter that forms the black hole. It lies a Planckian radial distance inside the apparent horizon (dashed line), at an infalling time when the apparent horizon is itself a Planckian radial distance inside the event horizon (solid line). Its generalized entropy is approximately $A_0/4G_N + S_\text{rad}$.}
\label{fig:bulgesurface}
\end{center}
\end{figure}

The bulge surface was obviously less of a focus in \cite{Penington:2019npb,Almheiri:2019psf}, because it is never the EW surface. However, the bulge surface that corresponds to the maximum size in the `forwards sweep' was briefly discussed in \cite{Almheiri:2019psf}. For a one-sided black hole formed from collapse, it lies inside the infalling matter that created the black hole. As the black hole forms, a classical apparent horizon appears that moves outwards in a spacelike direction towards the lightlike event horizon (which, being teleological, formed before the infalling matter even arrived). After the infalling matter has passed through and the black hole begins to evaporate, the classical apparent horizon ends up a Planckian radial distance \emph{outside} the event horizon. When no greybody factors are present, we can use Eqs.~89 and 90 from \cite{Penington:2019npb} to see that there will exist a quantum extremal surface, at sufficiently late times, at a radius
\begin{align} \label{eq:prediaryr}
r = r_s(v) - \frac{G_N c_\text{evap}}{3 (d-1) \Omega_{d-1} r_s^{d-2}},
\end{align}
where $r_s(v)$ is the Schwarzschild radius (and hence the radius of the apparent horizon) in the ingoing Vaidya metric describing the black hole, $c_\text{evap}$ is the number of two-dimensional bosonic modes (i.e.~number of angular momentum modes in higher dimensions) involved in the evaporation, and $\Omega_{d-1}$ is the volume of the unit $(d-1)$-sphere, and at an infalling time $v$ when
\begin{align} \label{eq:apparentevent}
r_\text{hor}(v) = r_s(v) + \frac{G_N c_\text{evap}}{3 (d-1) \Omega_{d-1} r_s^{d-2}}.
\end{align}
What is the infalling time $v$ when is Eq.~\ref{eq:apparentevent} satisfied? Since the apparent horizon $r_s(v)$ goes from far inside to a Planckian distance outside the event horizon $r_\text{hor}(v)$, a solution must exist (assuming the metric is smooth) somewhere inside the infalling matter. Generally it will be near the end of the infalling matter, when the apparent horizon has approached within a Planckian radial distance of the event horizon.

The early time bulge surface is shown in Fig.~\ref{fig:bulgesurface}. We note that its generalized entropy is indeed approximately equal to the initial Bekenstein-Hawking entropy of the black hole, plus the entropy of the Hawking radiation, as expected from our analysis of the maximal volume Cauchy slice (see \cite{Penington:2019npb, Almheiri:2019psf} for explicit calculations).

\begin{figure}[t]
\begin{center}
\vspace{-1.5cm}
\includegraphics[scale=.6]{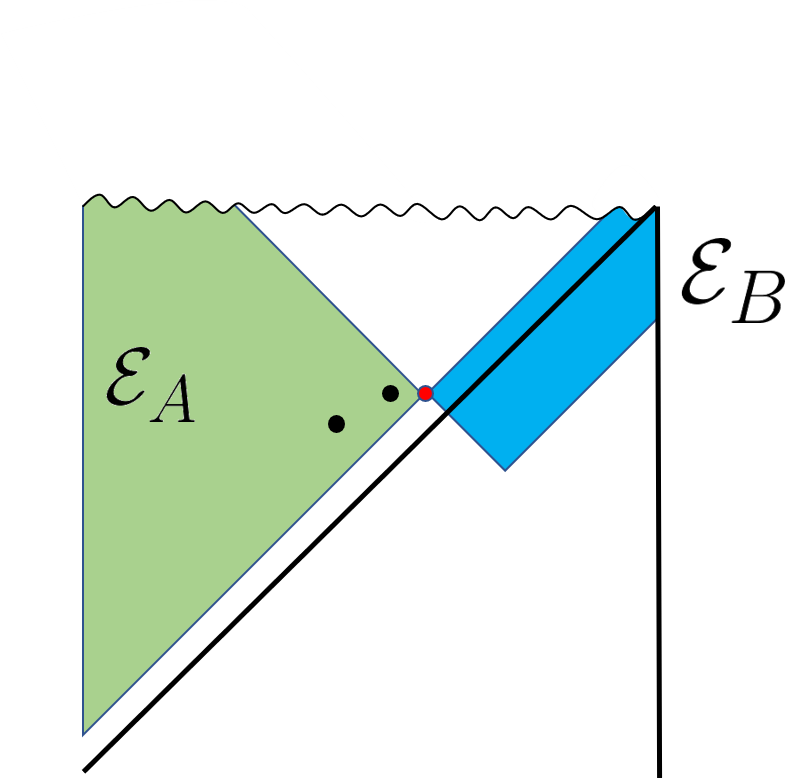}
\caption{The covariant bulge surface (black) at late times (i.e.~for the `reverse' sweep) consists of two spheres (or points in two dimensions). Both lie slightly inside the EW surface (red), which determines  the Hawking radiation entanglement wedge $\mathcal{E}_A$ (green) and the entanglement wedge $\mathcal{E}_B$ of the system containing the black hole (blue).}
\label{fig:latetimebulge}
\end{center}
\end{figure}

What about at late times, when our analysis of the maximal-volume slice suggested that a `reverse sweep' through the bridge-to-nowhere was optimal? In this case, we expect that the dominant bulge surface should consist of the union of two spheres, both close to the late-time horizon. In general, calculating the location of an extremal surface with this topology is considerably harder than finding extremal surfaces consisting of a single sphere. However, it is possible for JT gravity plus free Dirac fermions. We study this theory in Appendix \ref{app:explicitbulge}. As hoped, we find an extremal surface that consists of two points (or zero-spheres). Both points lie one scrambling time in the infalling past of the current boundary time, just like the late-time EW surface. However, as shown in Fig.~\ref{fig:latetimebulge}, both points are spacelike separated from, and further inside the horizon than, the EW surface. This is consistent with this surface being the maximum generalized entropy surface in a sweep through a Cauchy slice that goes from the empty surface to the late-time EW surface. Consistent with our expectations based on a single Cauchy slice, the generalized entropy of this surface is approximately $2 A_\text{hor}/4G_N + S_\text{rad}$.

\subsection{The Time Dependence of the Restricted Complexity}
\begin{figure}[t]
\begin{center}
\includegraphics[scale=.8]{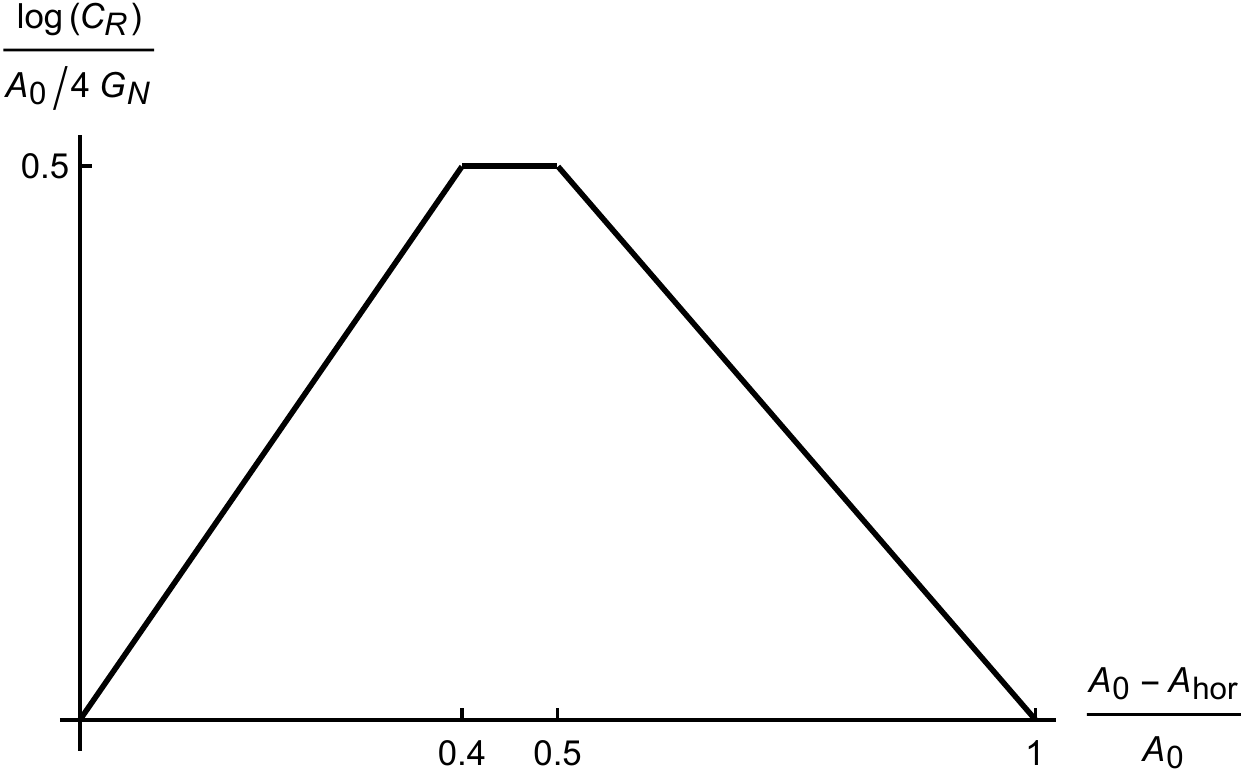}
\caption{A plot of our conjecture for $\log \, \mathcal{C}_R$ over the course of the black hole evaporation. For the purposes of this plot we have assumed $S_\text{rad} = 1.5 (A_0 - A_\text{hor})$.}
\label{fig:logcomp}
\end{center}
\end{figure}

Using our covariant restricted complexity conjecture from Sec. \ref{sec:covariant}, we can now make a precise conjecture about how the restricted complexity of the evaporating black hole state evolves over the course of the evaporation. We show a plot of $\log \mathcal{C}_R$ against $A_0 - A_\text{hor}$ over the course of the entire evaporation in Fig.~\ref{fig:logcomp}.

There are three distinct phases to this evolution. Before the Page time, the EW surface is empty, with generalized entropy $S_\text{rad}$, while the bulge surface is the forwards-sweep surface, with generalized entropy $S_\text{rad} + A_0 /4G_N$. Finally, the larger end surface has generalized entropy $A_\text{hor}/4G_N$. According to our conjecture in Eq.~\ref{conjCov},  the restricted complexity is controlled by the difference in generalized entropy between the bulge surface and the larger end surface. It is given by
\begin{align}
\mathcal{C}_R = O\Big(t \,\text{exp}\Big[\frac{1}{2} \Big(S_\text{rad} + \frac{A_0 - A_\text{hor}}{4 G_N}\Big)\Big]\Big).
\end{align}
The factor of $t$ here comes from the volume of the lunch, which controls $\CC_\textrm{TN}$ according to the conjectures of Refs. \cite{Susskind2014, Brown2015, Brown2015-1}.

The first phase transition happens at the Page time, where the EW surface becomes nonempty. This means that the larger end surface becomes the empty surface, with generalized entropy $S_\text{rad}$. The restricted complexity is therefore
\begin{align}
\mathcal{C}_R = O\Big(t \,\text{exp}\Big[\frac{A_0}{8 G_N}\Big]\Big),
\end{align}
and only changes linearly with time.

Finally, when $A_\text{hor} = A_0 / 2$, we have a second phase transition. This time, it is the bulge surface that changes. It becomes the reverse-sweep surface, with generalized entropy $S_\text{rad} + 2 A_\text{hor}/ 4 G_N$. The restricted complexity begins to decrease, and is given by
\begin{align}
\mathcal{C}_R = O\Big(t \,\text{exp}\Big[\frac{A_{\text{hor}}}{4 G_N}\Big]\Big).
\end{align}
Importantly, as the black hole completely evaporates the exponent tends to zero, and the restricted complexity becomes $O(t)$. This is exactly what we expect. The black hole has completely evaporated and so we have a one-sided system again. The restricted complexity will therefore be equal to the unrestricted complexity, which is $O(t)$.

  \section{Python's Lunches Beyond Black Hole Evaporation} \label{sec:otherlunches} \sc
Black hole evaporation is an inherently quantum mechanical phenomenon. It violates the null energy condition (NEC), for example, even when all the quantum fields in the theory would classically obey the NEC. One might therefore think that Python's Lunches themselves are an inherently quantum mechanical phenomenon and cannot exist in classical spacetimes. As we shall see, this is not at all true. Instead, there are numerous important examples, beyond black hole evaporation, of both classical and quantum lunches.

It is important to note that, in many of these examples, the  size of the lunch is fixed in the semiclassical limit. This means that we would not expect a tensor network toy model to fully scramble the degrees of freedom over the course of the lunch. We should therefore be somewhat circumspect in conjecturing that the restricted complexities are actually exponentially large in these cases.

\begin{figure}[t]
\begin{center}
\begin{subfigure}{.3\textwidth}
\centering
\includegraphics[scale=.4]{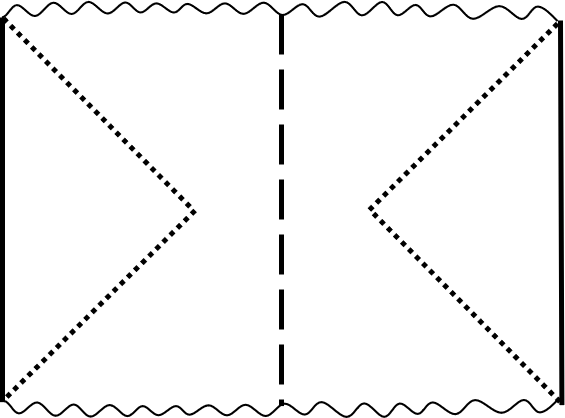} 
\caption{}
\end{subfigure}
\begin{subfigure}{.3\textwidth}
\centering
\includegraphics[scale=.3]{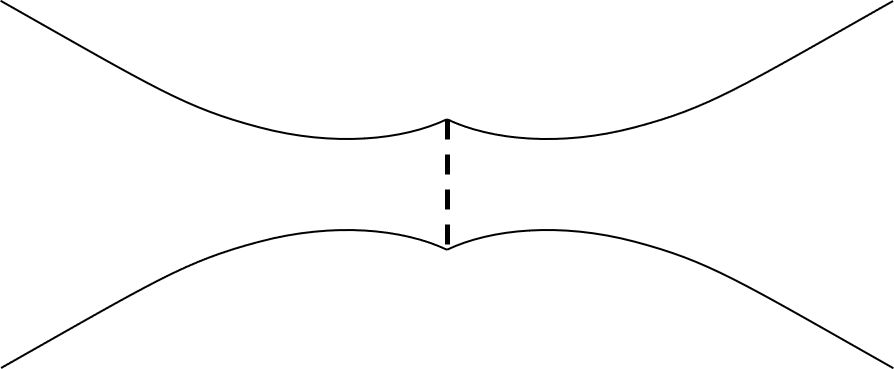} 
\caption{}
\end{subfigure}
\begin{subfigure}{.3\textwidth}
\centering
\includegraphics[scale=.5]{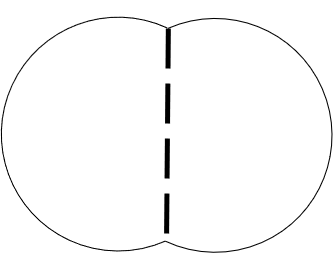} 
\caption{}
\end{subfigure}
\caption{ Adding a positive tension brane to the center of the Einstein-Rosen bridge separates the left and right horizons, creating a Python's Lunch. The Lorentzian geometry of such a state in shown in (a). A Cauchy slice is shown in (b). Such a state can be easily constructed using a Euclidean path integral with a heavy operator insertion halfway between the left and right boundaries, as shown in (c). However it cannot be easily constructed using unitary evolution from the thermofield-double state.}
\label{fig:bhbrane}
\end{center}
\end{figure}

The first, and simplest, example of a Python's Lunch is a two-sided black hole with a heavy brane, sitting in the Einstein-Rosen bridge, as shown in Fig.~\ref{fig:bhbrane}. The backreaction of the brane on the spacetime separates the left and right horizons, creating a Python's Lunch. The two end surfaces lie on the left and right bifurcation surfaces, while the bulge surface lies at the intersection of the brane world line with the static slice.\footnote{Formally, in the classical spacetime, we would need to smear the energy of the brane out slightly in order to have a smooth spacetime and hence an actual extremal surface here. This will inevitably happen once we include quantum effects.} Unlike an evaporating black hole, this spacetime does not violate the null energy condition. However, it cannot be created from the thermofield-double state using semiclassical unitary Lorentzian evolution, because the entire worldline of the brane lies behind the black hole horizon. It can be created using a simple Euclidean path integral, as shown in Fig.~\ref{fig:bhbrane}, but Euclidean evolution is not unitary. Both facts are consistent with the restricted complexity of the state being very large.\footnote{In this case, it is not obvious that even the \emph{unrestricted} unitary complexity of the state should be small. In fact, we should probably expect it to be large, because there is a nonempty extremal surface even for the combined left and right boundaries, see Sec. \ref{sec:onesided}.}

\begin{figure}[t]
\begin{center}
\includegraphics[scale=.5]{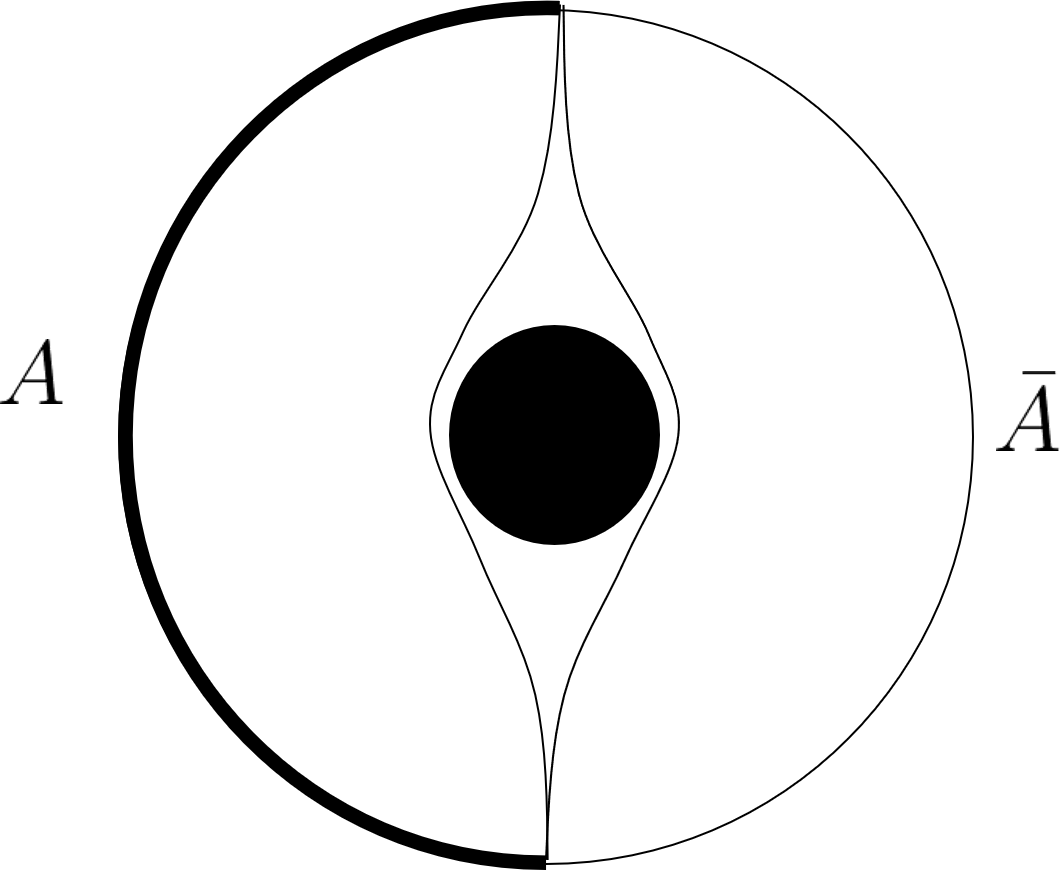}
\caption{A one-sided black hole forms a Python's Lunch between one half of the boundary, region $A$, and the other half, region $\bar A$. There is an extremal surface either side of the black hole, with the `bridge-to-nowhere' of the black hole forming the bulging lunch.}
\label{fig:bhtwohalves}
\end{center}
\end{figure}

A second classical example of a Python's lunch is a one-sided black hole, formed from collapse. This is shown in Fig.~\ref{fig:bhtwohalves}. The two ends of the python are just two halves of the single boundary. The bulging lunch in the middle is just the bridge-to-nowhere. On each side of the black hole there are classical extremal surfaces, which form the ends of the lunch. What about the bulge surface? At sufficiently early times, it seems likely that the bulge surface will go inside the horizon (for that matter, at sufficiently early times the end surfaces will also go inside the horizon). However, we would expect the area of any extremal surface going inside the horizon to grow with time. At late times, we should instead expect the bulge surface to `wrap around' the horizon. Indeed, for a BTZ black hole, it will be a self-intersecting geodesic that winds once around the horizon.

It is easy to see that the unrestricted complexity of this one-sided black hole state is small. It will be proportional to the time since the black hole first formed. However, because this time evolution couples the two halves of the boundary, then we should expect the restricted complexity to be very large, at least at late times. Again, the existence of a Python's Lunch corresponds to a large gap between restricted and unrestricted complexity.

A more quantum example, closer in spirit to the single-sided black hole evaporation studied in Sec. \ref{sec:evaporation}, but with a connected classical geometry everywhere, goes as follows. We start with the thermofield-double state, and then evolve it forwards in time, but with a small coupling between the left and right boundaries. This coupling will mix Hawking radiation between the two exteriors, causing Hawking radiation from the right to end up falling into the black hole from the left and vice versa. 

\begin{figure}[t]
\begin{center}
\includegraphics[scale=.4]{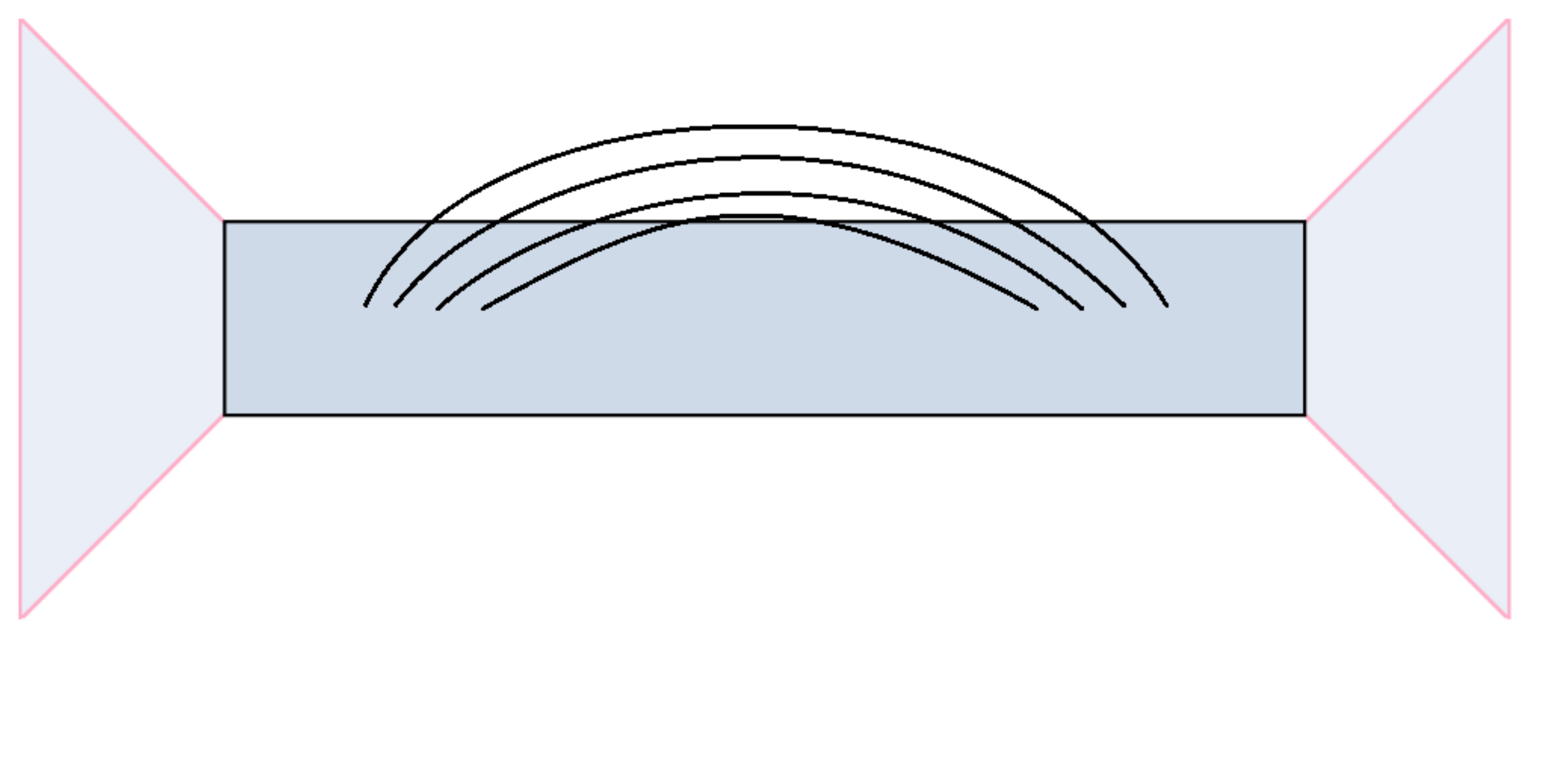}
\caption{Coupled evolution of the thermofield-double state creates a quantum lunch in a classical python. The classical cross-section of the wormhole remains constant in size, but there is long range entanglement between the two sides, which increases the generalized entropy of a cut through the middle of the wormhole.}
\label{fig:quantumlunch}
\end{center}
\end{figure}

Because the two sides of the black hole are in thermal equilibrium with one another, the size of the black hole will stay approximately constant. There will be no `classical' Python's Lunch. However the coupling between the two sides will create long-range entanglement between the quantum fields at each end of the wormhole, via Hawking radiation escaping one end and then falling into the other. This creates a quantum lunch (in a classical python). See Fig.~\ref{fig:quantumlunch}.

Again, the unrestricted complexity of the state should be small, because it was created by a simple unitary evolution from the thermofield-double state. However, this evolution coupled the two sides, and so the restricted complexity may well be large.

\begin{figure}[t]
\begin{center}
\includegraphics[scale=.5]{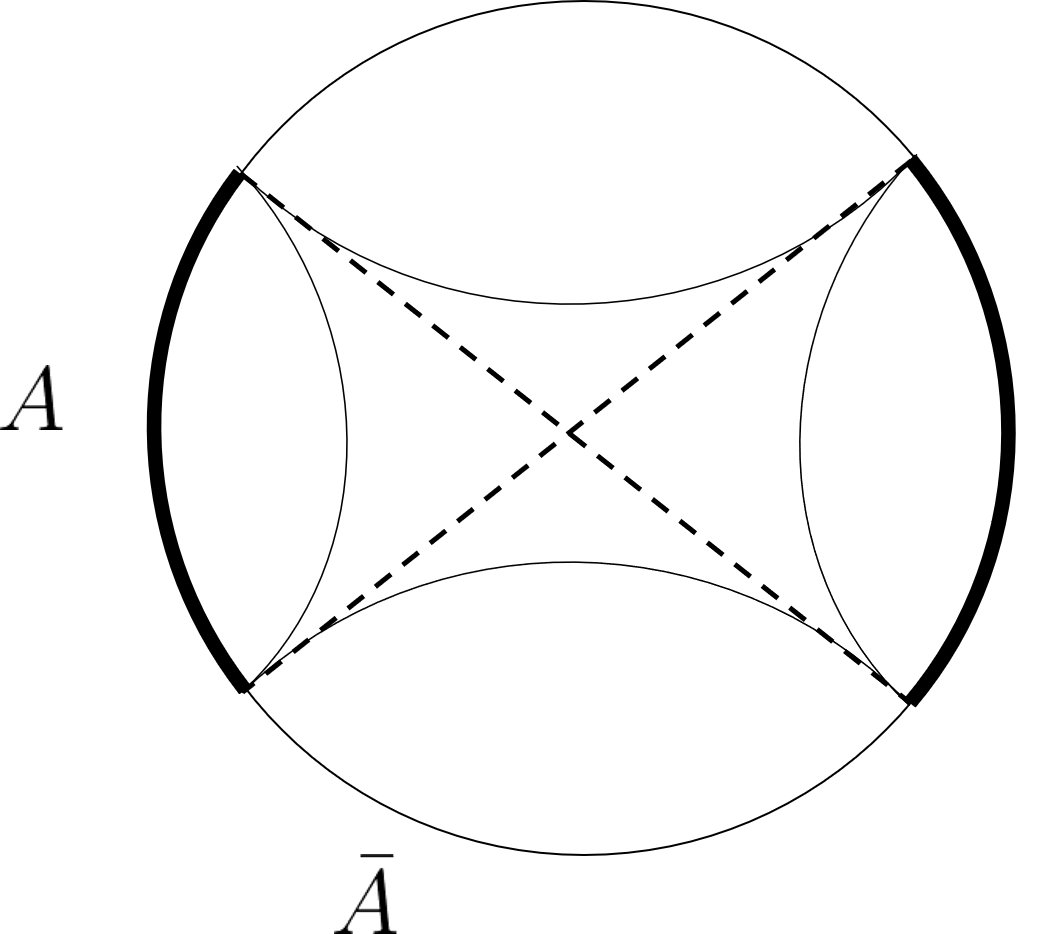}
\caption{Even the AdS$_3$ vacuum contains a Python's Lunch when viewed as a python from the boundary region $A$ consisting of two disjoint intervals to its complementary region $\bar A$.}
\label{fig:tworegionsvacuum}
\end{center}
\end{figure}

Our final example of a Python's Lunch is the AdS$_3$ vacuum. If we divide the boundary into two connected halves, as with the one-sided black hole discussed earlier in this section, there is no Python's Lunch. However, if each end of the python itself consists of two disconnected regions, as shown in Fig.~\ref{fig:tworegionsvacuum}, a lunch appears. There are two topologically distinct end extremal surfaces, plus a self-intersecting bulge surface in the middle. This suggests that the restricted complexity of constructing the vacuum state without coupling these two complementary disjoint regions may be very large. However, the small volume of the lunch suggests that it won't be fully scrambling and so our restricted complexity conjecture \eqref{eq:bigconjecture} may not apply.

\section{What is Holographic Complexity?} \label{measureHawking} \sc
There have been various proposed definitions for the bulk quantity that is to be holographically dual to boundary complexity. The two most prominent proposals have been the volume of the maximal-volume slice \cite{Susskind2014}, and the action of the Wheeler-de Witt patch \cite{Brown2015, Brown2015-1}. In practice, these two proposals tend to give similar answers.

Dual to this abundance  of promising bulk quantities is the abundance of promising boundary quantities. Various definitions of boundary complexity have been considered. The original suggestion was that it should be unitary circuit complexity (the minimal number of simple gates, from a given primitive gate set, required to build the state from a simple starting point). There is also a `continuous' variant on the unitary circuit complexity, which is defined using the Nielsen geometry \cite{Nielsen2006}. Finally, there is the intuition that the volume of a slice is dual to the size of a tensor network required to build the state. As we have seen in this paper, this is different from the unitary circuit complexity, because a tensor network may contain non-unitary elements.

One way to make a highly complex state with a small volume/action is to have the state of the bulk fields be highly complex. This suggest that the holographic complexity should have a `quantum correction', similar to the quantum corrections to the Ryu-Takayanagi formula, given by the complexity of the state of the bulk fields.

As we have seen in this paper, even when the bulk fields are in a simple state, the  \emph{restricted} unitary circuit complexity may be much larger than the volume/action, if the geometry contains a Python's Lunch. This is not a contradiction with the conjectures of Refs.~\cite{Susskind2014,Brown2015, Brown2015-1}, since the unrestricted complexity is still small, and comparable to the volume/action. However, as we shall see in this section, even the \emph{unrestricted} unitary circuit complexity can be much larger than the volume/action if we consider states that are prepared using non-unitary processes and therefore may contain `one-sided Python's Lunches'. This is true even when the bulk state is very simple. This suggests that the correct dual quantity to the holographic complexity is the size of the tensor network required to make the quantum state, or equivalently the circuit complexity where non-unitary post-selections onto the outcomes of simple measurements are allowed.

\subsection{Measuring Radiation and One-sided Lunches} \label{sec:onesided}
Suppose we take a one-sided black hole, and allow it to evaporate as in Sec. \ref{sec:evaporation}. However, rather than storing the radiation in a second system, we instead measure it in some complete basis. This basis does not have to be complicated: it can be a product basis, for example. The black hole will now be in a pure state; in particular, the interior modes that were previously entangled with the Hawking radiation will now be in a pure state that depends on the measurement outcome.

\begin{figure}[t]
\begin{center}
\includegraphics[scale=.45]{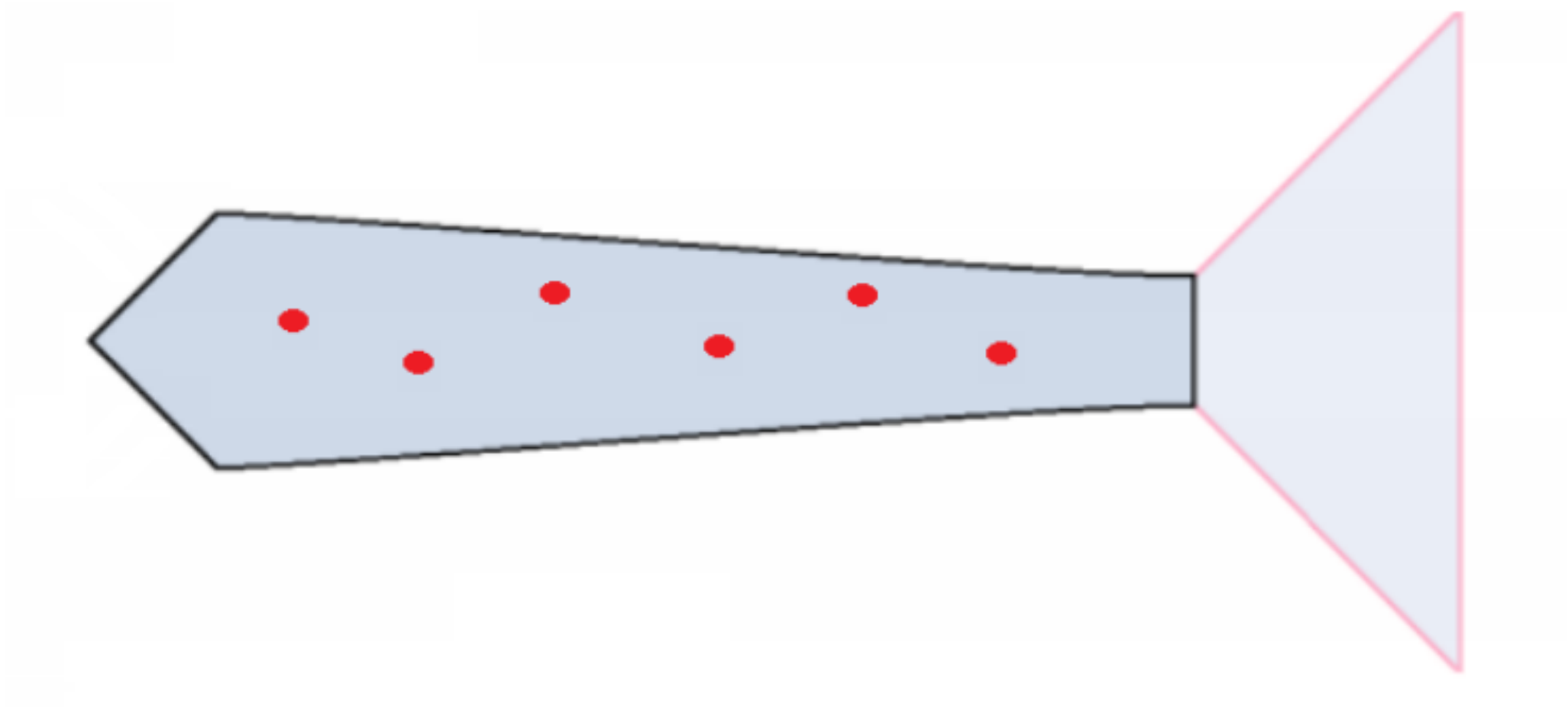}
\caption{ A one-sided black hole is allowed to evaporate, and then the Hawking radiation is measured. This produces a pure black hole microstate with interior modes that are in some simple state that depends on the measurement outcome. In the maximal volume slice, there is a local minimum of the generalized entropy near the horizon, where the area is smallest.}
\label{fig:onesidedlunch}
\end{center}
\end{figure}

\begin{figure}[t]
\begin{center}
\includegraphics[scale=.5]{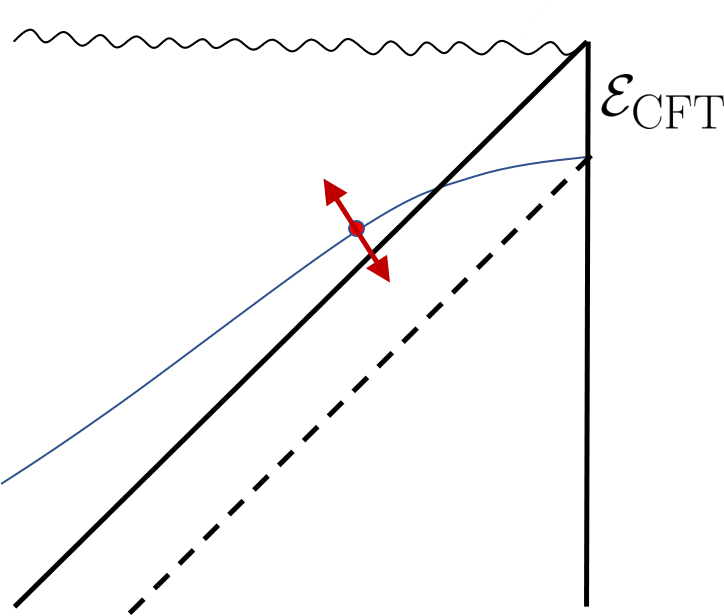}
\caption{To argue that a nonempty quantum extremal surface should exist for the pure state of a two-dimensional black hole, after measuring Hawking radiation, we consider Cauchy slices that asymptote to the maximal volume slice in the distant past, but which are allowed to vary elsewhere. In particular, we allow the slice to vary at infalling times approximately one scrambling time in the past of the current boundary time (this is where the non-empty extremal surface was found in \cite{Penington:2019npb, Almheiri:2019psf}). As shown in Fig. \ref{fig:onesidedlunch}, such slices should have a local minimum in their generalized entropy near the horizon. If we vary the Cauchy slice too far into the interior of the black hole, this local minimum will become small because the area becomes small. Conversely, if we push the slice too close to the past lightcone, the bulk entropy will become very small, which will also decrease the generalized entropy. By choosing our Cauchy slice to maximize the generalized entropy of this local minima, we would necessarily find a non-empty extremal surface, in a location similar to the surface found in \cite{Penington:2019npb, Almheiri:2019psf}.}
\label{fig:onesidedspacetime}
\end{center}
\end{figure}

The resulting bulk geometry can be thought of as a 'one-sided Python's Lunch', with a bridge to nowhere which is largest at the end and then becomes gradually smaller as one approaches a quantum extremal surface near the horizon, as shown in Fig.~\ref{fig:onesidedlunch}. The exact location of this quantum extremal surface is hard to calculate, but it is easy to show that it should exist, as argued in Fig. \ref{fig:onesidedspacetime}.

The volume of the maximal volume slice, and the action of the Wheeler-de Witt patch will grow linearly with the time the black hole was allowed to evaporate for. We expect that this will also be the size of the minimal tensor network needed to describe the state. However, if this tensor network resembles the bulk geometry, its cross-section will be largest at the end of the wormhole and then become smaller near the horizon. Such a tensor network cannot generically be produced by a unitary circuit of the same size, without allowing post-selection. 

Instead, applying our restricted complexity conjecture, in the special case where one system is trivial and so the restricted complexity is actually just the unrestricted complexity, we find that the unitary circuit complexity should be proportional to $\exp[(S_\textrm{max}^{\textrm{\textrm{(gen)}}} - S^{\textrm{\textrm{(gen)}}}_R)/2]$, where $S^{\textrm{(gen)}}_R \approx A_\text{hor}/4 G_N$ is the generalized entropy of the quantum extremal surface that we just discussed and $S_\textrm{max}^{\textrm{(gen)}} \approx A_0/4 G_N$ is the generalized entropy of a bulge surface inside the infalling matter that first formed the black hole.

This is exactly what we should expect. The state was prepared using a measurement, which can only be reproduced deterministically by using post-selection, or by using Grover search, as discussed in Sec.~\ref{sec:PL}  and Appendix~\ref{proj-complexity}. The complexity of this process is indeed exponential in the number of post-selected qubits.

\begin{figure}[t]
\begin{center}
\begin{subfigure}{.48\textwidth}
\centering
\includegraphics[scale=.4]{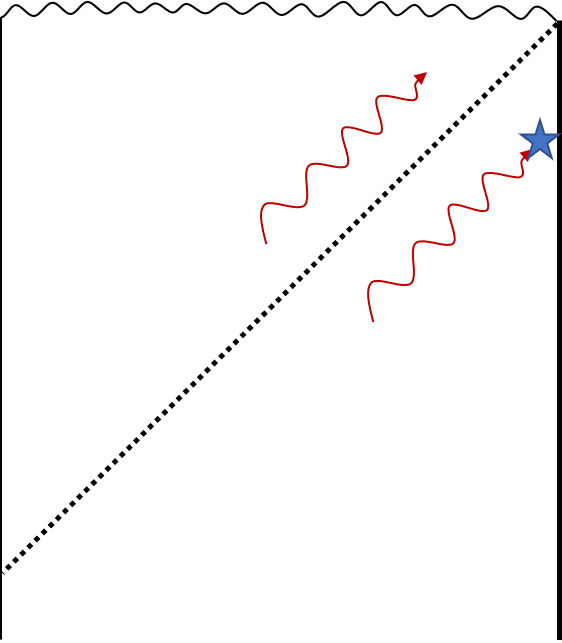} 
\caption{}
\end{subfigure}
\begin{subfigure}{.48\textwidth}
\centering
\includegraphics[scale=.5]{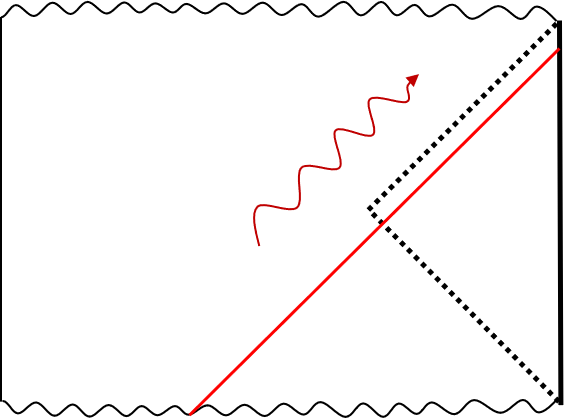} 
\caption{}
\end{subfigure}
\caption{(a) Penrose diagram for the spacetime of a one-sided black hole, with the Hawking radiation extracted and then measured. (b) If we reverse the time evolution after the measurement is done, the black hole won't disappear as one might naively expect. Instead, at one scrambling time in the past, the backreaction from the measurement (shown in red as a small shockwave) will create a white hole. The interior partners of the Hawking radiation will therefore never escape from behind the horizon.}
\label{fig:whitehole}
\end{center}
\end{figure}

One might worry that this conclusion seems wrong: what about if we just reversed time, ignoring the fact that the measurement had happened, until we got back to a time before the black hole ever formed? Once there was no black hole, it would presumably be easy to get back to a simple state using a simple circuit. The answer (see \cite{Kourkoulou:2017zaj, Almheiri:2018xdw, Penington:2019npb, Almheiri:2019psf} for similar discussion) is that, by measuring the Hawking radiation, we necessarily create a small positive-energy localized shock that approaches the horizon of the black hole as we go backwards in time. At an infalling time approximately the scrambling time in the past, the backreaction of this shock becomes significant and it turns the black hole into a white hole. If we continue to evolve the boundary backwards in time, the black hole will never disappear and the interior modes will never escape. Instead we will just see a time-reversed version of Hawking radiation coming out of the newly created white hole. This is shown in Fig.~\ref{fig:whitehole}.

How does this correspond to a tensor network model? As we evolve time backwards, the simplest tensor network describing the state initially becomes smaller, as the reverse time evolution undoes tensors that were previously added to the network by the forward evolution. However, we cannot `undo' the projections created by the measurement of Hawking radiation. At one scrambling time in the past, these projections will have infected the entire cross section of the network. Further backwards time evolution cannot remove any more tensors from the network, and so instead will have to add new tensors: the size of the simplest tensor network therefore `bounces' from this minimal size and begins to increase. This corresponds to the appearance of the white hole horizon.

Another example of a one-sided Python's Lunch is the brane-in-a-wormhole state discussed at the start of Sec.~\ref{sec:otherlunches}. In that section, we used it as an example of a two-sided lunch. However, the union of the left and right extremal surfaces is homologous to the union of the left and right boundaries. We therefore have a non-empty classical extremal surface for the union of the two boundaries, which creates a one-sided Python's Lunch. This suggests that even the unrestricted unitary complexity of the state may be very large. This is perfectly consistent, since we only know how to prepare such a state using a non-unitary Euclidean evolution.

We have argued that states with a non-empty extremal surface for the entire global boundary (i.e.~a one-sided Python's Lunch) have high unrestricted unitary complexity. A good consistency check on this  claim is that such states cannot be created by semiclassical Lorentzian evolution from states with a Cauchy slice that is entirely within the causal wedge of the boundary. This is indeed true: the entanglement wedge, defined using \emph{any} extremal surface (not just the HRT surface) must contain the causal wedge, and hence the entire spacetime, by standard focussing arguments (this requires the generalized second law or quantum focussing conjecture \cite{Bousso:2015mna} in the case of quantum extremal surfaces). It is therefore entirely consistent that they should always have very high unitary circuit complexity.
 
\subsection{Post-selected State Complexity} \label{sec:postcomplexity}
The black hole with measured Hawking radiation appears to be a counterexample to the idea that unitary circuit complexity equals volume/action. The volume and action only grow linearly with the time the black hole is allowed to evaporate for, but the size of the simplest unitary circuit required to produce the state appears to grow exponentially. One response to this problem is to note that this state could only be produced by measuring Hawking radiation, and that a measurement really corresponds to entangling the state with an ancilla measurement apparatus. If we include the measurement apparatus, the unrestricted complexity of the state (including the measurement apparatus) will still be small.

This argument is somewhat unsatisfying. It would be nice to have a boundary quantity (such as complexity) that corresponds to volume/action even for states that can only be produced using post-selection, or by Euclidean path integrals as with the first example from Sec. \ref{sec:otherlunches}. 

There is an obvious candidate quantity. We just redefine the notion of state complexity to allow post-selection onto simple states (say $|0\rangle$). Equivalently, we define it as the size of the smallest tensor network required to make the state.

With the usual definitions the complexity of creating a state from a simple state is the same as the complexity of starting with the state in question and returning to the simple state. But when post-selections are allowed, this changes.\footnote{Note that $\CC_p(I, \psi) \neq \CC_p( \psi, I)$, where $\CC_p(\cdot,\cdot)$ is relative state complexity, when projection into simple states is allowed. In fact,  going from $\ket{\psi}$ to $\ket{I}$ is trivial if post-selection is allowed, with $\CC_p( \psi, I) \leq N$.} The relevant complexity is $\CC_p(I, \psi)$ i.e.~the complexity of creating the state $\ket{\psi}$, from, the simple state $\ket{0}^{\otimes n}$, using both unitary gates and simple projections. 

 For specific states $|\psi\ra$, allowing projections may dramatically reduce the number of operations required, as we have seen. However, for typical states, projections don't buy you much. As we shall see below with a counting argument, it still takes $\sim 4^N$ operations to get to the most general state. 
\iffalse
Thus there are two kinds of complexities, when post-selection is allowed: the number of operations to get  from $\TFD$ to $|\psi\ra $ or ``forward complexity" and the number of operations operations to get from $|\psi\ra $ back to $\TFD,$ or ``backward complexity". Backward complexity is bounded by $N$ and is not very interesting. On the other hand forwards complexity can grow exponentially large and appears to be a strong candidate for the boundary dual of either volume or action.
\fi

As an aside, it is worth noting that it doesn't significantly matter whether we allow post-selection at arbitrary intermediate points in the state preparation process, or only at the end after all the unitary gates have been applied. This is because we can always implement the desired post-selection by using a unitary operator that `measures' the relevant register into an ancilla quantum register (this is sometimes called a von Neumann measurement of the first kind) and then post-selecting the ancilla register after all the other unitaries have been applied.

\subsection{Post-selected Complexity can be Exponential} 
The maximal unitary circuit complexity of an $N$ qubit state scales exponentially with $N$. This can be established with a counting argument. On the one hand Hilbert space is double-exponentially huge
\begin{equation}
\textrm{number of $\epsilon$-balls in $N$-qubit Hilbert space} \sim 2^{2^N}. 
\end{equation}
On the other hand, the number of states that can be made with $\mathcal{C}$ gates (for definiteness, let's say we have a universal $2$-local gate together with a $1$-local phase) is merely exponentially large. At each step, we can apply our $2$-local gate in one of ${N \choose 2}$ places, or apply our 1-local phase in one of ${N \choose 1}$ places, so
\begin{equation}
\# \textrm{states}  \ \, \lsim \ \left( {N \choose 2} + {N \choose 1}  \right)^\mathcal{C} . 
\end{equation}
In order to reach all the states, $\mathcal{C}$ must be exponentially big. \\

\noindent However we have now changed our definition of `complexity'. We have given ourselves the power not only to apply $2$-local gates, but also to project the first $m$-qubits to $|00000000\rangle$. Since this increases our power, it decreases the complexity. This gives rise to the state synthesis version of the PostBQP complexity class, and as we saw in Sec.~\ref{sec:superpolynomial} this is fantastically powerful -- there are many quantum states that would normally be exponentially hard to make that are now easy. Are they, in fact, \emph{all} now easy? 

Let's prove that the answer to this question is `no'. We will argue that even granting ourselves the power to post-select, there are still states that are exponentially hard to make. This can again be established with a counting argument. We have more options than before. At each step, as well as applying our $2$-local gate or our 1-local phase, we can also project on the first $m$ qubits\footnote{Notice that it would {not} have made a difference had we given ourselves the power to post-select \emph{any} $m$ qubits in the computational basis, since using just $m$ swap gates we can always move the desired qubits to the front. On the other hand,  it {would} have made a huge difference had we given ourselves the power to post-select in \emph{any} basis -- then all state synthesis is trivial, since we just post-select onto the desired state!}
 for any $1 \leq m \leq N$. Thus the number of different states we can make is 
\begin{equation}
\# \textrm{states}  \ \lsim \  \left( {N \choose 2}  +  {N \choose 1} + N \right)^\mathcal{C} ,
\end{equation}
which -- the point is -- is still only exponentially big. We still need exponentially large $\mathcal{C}$ to hit all of the double-exponentially numerous $\epsilon$-balls.\\

It would be nice to have a definition of post-selected state complexity that did not rely on a choice of discretization of the Hilbert space into $\epsilon$-balls. For unitary state complexity, one such definition is the smallest geodesic distance from the identity to a unitary taking one state to the other in the so-called Nielsen geometry  \cite{Nielsen2006}. This is a right-invariant (but not left-invariant) metric on the space of unitaries where distances are much smaller in simple directions (generated by $k$-local Hamiltonians) than in other directions.

However, if we allow post-selection on a single qubit, then any state can be prepared using a unitary that has arbitrarily small complexity as measured by the Nielsen geometry. The reason for this is that the Nielsen metric (like any metric) is continuous. Hence the complexity of a unitary mapping
\begin{align}
|0\ra |\psi\ra \to |0\ra |\psi\ra + \varepsilon |1\ra |\phi\ra
\end{align}
can be made arbitrarily small by making $\varepsilon$ sufficiently small. However, we can always post-select onto $|1\ra$ and produce $|\phi\ra$, no matter how small $\varepsilon$ is.

Instead, it seems like the right continuous measure of post-selected state complexity would be to make the cost of the post-selection be $\log p$, where $p$ is the probability of obtaining the correct measurement outcome. For typical states where the amplitudes of the post-selected outcomes are not exceptionally large or small, this still corresponds to having an $O(1)$ cost for each post-selected qubit.

\section{Summary} \sc

This paper has addressed an apparent inconsistency between the holographic complexity conjectures \cite{Susskind2014,Brown2015, Brown2015-1} and the Harlow-Hayden result \cite{HarlowHayden}. The inconsistency is manifest in an evaporating black hole slightly after the Page time: on the one hand, the volume or action of the black hole is only polynomial in the entropy $S$, and thus the holographic complexity must be moderate; on the other hand,  Harlow \& Hayden argue that the complexity of decoding the Hawking radiation must be exponentially large. The difference arises from using different definitions of complexity. The holographic complexity conjectures relate the volume/action of the geometry to {\it unrestricted} complexity, which allows gates that span the entire system; whereas Harlow \& Hayden's result is about {\it restricted} complexity, which forbids gates that couple the interior of the black hole to the previously emitted Hawking radiation. 

This distinction motivated us to ask: if action or volume are the geometric duals of unrestricted complexity \cite{Susskind2014,Brown2015, Brown2015-1}, what is the geometric dual of restricted complexity? We conjectured an answer. Exponentially large restricted complexity corresponds to the existence of a geometrical feature that we call a ``Python's lunch''. In a Python's lunch, the cross-sectional area of the wormhole grows and then shrinks again, in a min-max-min pattern. The restricted complexity, we conjectured in Eq.~\ref{eq:bigconjecture}, is  given by the exponential of the difference between the area of the maximum and the area of the larger of the two flanking minima. 

We tested this conjecture in a toy tensor-network model, and found agreement with the Harlow-Hayden estimate. 
We then made a covariant version of our conjecture, Eq.~\ref{conjCov}, by replacing the min and max areas of the Python's Lunch with generalized entropies of appropriate quantum extremal surfaces. With this generalization, we studied several examples of the Python's Lunch and estimated the restricted complexities in each case, including evaporating black holes, one-sided pure-state black holes, and empty AdS with two disjoint intervals.  In all cases where we were able to test our conjecture, the restricted complexity was consistent with the size of the Python's Lunch. 

Lastly, in Sec.~\ref{measureHawking} we returned to the subject of unrestricted complexity. We studied the example of black holes that have had all their Hawking radiation measured, and which therefore have been rendered pure. Using this example, we reconsidered exactly which boundary quantity it is that is holographically dual to the volume or action of the wormhole. In Refs.~\cite{Susskind2014, Brown2015, Brown2015-1} it was conjectured that this quantity is the unrestricted unitary circuit complexity, which means the allowed primitive gates are all unitary. Instead, we argued that the definition of unrestricted holographic complexity should also permit non-unitary post-selection -- holographic complexity should allow projections onto simple states.

  \section*{Acknowledgements}

We thank Andras Gilyen, Patrick Hayden, John Preskill, Stephen Shenker for fruitful discussions. H.G. is supported by the Simons Foundation through the It from Qubit collaboration. G.P. is supported in part by AFOSR award FA9550-16-1-0082 and DOE award {DE-SC0019380}. L.S. is supported by NSF Award Number 1316699.

%\section*{Appendix}

\appendix

\section{Complexity of Post-selection} \label{proj-complexity} \sc
A lemma we will use repeatedly in this appendix is that if an $N$-qubit state $|\psi \rangle$ is simple, then so too is the unitary
 \begin{equation}
 | \psi \rangle \textrm{ is simple }  \rightarrow  \hspace{3pt} U_{\psi } \equiv \mathds{1} - 2 | \psi \rangle \langle  \psi | \textrm{ is simple }. 
\end{equation}
This unitary flips the sign of the $| \psi \rangle$ component of a wavefunction, while leaving all orthogonal components unchanged. Let's explicitly construct a simple circuit that does this. First, note that what it means for $| \psi \rangle$ to be simple is precisely that there is a simple unitary ${U}_{| \bar{0} \rangle \rightarrow |\psi \rangle}$ that connects it to the reference state, $|\psi \rangle =  {U}_{| \bar{0} \rangle \rightarrow |\psi \rangle} | \bar{0} \rangle$. We can therefore write 
\begin{equation}
U_{\psi } =  {U}_{| \bar{0} \rangle \rightarrow |\psi \rangle}  \ U_{\bar{0} }  \  {U}_{| \bar{0} \rangle \rightarrow |\psi \rangle}^{\dagger} ;
\end{equation}
 this first transforms to a basis in which the $| \psi \rangle$ component of the wavefunction becomes the $| \bar{0} \rangle$ component, then flips the phase of the $| \bar{0} \rangle$ component, then transforms back again. This construction upper bounds the complexity 
\begin{equation}
\mathcal{C}\Bigl[\, U_{\psi }\, \Bigl] \, \leq  \, 2 \, \mathcal{C}\Bigl[ \, {U}_{| \bar{0} \rangle \rightarrow |\psi \rangle}  \Bigl] +\, \mathcal{C}\Bigl[\, U_{\bar{0} }\, \Bigl] \,  =  \, 2 \, \mathcal{C}\Bigl[ | \psi \rangle \Bigl] + \, \mathcal{C}\Bigl[\, U_{\bar{0} }\, \Bigl] \,  . \label{eq:projectionvsstatecost}
\end{equation}
We can understand the factor of 2 in this equation as arising from the fact that to make $\mathds{1} - 2 | \psi \rangle \langle \psi |$ the protocol sweeps twice over the circuit that  manufactures the state, first to unmake it (the $\langle \psi |$ part) then to make it again (the $| \psi \rangle$ part). 

\subsection{Projecting on a ququit} \label{sec:projectingonququit}
Suppose we have a simple unitary $U$ that maps a simple state to a superposition 
\begin{equation}
U  | s \rangle | 1 \rangle  =  \frac{| \alpha \rangle | 1 \rangle+ |\beta \rangle | 2 \rangle +| \gamma \rangle | 3 \rangle+ |\delta \rangle | 4 \rangle  }{2} .
\end{equation}
Since $U$ is simple, and $|s \rangle |1 \rangle$ is easy to make, the right-hand side must also be easy to make. 
\begin{itemize}
\item   { Question:  how complex can it be to make $ |\alpha \rangle |1 \rangle$}? 
\end{itemize}
As we will see, the answer is ``not  complex''. \\

\noindent One simple strategy to make $ |\alpha \rangle |1 \rangle$ is just to make $U |s \rangle |1 \rangle$ and then measure the last ququit. Sometimes we'll find the last ququit to be $|2 \rangle$, $|3 \rangle$, or $|4 \rangle$; if that happens we throw the state away and start over. Other times we'll find the last ququit to be $|1 \rangle$ and can declare victory. In this way we can make an $|\alpha \rangle$-factory that has efficiency $\frac{1}{4}$.

For some situations, this simple strategy suffices. But if the projection is a step buried deep within a larger circuit, starting again might not be so easy. Or if our initial state is entangled with another state that we do not control, then starting again might be impossible. And if we want to project not 1 ququit but $N$ ququits, the probability of success falls like $2^{-2N}$. \\

\noindent The solution to this problem is to use a close cousin of Grover's algorithm, as we will now explore. First, let's define the `pre-image' of the four possible answers we could have gotten
\begin{equation}
| \tilde{\alpha} \rangle \equiv U^\dagger | \alpha \rangle | 1 \rangle  \ \ ; \ \ | \tilde{\beta} \rangle \equiv U^\dagger | \beta \rangle | 2 \rangle  \ \ ; \ \ | \tilde{\gamma} \rangle \equiv U^\dagger | \gamma \rangle | 3 \rangle  \ \ ; \ \ | \tilde{\delta} \rangle \equiv U^\dagger | \delta \rangle | 4 \rangle  ,
\end{equation}
so that $| s \rangle |1 \rangle = \frac{1}{2} ( |\tilde{\alpha} \rangle + |\tilde{\beta} \rangle  + |\tilde{\gamma} \rangle  + |\tilde{\delta} \rangle )$. Next observe that  it is simple to make the operator $U_{\tilde{\alpha}}$ that flips the sign of the $|\tilde{\alpha} \rangle$ term in a wavefunction while leaving the other terms invariant. We can do this by evolving the state with $U$, doing a sign-flip controlled on the last ququit being $|1 \rangle$, and then evolving back with $U^\dagger$
\begin{equation}
V \equiv 1 - 2 |\tilde{\alpha} \rangle \langle \tilde{\alpha}| =  U^\dagger \Big( \mathds{1} \otimes (\mathds{1} - 2 | 1 \rangle \langle 1 |) \Big) U .
\end{equation}
Finally, since $| s \rangle |1 \rangle$ is by assumption easy to build, it must also be easy to build the operator that flips everything except  $| s \rangle |1 \rangle$,
\begin{equation}
U_{s} \equiv   2 | s \rangle   \langle s | \otimes | 1 \rangle \langle 1 | - \mathds{1} . 
\end{equation}
Now we concatenate these easy operations to give the desired projection
\begin{eqnarray}
V |s \rangle|1 \rangle &=& \frac{1}{2} ( - |\tilde{\alpha} \rangle + |\tilde{\beta} \rangle  + |\tilde{\gamma} \rangle  + |\tilde{\delta} \rangle )\\
U_{s} V|s \rangle | 1 \rangle &=&  | \tilde{\alpha} \rangle \\
U \, \Big[ U_{s} V \Big]  |s \rangle|1\rangle &=&   | {\alpha} \rangle |1 \rangle
\end{eqnarray}

\vspace{5mm}

\noindent Note that the only reason this worked was because there was an easily addressed ququit to diagnose the final branch. If we just had $U | s \rangle = |\alpha \rangle + | \beta \rangle + | \gamma \rangle + | \delta \rangle$ then this method wouldn't have worked, and indeed couldn't have worked since there is such a decomposition for \emph{any} $| \alpha \rangle$, even if $U$ is the identity. 

\subsection{Projecting on a qu$d$it} \label{subsec:projectonqudit}

The case of the last subsection was misleadingly easy: we needed only a single implementation of $U_{s} V$ to hit the target state exactly. More generally, we may wish to project onto the value of a qu$d$it, where we can think of $d=2^m$ for $m$ qubits or $m/2$ ququits. And it may be that the amplitudes of the states are not evenly distributed. Let's start with an initial state $\ket{s} \ket{0}^{\otimes n}$ on total of $k$ qubits with $n$-ancillary qubits at state $\ket{0}$. Then, we wish to project (post-select) onto the outcome of $m$ qubits.  The output state after the unitary $U$ is applied can be expressed  as
\begin{equation}
U | s \rangle | 0 \rangle^{\otimes n} = \sin \theta |\alpha \rangle  |0 \rangle^{\otimes m} + \cos \theta | \beta \rangle \, , 
\end{equation}
where $| \beta \rangle$ is any normalized state with $ \langle 0 | ^{\otimes m} | \beta \rangle = 0$. Repeating the procedure of the last subsection gives 
\begin{equation}
U \Big[ U_{s} V\Big]  |s \rangle | 0 \rangle^{\otimes n}  = \sin 3 \theta |\alpha \rangle  |0 \rangle^{\otimes m} + \cos 3 \theta | \beta \rangle \ . \label{eq:tripleuptheangle}
\end{equation}
where we denote by  
\begin{eqnarray} \label{Us-definition}
U_s &=&  2 | s \rangle   \langle s | \otimes | 0 \rangle \langle 0 |^{\otimes n} -  \mathds{1}  \\
V &=& U^\dagger \Big( \mathds{1} \otimes (\mathds{1} - 2 | 0 \rangle \langle  |^{\otimes m}) \Big) U 
\end{eqnarray}
for $n$ ancillary and $m$ postselected qubits. 
We see that in the last subsection we got lucky, since $\sin \theta = \frac{1}{2} \rightarrow  \sin 3 \theta = \sin 3 \frac{\pi}{6}  =  1$. For more general $\theta$, a single iteration will not yield the desired projection. Iterating $l$ times gives
\begin{equation}
U \Big[ U_{s} V \Big]^{l}  |s \rangle | 0 \rangle^{\otimes n}  = \sin [ (2l+1) \theta] |\alpha \rangle  |0 \rangle^{\otimes m} + \cos  [ (2l+1) \theta]   | \beta \rangle \ . 
\end{equation}
Thus the number of iterations to implement the projection is given by $(2l+1)\theta = \frac{\pi}{2}$ so 
\begin{equation}
\textrm{number of implementations of }\Big[ U_s V  \Big] =  \frac{\pi}{4 \theta }  - \frac{1}{2} . \label{eq:numberofimplementations}
\end{equation}
When $U |s \rangle  | 0 \rangle^{\otimes n}$ is an equal superposition of $d (=2^m)$ states, we have $\theta = \arcsin[ 1/ \sqrt{d} ]$ and this gives the celebrated large-$d$ Grover scaling $l \sim \frac{\pi}{4} \sqrt{d}$. \\

\noindent Finally, we must confront the possibility that Eq.~\ref{eq:numberofimplementations} does not give a whole number. This is problematic since in general even if a ${U}$ is easy to implement, $\sqrt{U}$ may be hard. We could lower our ambitions by implementing the integer part of $n$ and settling for being approximate. But we can do better. We first introduce a fresh qubit, and then use it to bleed some of amplitude out of  $\sin \theta | \alpha \rangle  | 0 \rangle^{\otimes m}$
\begin{equation}
| \bar{s} \rangle = U_\phi  U  |s \rangle | 0 \rangle^{\otimes n} | 0 \ra  = \cos \phi \,\sin \theta | \alpha \rangle  | 0 \rangle^{\otimes m} |0\ra + \sin \phi \,\sin \theta | \alpha \rangle  | 0 \rangle^{\otimes m} |1\ra   + | \textrm{other} \rangle.
\end{equation}
Then we repeat the iterative procedure using $|\bar{s} \rangle $ instead of $| s \rangle |0 \rangle^{\otimes n} $ and $I$ instead of $U$, and carefully choose $\phi$ to land on the next integer greater than $\frac{\pi}{4 \theta }  - \frac{1}{2}$,
\begin{equation}
\boxed{\textrm{number of implementations of }  U_{\bar{s}} V =  \Bigl\lceil \frac{\pi}{4 \theta }  - \frac{1}{2}  \Bigl\rceil } \  . \label{eq:exactnumberofimplementations}
\end{equation}
Even though it is in general not easy to implement a fractional power of an easy unitary, for the specific unitaries we are considering it is.\\

\noindent Let us see what this analysis means for the simplest possible case, that of an evenly split qu$b$it, with $\theta = \frac{\pi}{4}$. Equation~\ref{eq:tripleuptheangle} made the situation look hopeless -- we would just cycle in a loop forever, $\theta = \frac{\pi}{4} \rightarrow  \frac{3\pi}{4} \rightarrow \frac{\pi}{4} \rightarrow  \frac{3\pi}{4} \rightarrow \ldots$, never getting any closer to the target state. But now we see that the correct procedure it to first add an extra qubit, and then use $\phi = \frac{\pi}{4}$ to transform the pair of qubits to an even superposition of a ququit, returning us to the exactly implementable example of Sec.~\ref{sec:projectingonququit}.

\subsection{Projecting on very unlikely outcomes}
Suppose we wish to project onto a final state that has tiny amplitude, $\theta \ll 1$. For example, we may have an equal superposition over $m$ qubits with large $m$, giving $\theta = 2^{-m/2}$. How complex is this projection? Let us examine three possible methods:
\begin{itemize}
\item Measure-and-pray. If we measure the qubits and hope for the right answer, the probability that we get lucky is $\theta^2$, giving
\begin{equation}
\textrm{measure-and-hope method:} \ \ \ \textrm{complexity} \sim  {\theta^{-2} } =  d =  2^{m} \ . 
\end{equation} 
\item All-at-once Grover-style projection. Using the method of Sec.~\ref{subsec:projectonqudit}, we saw in Eq.~\ref{eq:exactnumberofimplementations} that we can effect a square-root speed up, 
\begin{equation}
\textrm{all-at-once Grover:} \ \ \  \textrm{complexity} \sim \theta^{-1} = \sqrt{d} =  2^{m/2} \ .  \label{eq:allatoncegrover}
\end{equation} 
\item Step-by-step Grover-style projection. In Eq.~\ref{eq:allatoncegrover}, we simultaneously projected onto the values of all the target qubits. An alternative strategy would be to project on each target ququit in turn, so that each individual projection is then onto a state that is not particularly unlikely. As discussed in  Appendix \ref{sec:projectingonququit}, the complexity of projecting onto the first ququit is $3 \mathcal{C}(U)$, because we have to implement the unitary $U$ three times to do the post-selection. What about the projection onto the second ququit? Now the unitary $U$ has been replaced by the unitary $U'$ that also does the projection onto the first ququit. So the complexity of this second projection is $3 \mathcal{C}(U') = 9 \mathcal{C}(U)$. It should be clear that the complexity of projecting onto each ququit grows exponentially. The complexity of the full projection is therefore
\begin{equation}
\textrm{step-by-step Grover:} \ \ \  \textrm{complexity} \sim 3^{m/2} \ ,  \label{eq:stepbystepgrover}
\end{equation} 
since there are $m/2$ ququits.
\end{itemize}

\noindent Summary: to project on $m$ qubits, guess-and-check costs $2^m$, all-at-once-Grover costs O($2^{m/2}$), and one-by-one-Grover costs O($3^{m/2}$). The winning strategy is to project on all the qubits at once, as in Sec.~\ref{subsec:projectonqudit}. The complexity of doing so is O($2^{m/2}$). 
% Using this method, to project on $2m$ qubits costs O($2^m$). This achieves a squareroot speed up over the guess-and-check method. 

\vspace{2mm}
\subsection{Removing the state dependence} \label{app:stateindep}
So far we have only tried to construct a unitary that produces a single output state $| \alpha \rangle$, given input $\ket{s} \ket{0}^{\otimes n}$  and unitary $U$. We used a unitary sequence 
\begin{equation}
U \Big[ \Big(2 | s \rangle   \langle s | \otimes | 0 \rangle \langle 0 |^{\otimes n} -  \mathds{1} \Big) U^\dagger \Big( \mathds{1} \otimes (\mathds{1} - 2 | 0 \rangle \langle 0 |^{\otimes m}) \Big) U \Big]^{l} 
\end{equation}
that works only for the particular input state $\ket{s}$. 

Our actual task is somewhat more complicated. We need to construct a unitary circuit that produces the same output as our post-selected circuit for \emph{any} input state $\ket{s}$. In other words, we want to find a unitary $\widetilde{U}$, such that for any input state $| s \rangle$ we have
\begin{align} \label{eq:UU}
\widetilde{U} \ket{s} \ket{0}^{\otimes n} =  C \ket{0}^{\otimes m} \bra{0}^{\otimes m} U \ket{s} \ket{0}^{\otimes n},
\end{align}
where $U$ is a simple unitary, $C$ is a numerical constant. It turns out that we can easily adapt our construction from previous section to produce such a unitary if such a unitary $\widetilde{U} $ exists at all. Our construction is very closely related to {\it robust oblivious amplitude amplification}, which was independently introduced in \cite{BCC15, Gilyen2017, Gilyen2019}. We became aware of this work after this section of the manuscript had been completed. 

Importantly, for the moment, we shall assume that an \emph{exact} unitary $\widetilde{U}$ exists that exactly satisfies \eqref{eq:UU}. We shall discuss what happens when $\widetilde{U}$ can only approximately satisfy \eqref{eq:UU} at the end of this subsection.

Let $\ket{i}$ be a computational basis for the input Hilbert space. By our assumption that there exists some unitary $\widetilde{U} $ satisfying \eqref{eq:UU}, it follows that 
\begin{align}
 \bra{i} \bra{0}^{\otimes n} U^\dagger \ket{0}^{\otimes m} \bra{0}^{\otimes m} U \ket{j} \ket{0}^{\otimes n} = \frac{1}{C^2} \delta_{ij}
\end{align}
This implies that 
\begin{align} \label{eq:gjdsf}
U \ket{i} \ket{0}^{\otimes n} = \sin \theta \ket{\alpha_i} \ket{0}^{\otimes m} + \cos \theta \ket{\beta_i},
\end{align}
where $\theta$ is independent of $i$ and $\sin \theta = 1/\sqrt{C}$, $\braket{\alpha_i | \alpha_j} = \braket{\beta_i | \beta_j} =  \delta_{ij}$ and
\begin{align}
\bra{0}^{\otimes m} \ket{\beta_i} = 0
\end{align}
for all $i$. We can get rid of state dependence of the original protocol by replacing $U_s$ in \ref{Us-definition} with 
\begin{align}
W =  \mathds{1} \otimes \Big(2 | 0 \rangle \langle 0 |^{\otimes n} -  \mathds{1} \Big).
\end{align}
This is independent of the input state  $\ket{s}$. For an arbitrary state $\ket{s}$, we again have 
\begin{equation}
U | s \rangle | 0 \rangle^{\otimes n} = \sin \theta |\alpha \rangle  |0 \rangle^{\otimes m} + \cos \theta | \beta \rangle \, , 
\end{equation}
and one can check that
\begin{equation}
U \Big[ W V\Big]  |s \rangle | 0 \rangle^{\otimes n}  = \sin 3 \theta |\alpha \rangle  |0 \rangle^{\otimes m} + \cos 3 \theta | \beta \rangle \ . \label{eq:tripleuptheangle-gen}
\end{equation}
The key step is that \eqref{eq:gjdsf} implies
\begin{align}
\bra{0}^{\otimes n} U^\dagger \left[\cos \theta |\alpha \rangle  |0 \rangle^{\otimes m} - \sin \theta | \beta \rangle \right] = 0 \, .
\end{align}
Repeating the process $N$ times gives
\begin{equation}
U \Big[ W V \Big]^{l}  |s \rangle | 0 \rangle^{\otimes n}  = \sin [ (2l+1) \theta] |\alpha \rangle  |0 \rangle^{\otimes m} + \cos  [ (2l+1) \theta]   | \beta \rangle \ . 
\end{equation}
The rest of the argument is identical to that in Appendix \ref{subsec:projectonqudit}, proving that 
\begin{equation}\label{stateIndepProt}
U \Big[\Big( \mathds{1} \otimes (2 | 0 \rangle \langle 0 |^{\otimes n} -  \mathds{1} ) \Big)  U^\dagger \Big( \mathds{1} \otimes (\mathds{1} - 2 | 0 \rangle \langle 0 |^{\otimes m}) \Big) U \Big]^{l} 
\end{equation}
will output state $\ket{\alpha} \ket{0}^{m}$ with very high probability for any $\ket{s}$ and $l=\pi / 4 \theta$ iterations of the Grover step. For typical values of $\theta \sim 2^{-m/2}$, the complexity is $O[\mathcal{C}(U)2^{-m/2}]$ as before.

It is important to note that the protocol we just constructed relied crucially on the assumption that there exists an exact unitary  $\widetilde{U}$ satisfying Eq.~\ref{eq:UU}. If it is instead only approximate (a more realistic assumption), we can run into problems because we are applying an exponentially long circuit: small errors at each stage can add up to become very large.

We now argue that this will not be the case so long as $n \gg m \gg 1$ and $U$ is scrambling. (If $U$ is not scrambling we expect that much more efficient circuits may well exist anyway.) Scrambling unitaries are well modeled by typical elements of unitary $2$-designs. We want to show that a Grover search using \eqref{stateIndepProt} and a fixed number of repetitions (independent of $\ket{\psi}$) will work for an arbitrary input state $\ket{\psi}$. This implies that the first and second moment calculations will be exactly given by Haar averages using Weingarten coefficients \cite{Collins2003}. Specifically, we first compute the mean of the $\sin^2 \theta$, 
\begin{align}
\overline{\sin^ 2\theta}  =\EE_{U \in Haar} \Big{|}  \Big{|}  \bra{0}^{\otimes m} U \ket{s} \ket{0}^{\otimes n} \Big{|}  \Big{|}_2 = \frac{1}{2^m} \ . 
\end{align}
In addition, we would like to estimate the variance of the $\sin^ 2\theta$. We can also compute this quantity since it only involves two copies of $U$ and two copies of $U^\dagger$, 
\begin{align}
\Delta_\theta  = \text{var} [ \sin^2 \theta] = \EE_{U \in Haar} \Big{|}  \Big{|}  \bra{0}^{\otimes m} U \ket{s} \ket{0}^{\otimes n} \Big{|}  \Big{|}_2^2  -   \frac{1}{4^m} = \frac{1}{2^{k+m}}
\end{align}
where $k$ is the total number of qubits. 
 This implies that, 
\begin{equation}
\sin^2 \theta  \approx  2^{-m} + O (2^{-(k+m)/2}) \, , 
\end{equation}
where $k$ is the total number of qubits. It follows that we can assume $\theta^2 = 2^{-m}$ for the state with exponentially small error. 

In the state-independent protocol, we use $W$ instead of the state-dependent projector $U_{s}$. This results in additional errors  if $\widetilde{U}$ is not an exact unitary. However, this error in each Grover step can be bounded. The error for a fixed unitary $U$ is 
\begin{align}\nonumber 
\epsilon_U = \Big\lVert  \Big((\ket{s}\bra{s} - \mathds{1} ) \otimes \ket{0}\bra{0}^{\otimes n} \Big)  U^\dagger \Big( \mathds{1} \otimes (\mathds{1} - 2 | 0 \rangle \langle 0 |^{\otimes m}) \Big) U \ket{s} \ket{0}^{\otimes n} \Big\rVert _2. 
\end{align}
Again we can use that $U$ is an element of a 2-design and calculate the mean of the $\epsilon_U$. Again, one can do those integrals using Weingarten coefficients \cite{Collins2003} and arrive at, 
\begin{align}
\overline  \epsilon   = \EE_{U \in Haar} \Big( \epsilon_U \Big) = 
 %&\EE_{U}  \Big\lVert  2\Big((\ket{s}\bra{s} - \mathds{1} ) \otimes \ket{0}\bra{0}^{\otimes n} \Big)  U^\dagger \Big( \mathds{1} \otimes (\mathds{1} - 2 | 0 \rangle \langle 0 |^{\otimes m}) \Big) U \ket{s} \ket{0}^{\otimes n} \Big\rVert _2  \\ \nonumber 
 \frac{1}{2^n}. 
\end{align}
The amplitude of being mapped into a wrong state in each Grover step is $\sqrt{\overline  \epsilon } = 2^{n/2}$. Our state independent protocol  requires application of \eqref{stateIndepProt} $2^{m/2}$ times, so the total accumulated error will be given by $\sqrt{2^{m-n}}$, which is exponentially small in the limit of interest, when $n \gg m \gg 1$. This completes our argument.

\section{Maximinimax Prescription for the Bulge Surface} \label{app:maximin} \sc
In this appendix, we argue that spacetimes with more than one extremal surface generically contain Python's Lunches. To do so, we use a variant on the maximin arguments introduced by Wall in Ref.~\cite{Wall:2012uf}. Such arguments have numerous subtleties and require considerable effort to rule out as many edge cases as possible. We shan't worry too much about such details here; instead we will just  give physics-level arguments that justify our construction. We shall restrict our attention to classical spacetimes obeying the null energy condition (NEC), although it is possible to generalize maximin arguments to include quantum effects \cite{QuantumMaximin}.

Our starting assumption is the existence of two distinct extremal surfaces, the HRT surface $\chi_1$ and an additional surface $\chi_2$, homologous to the same boundary region. The HRT surface $\chi_1$ can be found by a maximin prescription, where one finds the globally minimal area surface within some Cauchy slice, and then maximises that area over all Cauchy slices. The second extremal surface $\chi_2$ must have equal or larger area. Generically it will have larger area  and we assume that this is indeed the case.

We first argue that no point on the second extremal surface $\chi_2$ can be timelike separated from any point on the HRT surface $\chi_1$. Let $C_1$ be a Cauchy slice in which $\chi_1$ is the unique minimal surface. We define a new Cauchy slice 
\begin{align}
C_2 = \partial(J^-[\chi_2] \cup (J^-[C_1] \cap \overline{J^+[\chi_2]})),
\end{align}
where $J^-[\chi_2]$ and $J^-[C_1]$ are the past of $\chi_2$ and $C_1$ respectively and $\overline{J^-[\chi_2]}$ is the complement of the future of $\chi_2$. Note that $C_2$ contains the surface $\chi_2$ and so is nowhere timelike separated from it.

But by standard focussing arguments, the minimal area surface in $C_2$ is at least as big as the minimal area surface on $C_1$, with equality if and only if $\chi_1$ is in $C_2$.\footnote{We assume $\chi_1$ is the unique minimal area surface in $C_1$.} This is because we can focus any surface in $C_2$ to a surface in $C_1 \cap C_2$ with no greater area.

Since we assume that $\chi_1$ is the unique maximin surface, we find that $\chi_1 \in C_2$, and so $\chi_1$ is not timelike separated from $\chi_2$. Assuming the spacetime is generic (i.e.~the NEC holds as a strict inequality) we can also use standard focussing arguments to argue that $\chi_1$ cannot be lightlike separated from $\chi_2$.

Let us temporarily assume that there exists a Cauchy slice, within which any sufficiently small deformation of $\chi_2$, which preserves the homology constraint but which is not necessarily local, will increase its area. (We shall consider the alternative possibility below.) If we deform $C_2$ in a sufficiently small neighborhood of $\chi_2$, we should then be able to find a new Cauchy slice, still containing $\chi_1 \cup \chi_2$, on which $\chi_2$ is minimal within a small neighborhood and $\chi_1$ is still globally minimal. This will be important later.

We define the Wheeler-de Witt patch $W_{1,2}$ as the bulk domain of dependence of any spacelike slice bounded by $\chi_1 \cup \chi_2$. We can then construct a new surface $\chi_3$ by the following maximinimax procedure. 

First we choose some Cauchy slice $C_3$ for the Wheeler-de Witt patch $W_{1,2}$.\footnote{Note that this is a Cauchy slice for the patch $W_{1,2}$ but not a Cauchy slice for the entire spacetime.} 

Next, we choose a smooth non-degenerate function $\phi_3: C_3 \to [0,1]$, where $\phi_3(\chi_1) = 0$ and $\phi_3(\chi_2) = 1$. Morally, the level sets of the function $\phi_3$ define a foliation of $C_3$. However, it is somewhat more general than this because the topology of the level set can change if $\phi_3$ has critical points. Formally, this is known as a `sweepout' of $C_3$. It is necessary both for physical reasons, since in general an extremal surface may have arbitrary topology, so long as it satisfies the homology constraint, and for mathematical reasons, to prevent the appearance of singularities in the surface if we insist that it have the `wrong' topology.

Finally, we choose the level set $\phi_3^{-1}(x_3)$ for $x_3 \in [0,1]$ of \emph{maximal} area. Note that the level set $\phi_3^{-1}(x)$ will be singular if $x$ is a critical value of $\phi_3$, but so long as $\phi_3$ is non-degenerate, the singularities will be at isolated points and the area of the surface should still be well defined. 

Having found the surface $\chi_3 = \phi_3^{-1}(x_3)$ of maximal area, we minimise that maximal area over all allowed functions $\phi_3$. Finally, we maximize that minimax area over all Cauchy slices for $W$. We call the resulting surface the maximinimax surface $\chi_3$. In other words, we have
\be
\chi_3 = \phi_3^{-1}(x_3),
\ee
where the level set $ \phi_3^{-1}(x_3)$ is defined by the maximinimaximization
\be
\max_{C_3} \min_{\phi_3} \max_{x_3} A(\phi_3^{-1}(x))
\ee
Provided a unique maximinimax surface $\chi_3$ exists\footnote{If the surface was non-unique, one would have to worry about stability issues.} and does not lie at the boundary of any of the spaces we are optimizing over, it will be extremal, since, at linear order, an arbitrary variation of the surface $\chi_3$ can be achieved by a linear combination of variations in $C_3$, $\phi_3$ and $x_3$.

We shall not try to rigorously prove that this will be true (even generically). After all, we have not even tried to rigorously define our construction -- the geometric measure theory \cite{federer2014geometric} required to do so would be well beyond the scope of this paper. However we shall make a few comments about why we expect that a maximinimax surface should exist and not lie at the various boundaries of the optimization space we are searching over.

So long as $\phi_3$ is smooth and non-degenerate, the area of the level set should be a continuous function on a compact interval, and so a maximal area surface should exist. 

One might worry that the minimization over functions $\phi_3$ could approach a function that is not smooth and non-degenerate, for which a maximal area level set is not well defined. Our understanding of the results of Almgrem-Pitts min-max theory \cite{Algrem1965, Pitts1081, Colding2003} is that this will not end up being the case. Instead, the minimax surface $\chi_3 \in C_3$ should be a well defined varifold and will be a smooth (possibly self-intersecting) submanifold if the spacetime dimension is less than seven. Intuitively, it is reasonable to expect that, as the function $\phi_3$ becomes more badly behaved, the maximal area surface should only increase, rather than decrease in area. 

Similarly, we expect any bad behavior in the Cauchy slice $C_3$ will tend to decrease the area of the minimax surface. Hence a maximinimax surface $\chi_3$ should exist and be well behaved. For more detailed arguments in this direction, see \cite{Wall:2012uf}. For known examples, involving timelike de Sitter boundaries, where maximin surfaces do not exist, see \cite{Fischetti2014}. 

How could the maximinimax surface $\chi_3$ end up on the boundary of the space of surfaces we are searching over? Firstly, the maximinimax surface $\chi_3$ could have nonzero intersection with $\chi_1 \cup \chi_2$. Suppose this intersection were not a connected component of $\chi_3$ (and $\chi_1 \cup \chi_2$). Since the surface $\chi_3$ cannot ever go outside the Wheeler-de Witt patch $W_{1,2}$ and $\chi_1$ and $\chi_2$ are extremal, there must be some point where $\chi_3$ has nonzero mean curvature, within the Cauchy slice, where it bends `inwards' into $W$. We could then decrease the max area of a level set in $\phi_3$ by deforming $\phi_3$ slightly to make the surface $\chi_3$ moves slightly inwards at this point, which gives the desired contradiction.

What about if the intersection is a connected component? In that case, the entire connected component will already be extremal and we don't need to worry about it. One might, of course, worry that $\chi_3$ could end up being the same as either $\chi_1$ or $\chi_2$, in which case we wouldn't have really found a new extremal surface. However this is impossible, since, by assumption, there exists a Cauchy slice where any small deformation of $\chi_1$ or $\chi_2$ will increase their area, and hence neither can have maximal area within any $\phi_3$. The min-max surface in this Cauchy slice will have a larger area than either $\chi_1$ or $\chi_2$, which rules out either $\chi_1$ or $\chi_2$ being maximinimax.

Finally, one might worry that points in $\chi_3$ might end up lightlike separated from either other points in $\chi_3$, or points in $\chi_1$ or $\chi_2$. The first cannot happen, because of arguments similar to those in \cite{Wall:2012uf}.  If a) the minimax surface contained a null segment, the area of the minimax could always be increased by a sufficiently small deformation of the Cauchy slice near this null segment. However, if b) the minimax surface did not contain a null segment, it could not be extremal within the Cauchy slice (using focussing arguments for generic spacetimes), which the minimax should be since its variation is unconstrained so long as it doesn't intersect $\chi_1$ or $\chi_2$. The second cannot happen because (using focussing in a generic spacetime) we could then increase the area of the minimax surface by deforming $\phi_3$ so that the level set is locally deformed along the lightcone towards $\chi_1$ or $\chi_2$.

We also note that the maximinimax construction automatically guarantees that the bulge surface $\chi_3$ has larger area than either $\chi_1$ or $\chi_2$.

Having shown that an intermediate bulge surface exists between $\chi_1$ and $\chi_2$ whenever $\chi_2$ is minimal with respect to any small deformation within some fixed Cauchy slice, we now consider the opposite case, where there exist small deformations of $\chi_2$ which decrease the area of $\chi_2$ within any Cauchy slice. In this case, $\chi_2$ cannot be an end surface and so must instead itself be the bulge surface. Without loss of generality, we assume that $\chi_1$ is contained in the interior $\text{Int}[\chi_2]$ of $\chi_2$.\footnote{Recall that the two surfaces cannot ever be timelike separated. They also can't intersect, because then generic perturbations would presumably make them be timelike separated.}

We shall also assume that, in any Cauchy slice, there exist small deformations of $\chi_2$ that a)  decrease the area and b) lie entirely in the exterior $\text{Ext}[\chi_2]$ (defined as the complement of the interior $\text{Int}[\chi_2]$). The alternative possibility, where there exist Cauchy slices where only deformations that enter the interior can decrease the area, but none where no deformations can decrease the area, should be non-generic and can be interpreted as the bulge surface and one end surface degenerating into one another. 

We can now use the usual maximin construction, to find a second end surface $\chi_3$. We simply constrain our search to Cauchy slices containing $\chi_2$, and to surfaces within that Cauchy slice that are entirely in the exterior $\text{Ext}[\chi_2]$. As before, to show extremality, we just need a) for the maximin surface to exist and b) for variations of the Cauchy slice be sufficient to freely vary the maximin surface. 

The arguments for both are essentially identical to those for the original maximin construction. The only new potential obstructions that need to be ruled out are the maximin surface either a) intersecting, or b) being lightlike separated from, the surface $\chi_2$. In the first case, if the intersection was not a connected component the maximin surface would have to bend inwards somewhere, which contradicts its minimality within the Cauchy slice. Any intersection on a connected component will be automatically extremal. Finally, the surface $\chi_3$ cannot simply be equal to $\chi_2$, since, by assumption, $\chi_2$ does not have globally minimal area within any Cauchy slice.

What about the possibility of lightlike separation from $\chi_2$? By focussing arguments in a generic spacetime, the change in area from a lightlike deformation of the surface $\chi_3$ towards $\chi_2$ (in direction $k^a$) would have to be positive at linear order. Since $\chi_3$ is minimal within the Cauchy slice, there must be some spacelike direction $r^a$ pointing away from the lightcone for which the change in area is nonnegative at linear order. However, this implies a deformation in a timelike direction $t^a$ (that makes $\chi_3$ spacelike separated from $\chi_2$) must increase the area at linear order, in contradiction with the maximality of the Cauchy slice.\footnote{We are ignoring issues about stability here.}

Finally, we note that, since $\chi_3$ is minimal within a slice containing $\chi_2$, it must have smaller area than $\chi_2$. We therefore conclude that the generic situation when more than one extremal surface exists if to have a Python's Lunch: three extremal surfaces, with the middle bulge surface having larger area than either end surface.

\section{Explicit Calculation of the Late-time Bulge Surface} \label{app:explicitbulge} \sc
In this appendix, we explicitly calculate the location of the extremal surface that forms the bulge surface at late times in a particular theory. The theory is JT gravity with $c$ Dirac fermions, and we consider a black hole that is evaporating using transparent boundary conditions, as in \cite{Almheiri:2019psf, Almheiri:2019hni}. Dirac fermions are the only conformal field theory for which the calculation is possible, since they are the only conformal field theory for which the two-interval von Neumann entropy is known analytically.

We note that this surface only becomes the bulge surface when the horizon area, which, in the case of JT gravity is the horizon dilaton value $\phi + \phi_0$, is less than half of its initial value.\footnote{In fact, if we started with a two-sided black hole, which is generally the case in JT gravity, then the horizon area needs to be less than half of the \emph{increase} in horizon area from the initial energy thrown into the two-sided black hole, before the evaporation began.} This means that the initial black hole cannot have been in the regime $\phi \ll \phi_0$ where JT gravity is justified as the dimensional reduction of a near extremal black hole. Nonetheless, a) there is no obvious problem (other than UV issues which are unimportant for this calculation) with defining JT gravity as a theory in its own right when $\phi \simgeq \phi_0$ and b) it provides a calculable example of an extremal surface that should also exist in more general examples of evaporating black holes.

The JT gravity action is given by
\begin{align}
\begin{split}
S = \frac{\phi_0}{16 \pi G_N}&\left[ \int_M d^2 x \sqrt{ -g} R + 2 \int_{\partial M} K \right] \\&+ \frac{1}{16 \pi G_N}\left[ \int_M d^2 x \sqrt{ -g} \phi (R + 2) + 2 \int_{\partial M} \phi_b K \right] + S_\mathrm{CFT} [g],
\end{split}
\end{align}

where the scalar field $\phi$ is called the dilaton and $S_\mathrm{CFT} [g]$ is the action for the CFT (in this case $c$ Dirac fermions) in the gravitational background. We also impose boundary conditions
\begin{align}
g_{tt} |_\text{bdy} = \frac{1}{\varepsilon^2} , \,\,\,\,\, \phi = \phi_b = \frac{\bar{\phi}_r}{\varepsilon},
\end{align}
where $t$ is the physical boundary time, $\bar{\phi}_r$ is the fixed renormalized boundary dilaton value and $\varepsilon$ is small.

The metric of a static black hole in JT gravity is given by
\begin{align}
ds^2 = - \frac{4 \pi^2\, T^2\, du\, dv}{ \sinh^2 [ \pi T (u - v)]},
\end{align}
with the dilaton profile given by
\begin{align}
\phi = 2 \bar{\phi}_r \pi \,T \coth [ \pi T (u - v0)].
\end{align}
Here $u$ and $v$ are the advanced and retarded times respectively. The Bekenstein-Hawking entropy $S_\mathrm{BH}$ is given by
\begin{align} \label{eq:SBH}
S_\mathrm{BH} = \frac{\phi_\mathrm{hor} + \phi_0}{4 G_N} = \frac{2 \bar{\phi}_r \pi \,T}{4 G_N} + S_0
\end{align}
where $\phi_\mathrm{hor}$ is the horizon dilaton value and $S_0 = \phi_0 / 4G_N$ is the extremal entropy.

To extend our coordinate system behind the horizon we simply define the Kruskal-like coordinate $U = - \exp(- 2 \pi T u)$. We find
\begin{align}
ds^2 = - e^{2 \pi T v} \frac{8 \pi T dU dv}{ (1 +  U e^{2 \pi T v})^2},
\end{align}
and
\begin{align}
\phi =  2 \bar{\phi}_r \pi T \frac{1 - U e^{2 \pi T v}}{1 + U e^{2 \pi T v}}.
\end{align}
In the near-horizon region $U e^{2 \pi T v} \ll 1$, these simplify to
\begin{align} \label{eq:nearhorizonmetric}
ds^2 = - e^{2 \pi T v} 8 \pi T dU dv,
\end{align}
and
\begin{align}
\phi =  2 \bar{\phi}_r \pi T\left[1 - 2 U e^{2 \pi T v}\right].
\end{align}
In the semiclassical limit, an evaporating black hole is well approximated by an ingoing Vaidya metric, where we simply promote the temperature $T$ to be a slowly varying function of the infalling time $v$. The change in temperature is determined by the rate of energy loss from the black hole, where we have
\begin{align}
\frac{2 \pi \bar{\phi}_r}{4G_N} \, \frac{d T} {d v } = \frac{d S_{\mathrm{BH}}}{dv} = \frac{1}{T} \, \frac{d M}{d v} = - \frac{ \pi c T}{12}.
\end{align}
Here the first equality uses \eqref{eq:SBH}, the second equality is the first law of black-hole thermodynamics and the last equality is the $(1+1)$-dimensional Stefan-Boltzmann law. It follows that, in the near horizon region, we have
\begin{align} \label{eq:phiv}
\frac{1}{4 G_N} \, \frac{\partial \phi}{\partial v} = - \frac{\bar{\phi}_r (2 \pi T)^2 2 U e^{2 \pi T v}}{4 G_N} - \frac{\pi T c}{12},
\end{align}
and
\begin{align} \label{eq:phiU}
\frac{1}{4 G_N} \, \frac{\partial \phi}{\partial U} = - \frac{4 \pi T \bar{\phi}_r e^{2 \pi T v}}{4 G_N}.
\end{align}

We now briefly review the calculation of the location of the nonempty `end surface', which is the EW surface after the Page time. This surface consists of a single point $(U_0, v_0)$. This calculation was previously done for JT gravity in \cite{Almheiri:2019psf}, although our strategy is more similar to \cite{Penington:2019npb}.

Let the `current' boundary time be $t = 0$. The outgoing modes are in the vacuum state with respect to $U$. Outgoing modes in the interval $U_0 \geq U \geq -1$ are included in the entanglement wedge of the boundary; modes with $U \geq U_0$ are outside the wedge, while modes with $U \leq -1$ have already escaped the spacetime. The entropy of the outgoing modes is therefore given by
\begin{align}
S_\text{bulk}^\text{(out)} = \frac{c}{6}\log \frac{U_0 + 1}{\sqrt{\varepsilon_U^\text{(ext)} \varepsilon_U^\text{(bdy)}}}
\end{align}
where $\varepsilon_U^\text{(ext)}$ and $ \varepsilon_U^\text{(bdy)}$ are the cut-offs in units of $U$ at the extremal surface $(U_0, v_0)$ and the boundary respectively. Since the near horizon metric \eqref{eq:nearhorizonmetric} is independent of $U$, we can choose $\varepsilon_U^\text{(ext)}$ and the entropy $S_\text{bulk}^\text{(in)}$ of the ingoing modes to both be independent of $U$. We therefore find
\begin{align}
\frac{\partial S_\text{bulk}}{\partial U_0} = \frac{c}{6 (U_0 + 1)}.
\end{align}
What about the ingoing modes? These are in the vacuum state with respect to the physical infalling time $v$. It follows that their entropy is given by
\begin{align}
S_\text{bulk}^\text{(in)} = \frac{c}{6}\log \frac{|v_0|}{\sqrt{\varepsilon_v^\text{(ext)} \varepsilon_v^\text{(bdy)}}},
\end{align}
where $\varepsilon_v^\text{(ext)}$ and $ \varepsilon_v^\text{(bdy)}$ are the cut-offs in units of $v$ on the ingoing modes at the extremal surface $(U_0, v_0)$ and boundary respectively. Since in the semiclassical limit we will have $v_0 \to - \infty$, one might think that the gradient $\partial S_\text{bulk}/ \partial v$ vanishes.

However, this argument would be naive. To correctly renormalize the entropy we need to keep the proper cut-off
\begin{align}
\varepsilon_\text{prop} = \sqrt{g(\varepsilon_U^\text{(ext)} \partial/ \partial U, \varepsilon_v^\text{(ext)} \partial/ \partial v)}
\end{align}
fixed as we vary the surface. From \eqref{eq:nearhorizonmetric}, we therefore need
\begin{align}
\varepsilon_U^\text{(ext)} \varepsilon_v^\text{(ext)} \propto e^{- 2 \pi T v}.
\end{align}
Hence
\begin{align}
\frac{\partial S_\text{bulk}}{\partial v} = \frac{\pi T c}{6}.
\end{align}
The location of the extremal surface now follows from extremizing the generalized entropy 
\begin{align}
S^{\textrm{(gen)}} = (\phi + \phi_0)/ 4G_N + S_{\text{bulk}}.
\end{align}
We find
\begin{align}
\frac{\partial S^{\textrm{(gen)}}}{\partial U} =  - \frac{4 \pi T \bar{\phi}_r e^{2 \pi T v}}{4 G_N} +  \frac{c}{6 (U_0 + 1)} = 0
\end{align}
and
\begin{align}
\frac{\partial S^{\textrm{(gen)}}}{\partial v} = - \frac{\bar{\phi}_r (2 \pi T)^2 2 U_0 e^{2 \pi T v}}{4 G_N} + \frac{\pi T c}{12}  = 0
\end{align}
which implies
\begin{align}
U_0 = \frac{1}{3} \,\,\,\, v_0 = -\frac{1}{2 \pi T} \log \frac{16 (S - S_0)}{c},
\end{align}
in agreement with the results from \cite{Almheiri:2019psf}.

We are now ready to attempt our actual task: calculating the location of the late-time bulge surface. This is the union of two points $(U_1, v_1)$ and $(U_2, v_2)$, where we assume $U_2 > U_1$ and $v_2 < v_1$ (since the points need to be spacelike separated). In fact, if we started with a two-sided black hole, the bulge surface really consists of three points, where the third point $(U_3, v_3)$ lies close to the other `end surface', near the initial bifurcation surface of the two-sided black hole. In the semiclassical limit we have $U_3  = \exp(O(1/G_N))$ and $v_3 = - O(1/G_N)$. The corrections to the generalized entropy gradient for $(U_3, v_3)$ from the existence of the additional points $(U_1, v_1)$ and $(U_2, v_2)$ are therefore highly suppressed and we can treat $(U_3, v_3)$ as lying exactly on the quantum extremal end surface.

The outgoing entropy is now the entropy of the union of the two intervals $[-1, U_1]$ and $[U_2, U_3]$, which for $c$ Dirac fermions is given by (see \cite{Casini2005}) 
\begin{align} \label{eq:Sout2}
S_\text{bulk}^\text{(out)} = \frac{c}{6} \log \left[ \frac{(U_3 + 1)(U_1 + 1)(U_3 - U_2)(U_2 - U_1)}{(U_3 - U_1)(U_2 + 1) \sqrt{\varepsilon_{U_3} \varepsilon_{U_2} \varepsilon_{U_1} \varepsilon_{U}^\text{(bdy)} }}\right],
\end{align}
where $\varepsilon_{U_3}$, $\varepsilon_{U_2}$, $\varepsilon_{U_1}$ and $\varepsilon_{U}^\text{(bdy)}$ are the outgoing mode cut-offs in units of $U$ at $U_3$, $U_2$, $U_1$ and the boundary respectively. Since $U_3 =  \exp(O(1/G_N))$, terms involving $U_3$ do not contribute to the gradient of the entropy.

Formally, the ingoing entropy should be calculated by a similar formula for the two intervals $[v_3,v_2]$ and $[v_1, 0]$. However, since $v_1 = - O(\log G_N)$ and $v_2 - v_3 = O(1/G_N)$, while we shall find that $v_1 - v_2 = O(1/T)$, for our purposes we can approximate
\begin{align}  \label{eq:Sin2}
S_\text{bulk}^\text{(in)} = \frac{c}{6} \log \frac{v_1 - v_2}{ \sqrt{\varepsilon_{v_2} \varepsilon_{v_1} }} + \dots,
\end{align}
where $\varepsilon_{v_1}$ and $\varepsilon_{v_2}$ are the ingoing cut-offs in units of $v$ at $v_1$ and $v_2$ respectively, and we have elided constant terms. As before, to correctly renormalize the entropy, we need
\begin{align}
\varepsilon_{v_i} \varepsilon_{U_i} \propto e^{2 \pi T v_i}.
\end{align}
Extremizing the generalized entropy, using \eqref{eq:phiv}, \eqref{eq:phiU}, \eqref{eq:Sout2} and \eqref{eq:Sin2}, we find
\begin{align}
&\frac{\partial S^{\textrm{(gen)}}}{\partial U_1} = - \frac{c}{6(U_2 - U_1)} + \frac{c}{6(U_1 + 1)} - 2 (S - S_0)e^{2 \pi T v_1} = 0, \\
&\frac{\partial S^{\textrm{(gen)}}}{\partial U_2} =  \frac{c}{6(U_2 - U_1)} - \frac{c}{6(U_2 + 1)} - 2 (S - S_0)e^{2 \pi T v_2} = 0, \\
&\frac{\partial S^{\textrm{(gen)}}}{\partial v_1} = - 2 (2 \pi T)(S - S_0) U_1 e^{2 \pi T v_1} + \frac{\pi T c}{12} + \frac{c}{6(v_1 - v_2)}= 0, \\
&\frac{\partial S^{\textrm{(gen)}}}{\partial v_2} = - 2 (2 \pi T)(S - S_0) U_2 e^{2 \pi T v_2} + \frac{\pi T c}{12} - \frac{c}{6(v_1 - v_2)}= 0.
\end{align}
This set of equations can be solved numerically. We find
\begin{align}
(U_1, v_1) = (0.874, v_0 - \frac{0.410}{2 \pi T}), \,\,\,\,\,\,\, (U_2, v_2) = (28.8, v_0 - \frac{5.81}{2 \pi T}).
\end{align}
As expected, we have $U_2 > U_1 > U_0$ and $v_2 < v_1 < v_0$.

Finally, we note that the classical contribution to the generalized entropy for this surface is $2 (\phi_0 + 2 \pi T \bar{\phi}_r)$.\footnote{We will not worry about the time-dependence of the temperature here, since we are only interested in the answer at leading order.} Meanwhile, up to subleading corrections, \eqref{eq:Sout2} is equal to $c/6 \log U_3$, which is the entropy of the interior partners of the Hawking radiation. Hence, at leading order, the bulk von Neumann entropy is simply the entropy $S_\text{rad}$ of the Hawking radiation. We therefore find that the total generalized entropy at leading order is $2 S_{BH} + S_\text{rad}$, where $S_{BH}$ is the final Bekenstein-Hawking entropy of the black hole.

{\footnotesize
\bibliography{references}

\providecommand{\href}[2]{#2}\begingroup\raggedright\begin{thebibliography}{10}

\bibitem{HarlowHayden}
D.~Harlow and P.~Hayden, ``{Quantum Computation vs. Firewalls},''
  \href{http://dx.doi.org/10.1007/JHEP06(2013)085}{{\em JHEP} {\bfseries 06}
  (2013) 085},
\href{http://arxiv.org/abs/1301.4504}{{\ttfamily arXiv:1301.4504 [hep-th]}}.
%%CITATION = ARXIV:1301.4504;%%.

\bibitem{Susskind2014}
L.~Susskind, ``{Computational Complexity and Black Hole Horizons},''
  \href{http://dx.doi.org/10.1002/prop.201500093, 10.1002/prop.201500092}{{\em
  Fortsch. Phys.} {\bfseries 64} (2016) 44--48},
  \href{http://arxiv.org/abs/1403.5695}{{\ttfamily arXiv:1403.5695 [hep-th]}}.
[Fortsch. Phys.64,24(2016)].
%%CITATION = ARXIV:1403.5695;%%.

\bibitem{Brown2015}
A.~R. Brown, D.~A. Roberts, L.~Susskind, B.~Swingle, and Y.~Zhao,
  ``{Complexity, action, and black holes},''
  \href{http://dx.doi.org/10.1103/PhysRevD.93.086006}{{\em Phys. Rev.}
  {\bfseries D93} no.~8, (2016) 086006},
\href{http://arxiv.org/abs/1512.04993}{{\ttfamily arXiv:1512.04993 [hep-th]}}.
%%CITATION = ARXIV:1512.04993;%%.

\bibitem{Brown2015-1}
A.~R. Brown, D.~A. Roberts, L.~Susskind, B.~Swingle, and Y.~Zhao, ``Complexity
  equals action,'' \href{http://arxiv.org/abs/arXiv:1509.07876}{{\ttfamily
  arXiv:1509.07876}}.

\bibitem{Maldacena2013}
J.~Maldacena and L.~Susskind, ``{Cool horizons for entangled black holes},''
  \href{http://dx.doi.org/10.1002/prop.201300020}{{\em Fortsch. Phys.}
  {\bfseries 61} (2013) 781--811},
\href{http://arxiv.org/abs/1306.0533}{{\ttfamily arXiv:1306.0533 [hep-th]}}.
%%CITATION = ARXIV:1306.0533;%%.

\bibitem{Nahum:2017yvy}
A.~Nahum, S.~Vijay, and J.~Haah, ``{Operator Spreading in Random Unitary
  Circuits},'' \href{http://dx.doi.org/10.1103/PhysRevX.8.021014}{{\em Phys.
  Rev.} {\bfseries X8} no.~2, (2018) 021014},
\href{http://arxiv.org/abs/1705.08975}{{\ttfamily arXiv:1705.08975
  [cond-mat.str-el]}}.
%%CITATION = ARXIV:1705.08975;%%.

\bibitem{Khemani2017}
V.~Khemani, A.~Vishwanath, and D.~A. Huse, ``Operator spreading and the
  emergence of dissipative hydrodynamics under unitary evolution with
  conservation laws,'' \href{http://arxiv.org/abs/arXiv:1710.09835}{{\ttfamily
  arXiv:1710.09835}}.

\bibitem{Gharibyan:2018jrp}
H.~Gharibyan, M.~Hanada, S.~H. Shenker, and M.~Tezuka, ``{Onset of Random
  Matrix Behavior in Scrambling Systems},''
  \href{http://dx.doi.org/10.1007/JHEP02(2019)197,
  10.1007/JHEP07(2018)124}{{\em JHEP} {\bfseries 07} (2018) 124},
  \href{http://arxiv.org/abs/1803.08050}{{\ttfamily arXiv:1803.08050
  [hep-th]}}.
[Erratum: JHEP02,197(2019)].
%%CITATION = ARXIV:1803.08050;%%.

\bibitem{Hayden:2016cfa}
P.~Hayden, S.~Nezami, X.-L. Qi, N.~Thomas, M.~Walter, and Z.~Yang,
  ``{Holographic duality from random tensor networks},''
  \href{http://dx.doi.org/10.1007/JHEP11(2016)009}{{\em JHEP} {\bfseries 11}
  (2016) 009},
\href{http://arxiv.org/abs/1601.01694}{{\ttfamily arXiv:1601.01694 [hep-th]}}.
%%CITATION = ARXIV:1601.01694;%%.

\bibitem{Gao2016}
P.~Gao, D.~L. Jafferis, and A.~C. Wall, ``{Traversable Wormholes via a Double
  Trace Deformation},'' \href{http://dx.doi.org/10.1007/JHEP12(2017)151}{{\em
  JHEP} {\bfseries 12} (2017) 151},
\href{http://arxiv.org/abs/1608.05687}{{\ttfamily arXiv:1608.05687 [hep-th]}}.
%%CITATION = ARXIV:1608.05687;%%.

\bibitem{Maldacena2017}
J.~Maldacena, D.~Stanford, and Z.~Yang, ``{Diving into traversable
  wormholes},'' \href{http://dx.doi.org/10.1002/prop.201700034}{{\em Fortsch.
  Phys.} {\bfseries 65} no.~5, (2017) 1700034},
\href{http://arxiv.org/abs/1704.05333}{{\ttfamily arXiv:1704.05333 [hep-th]}}.
%%CITATION = ARXIV:1704.05333;%%.

\bibitem{Aaronson2016}
S.~Aaronson, ``{The Complexity of Quantum States and Transformations: From
  Quantum Money to Black Holes},''
\newblock 2016.
\newblock
\href{http://arxiv.org/abs/1607.05256}{{\ttfamily arXiv:1607.05256
  [quant-ph]}}.
\newblock
%%CITATION = ARXIV:1607.05256;%%.

\bibitem{Yoshida:2017non}
B.~Yoshida and A.~Kitaev, ``{Efficient decoding for the Hayden-Preskill
  protocol},''
\href{http://arxiv.org/abs/1710.03363}{{\ttfamily arXiv:1710.03363 [hep-th]}}.
%%CITATION = ARXIV:1710.03363;%%.

\bibitem{Grover-1}
L.~K. Grover, \href{http://dx.doi.org/10.1145/237814.237866}{``A fast quantum
  mechanical algorithm for database search,''} in {\em Proceedings of the
  Twenty-eighth Annual ACM Symposium on Theory of Computing}, STOC '96,
  pp.~212--219.
\newblock ACM, New York, NY, USA, 1996.
\newblock \url{http://doi.acm.org/10.1145/237814.237866}.

\bibitem{Grover-2}
L.~Grover, ``Quantum mechanics helps in searching for a needle in a haystack,''
  \href{http://dx.doi.org/10.1103/PhysRevLett.79.325}{{\em Physical Review
  Letters} {\bfseries 79} no.~2, (7, 1997) }.

\bibitem{PostBQP}
S.~Aaronson, ``Quantum computing, postselection, and probabilistic
  polynomial-time,''
  \href{http://arxiv.org/abs/arXiv:quant-ph/0412187}{{\ttfamily
  arXiv:quant-ph/0412187}}.

\bibitem{Bennett:1996iu-1}
C.~H. Bennett, E.~Bernstein, G.~Brassard, and U.~Vazirani, ``Strengths and
  weaknesses of quantum computing,''
  \href{http://dx.doi.org/10.1137/S0097539796300933}{{\em SIAM Journal on
  Computing} {\bfseries 26} no.~5, (1997) 1510--1523},
  \href{http://arxiv.org/abs/https://doi.org/10.1137/S0097539796300933}{{\ttfamily
  https://doi.org/10.1137/S0097539796300933}}.
  \url{https://doi.org/10.1137/S0097539796300933}.

\bibitem{Bennett:1996iu-2}
M.~Boyer, G.~Brassard, P.~Høyer, and A.~Tapp, ``Tight bounds on quantum
  searching,''
  \href{http://dx.doi.org/10.1002/(SICI)1521-3978(199806)46:4/5<493::AID-PROP493>3.0.CO;2-P}{{\em
  Fortschritte der Physik} {\bfseries 46} no.~4‐5, (1998) 493--505},
  \href{http://arxiv.org/abs/https://onlinelibrary.wiley.com/doi/pdf/10.1002/\%28SICI\%291521-3978\%28199806\%2946\%3A4/5\%3C493\%3A\%3AAID-PROP493\%3E3.0.CO\%3B2-P}{{\ttfamily
  https://onlinelibrary.wiley.com/doi/pdf/10.1002/\%28SICI\%291521-3978\%28199806\%2946\%3A4/5\%3C493\%3A\%3AAID-PROP493\%3E3.0.CO\%3B2-P}}.
  \url{https://onlinelibrary.wiley.com/doi/abs/10.1002/\%28SICI\%291521-3978\%28199806\%2946\%3A4/5\%3C493\%3A\%3AAID-PROP493\%3E3.0.CO\%3B2-P}.

\bibitem{Bennett:1996iu-3}
C.~Zalka, ``Grover's quantum searching algorithm is optimal,''
  \href{http://dx.doi.org/10.1103/PhysRevA.60.2746}{{\em Phys. Rev. A}
  {\bfseries 60} (Oct, 1999) 2746--2751}.
  \url{https://link.aps.org/doi/10.1103/PhysRevA.60.2746}.

\bibitem{BCC15}
D.~W. Berry, A.~M. Childs, R.~Cleve, R.~Kothari, and R.~D. Somma, ``Simulating
  hamiltonian dynamics with a truncated taylor series.,'' {\em Physical review
  letters} {\bfseries 114 9} (2015) 090502.

\bibitem{Gilyen2017}
A.~Gily\'{e}n, S.~Arunachalam, and N.~Wiebe, ``Optimizing quantum optimization
  algorithms via faster quantum gradient computation,''
  \href{http://arxiv.org/abs/arXiv:1711.00465}{{\ttfamily arXiv:1711.00465}}.

\bibitem{Gilyen2019}
A.~Gily{\'e}n, Y.~Su, G.~H. Low, and N.~Wiebe,
  \href{http://dx.doi.org/10.1145/3313276.3316366}{``Quantum singular value
  transformation and beyond: Exponential improvements for quantum matrix
  arithmetics,''} in {\em Proceedings of the 51st Annual ACM SIGACT Symposium
  on Theory of Computing}, STOC 2019, pp.~193--204.
\newblock ACM, New York, NY, USA, 2019.
\newblock \url{http://doi.acm.org/10.1145/3313276.3316366}.

\bibitem{Ryu:2006bv}
S.~Ryu and T.~Takayanagi, ``{Holographic derivation of entanglement entropy
  from AdS/CFT},'' \href{http://dx.doi.org/10.1103/PhysRevLett.96.181602}{{\em
  Phys. Rev. Lett.} {\bfseries 96} (2006) 181602},
\href{http://arxiv.org/abs/hep-th/0603001}{{\ttfamily arXiv:hep-th/0603001
  [hep-th]}}.
%%CITATION = HEP-TH/0603001;%%.

\bibitem{Hubeny:2007xt}
V.~E. Hubeny, M.~Rangamani, and T.~Takayanagi, ``{A Covariant holographic
  entanglement entropy proposal},''
  \href{http://dx.doi.org/10.1088/1126-6708/2007/07/062}{{\em JHEP} {\bfseries
  07} (2007) 062},
\href{http://arxiv.org/abs/0705.0016}{{\ttfamily arXiv:0705.0016 [hep-th]}}.
%%CITATION = ARXIV:0705.0016;%%.

\bibitem{Lewkowycz:2013nqa}
A.~Lewkowycz and J.~Maldacena, ``{Generalized gravitational entropy},''
  \href{http://dx.doi.org/10.1007/JHEP08(2013)090}{{\em JHEP} {\bfseries 08}
  (2013) 090},
\href{http://arxiv.org/abs/1304.4926}{{\ttfamily arXiv:1304.4926 [hep-th]}}.
%%CITATION = ARXIV:1304.4926;%%.

\bibitem{Engelhardt:2014gca}
N.~Engelhardt and A.~C. Wall, ``{Quantum Extremal Surfaces: Holographic
  Entanglement Entropy beyond the Classical Regime},''
  \href{http://dx.doi.org/10.1007/JHEP01(2015)073}{{\em JHEP} {\bfseries 01}
  (2015) 073},
\href{http://arxiv.org/abs/1408.3203}{{\ttfamily arXiv:1408.3203 [hep-th]}}.
%%CITATION = ARXIV:1408.3203;%%.

\bibitem{Dong:2017xht}
X.~Dong and A.~Lewkowycz, ``{Entropy, Extremality, Euclidean Variations, and
  the Equations of Motion},''
  \href{http://dx.doi.org/10.1007/JHEP01(2018)081}{{\em JHEP} {\bfseries 01}
  (2018) 081},
\href{http://arxiv.org/abs/1705.08453}{{\ttfamily arXiv:1705.08453 [hep-th]}}.
%%CITATION = ARXIV:1705.08453;%%.

\bibitem{Wall:2012uf}
A.~C. Wall, ``{Maximin Surfaces, and the Strong Subadditivity of the Covariant
  Holographic Entanglement Entropy},''
  \href{http://dx.doi.org/10.1088/0264-9381/31/22/225007}{{\em Class. Quant.
  Grav.} {\bfseries 31} no.~22, (2014) 225007},
\href{http://arxiv.org/abs/1211.3494}{{\ttfamily arXiv:1211.3494 [hep-th]}}.
%%CITATION = ARXIV:1211.3494;%%.

\bibitem{QuantumMaximin}
C.~Akers, N.~Engelhardt, G.~Penington, and M.~Usatyuk, ``{Quantum Maximin
  Surfaces},'' {\em To appear} .

\bibitem{Page:1993df}
D.~N. Page, ``{Average entropy of a subsystem},''
  \href{http://dx.doi.org/10.1103/PhysRevLett.71.1291}{{\em Phys. Rev. Lett.}
  {\bfseries 71} (1993) 1291--1294},
\href{http://arxiv.org/abs/gr-qc/9305007}{{\ttfamily arXiv:gr-qc/9305007
  [gr-qc]}}.
%%CITATION = GR-QC/9305007;%%.

\bibitem{Almheiri:2012rt}
A.~Almheiri, D.~Marolf, J.~Polchinski, and J.~Sully, ``{Black Holes:
  Complementarity or Firewalls?},''
  \href{http://dx.doi.org/10.1007/JHEP02(2013)062}{{\em JHEP} {\bfseries 02}
  (2013) 062},
\href{http://arxiv.org/abs/1207.3123}{{\ttfamily arXiv:1207.3123 [hep-th]}}.
%%CITATION = ARXIV:1207.3123;%%.

\bibitem{Hayden:2007cs}
P.~Hayden and J.~Preskill, ``{Black holes as mirrors: Quantum information in
  random subsystems},''
  \href{http://dx.doi.org/10.1088/1126-6708/2007/09/120}{{\em JHEP} {\bfseries
  09} (2007) 120},
\href{http://arxiv.org/abs/0708.4025}{{\ttfamily arXiv:0708.4025 [hep-th]}}.
%%CITATION = ARXIV:0708.4025;%%.

\bibitem{Almheiri:2018xdw}
A.~Almheiri, ``{Holographic Quantum Error Correction and the Projected Black
  Hole Interior},''
\href{http://arxiv.org/abs/1810.02055}{{\ttfamily arXiv:1810.02055 [hep-th]}}.
%%CITATION = ARXIV:1810.02055;%%.

\bibitem{Hayden:2018khn}
P.~Hayden and G.~Penington, ``{Learning the Alpha-bits of Black Holes},''
\href{http://arxiv.org/abs/1807.06041}{{\ttfamily arXiv:1807.06041 [hep-th]}}.
%%CITATION = ARXIV:1807.06041;%%.

\bibitem{Penington:2019npb}
G.~Penington, ``{Entanglement Wedge Reconstruction and the Information
  Paradox},''
\href{http://arxiv.org/abs/1905.08255}{{\ttfamily arXiv:1905.08255 [hep-th]}}.
%%CITATION = ARXIV:1905.08255;%%.

\bibitem{Almheiri:2019psf}
A.~Almheiri, N.~Engelhardt, D.~Marolf, and H.~Maxfield, ``{The entropy of bulk
  quantum fields and the entanglement wedge of an evaporating black hole},''
\href{http://arxiv.org/abs/1905.08762}{{\ttfamily arXiv:1905.08762 [hep-th]}}.
%%CITATION = ARXIV:1905.08762;%%.

\bibitem{Almheiri:2019hni}
A.~Almheiri, R.~Mahajan, J.~Maldacena, and Y.~Zhao, ``{The Page curve of
  Hawking radiation from semiclassical geometry},''
\href{http://arxiv.org/abs/1908.10996}{{\ttfamily arXiv:1908.10996 [hep-th]}}.
%%CITATION = ARXIV:1908.10996;%%.

\bibitem{PSSY}
G.~Penington, S.~H. Shenker, D.~Stanford, and Z.~Yang, ``Replica wormholes and
  the black hole interior,''
  \href{http://arxiv.org/abs/arXiv:1911.11977}{{\ttfamily arXiv:1911.11977}}.

\bibitem{AHMST}
A.~Almheiri, T.~Hartman, J.~Maldacena, E.~Shaghoulian, and A.~Tajdini,
  ``Replica wormholes and the entropy of hawking radiation,''
  \href{http://arxiv.org/abs/arXiv:1911.12333}{{\ttfamily arXiv:1911.12333}}.

\bibitem{Headrick:2014cta}
M.~Headrick, V.~E. Hubeny, A.~Lawrence, and M.~Rangamani, ``{Causality \&
  holographic entanglement entropy},''
  \href{http://dx.doi.org/10.1007/JHEP12(2014)162}{{\em JHEP} {\bfseries 12}
  (2014) 162},
\href{http://arxiv.org/abs/1408.6300}{{\ttfamily arXiv:1408.6300 [hep-th]}}.
%%CITATION = ARXIV:1408.6300;%%.

\bibitem{Czech:2012bh}
B.~Czech, J.~L. Karczmarek, F.~Nogueira, and M.~Van~Raamsdonk, ``{The Gravity
  Dual of a Density Matrix},''
  \href{http://dx.doi.org/10.1088/0264-9381/29/15/155009}{{\em Class. Quant.
  Grav.} {\bfseries 29} (2012) 155009},
\href{http://arxiv.org/abs/1204.1330}{{\ttfamily arXiv:1204.1330 [hep-th]}}.
%%CITATION = ARXIV:1204.1330;%%.

\bibitem{Jafferis:2015del}
D.~L. Jafferis, A.~Lewkowycz, J.~Maldacena, and S.~J. Suh, ``{Relative entropy
  equals bulk relative entropy},''
  \href{http://dx.doi.org/10.1007/JHEP06(2016)004}{{\em JHEP} {\bfseries 06}
  (2016) 004},
\href{http://arxiv.org/abs/1512.06431}{{\ttfamily arXiv:1512.06431 [hep-th]}}.
%%CITATION = ARXIV:1512.06431;%%.

\bibitem{Dong:2016eik}
X.~Dong, D.~Harlow, and A.~C. Wall, ``{Reconstruction of Bulk Operators within
  the Entanglement Wedge in Gauge-Gravity Duality},''
  \href{http://dx.doi.org/10.1103/PhysRevLett.117.021601}{{\em Phys. Rev.
  Lett.} {\bfseries 117} no.~2, (2016) 021601},
\href{http://arxiv.org/abs/1601.05416}{{\ttfamily arXiv:1601.05416 [hep-th]}}.
%%CITATION = ARXIV:1601.05416;%%.

\bibitem{Cotler:2017erl}
J.~Cotler, P.~Hayden, G.~Penington, G.~Salton, B.~Swingle, and M.~Walter,
  ``{Entanglement Wedge Reconstruction via Universal Recovery Channels},''
  \href{http://dx.doi.org/10.1103/PhysRevX.9.031011}{{\em Phys. Rev.}
  {\bfseries X9} no.~3, (2019) 031011},
\href{http://arxiv.org/abs/1704.05839}{{\ttfamily arXiv:1704.05839 [hep-th]}}.
%%CITATION = ARXIV:1704.05839;%%.

\bibitem{Faulkner:2013ana}
T.~Faulkner, A.~Lewkowycz, and J.~Maldacena, ``{Quantum corrections to
  holographic entanglement entropy},''
  \href{http://dx.doi.org/10.1007/JHEP11(2013)074}{{\em JHEP} {\bfseries 11}
  (2013) 074},
\href{http://arxiv.org/abs/1307.2892}{{\ttfamily arXiv:1307.2892 [hep-th]}}.
%%CITATION = ARXIV:1307.2892;%%.

\bibitem{Almheiri:2014lwa}
A.~Almheiri, X.~Dong, and D.~Harlow, ``{Bulk Locality and Quantum Error
  Correction in AdS/CFT},''
  \href{http://dx.doi.org/10.1007/JHEP04(2015)163}{{\em JHEP} {\bfseries 04}
  (2015) 163},
\href{http://arxiv.org/abs/1411.7041}{{\ttfamily arXiv:1411.7041 [hep-th]}}.
%%CITATION = ARXIV:1411.7041;%%.

\bibitem{Susskind:1993if}
L.~Susskind, L.~Thorlacius, and J.~Uglum, ``{The Stretched horizon and black
  hole complementarity},''
  \href{http://dx.doi.org/10.1103/PhysRevD.48.3743}{{\em Phys. Rev.} {\bfseries
  D48} (1993) 3743--3761},
\href{http://arxiv.org/abs/hep-th/9306069}{{\ttfamily arXiv:hep-th/9306069
  [hep-th]}}.
%%CITATION = HEP-TH/9306069;%%.

\bibitem{Maldacena:2013xja}
J.~Maldacena and L.~Susskind, ``{Cool horizons for entangled black holes},''
  \href{http://dx.doi.org/10.1002/prop.201300020}{{\em Fortsch. Phys.}
  {\bfseries 61} (2013) 781--811},
\href{http://arxiv.org/abs/1306.0533}{{\ttfamily arXiv:1306.0533 [hep-th]}}.
%%CITATION = ARXIV:1306.0533;%%.

\bibitem{Page:2013dx}
D.~N. Page, ``{Time Dependence of Hawking Radiation Entropy},''
  \href{http://dx.doi.org/10.1088/1475-7516/2013/09/028}{{\em JCAP} {\bfseries
  1309} (2013) 028},
\href{http://arxiv.org/abs/1301.4995}{{\ttfamily arXiv:1301.4995 [hep-th]}}.
%%CITATION = ARXIV:1301.4995;%%.

\bibitem{Nielsen2006}
M.~A. Nielsen, M.~R. Dowling, M.~Gu, and A.~C. Doherty, ``Quantum computation
  as geometry,'' \href{http://dx.doi.org/10.1126/science.1121541}{{\em Science}
  {\bfseries 311} no.~5764, (2006) 1133--1135},
  \href{http://arxiv.org/abs/https://science.sciencemag.org/content/311/5764/1133.full.pdf}{{\ttfamily
  https://science.sciencemag.org/content/311/5764/1133.full.pdf}}.
  \url{https://science.sciencemag.org/content/311/5764/1133}.

\bibitem{Kourkoulou:2017zaj}
I.~Kourkoulou and J.~Maldacena, ``{Pure states in the SYK model and
  nearly-$AdS_2$ gravity},''
\href{http://arxiv.org/abs/1707.02325}{{\ttfamily arXiv:1707.02325 [hep-th]}}.
%%CITATION = ARXIV:1707.02325;%%.

\bibitem{Bousso:2015mna}
R.~Bousso, Z.~Fisher, S.~Leichenauer, and A.~C. Wall, ``{Quantum focusing
  conjecture},'' \href{http://dx.doi.org/10.1103/PhysRevD.93.064044}{{\em Phys.
  Rev.} {\bfseries D93} no.~6, (2016) 064044},
\href{http://arxiv.org/abs/1506.02669}{{\ttfamily arXiv:1506.02669 [hep-th]}}.
%%CITATION = ARXIV:1506.02669;%%.

\bibitem{Collins2003}
B.~Collins, ``Moments and cumulants of polynomial random variables on unitary
  groups, the itzykson-zuber integral and free probability,''
  \href{http://arxiv.org/abs/arXiv:math-ph/0205010}{{\ttfamily
  arXiv:math-ph/0205010}}.

\bibitem{federer2014geometric}
H.~Federer, {\em Geometric measure theory}.
\newblock Springer, 2014.

\bibitem{Algrem1965}
W.~K. Allard, ``Notes on the theory of varifolds,'' in {\em Th\'eorie des
  vari\'et\'es minimales et applications}, no.~154-155 in Ast\'erisque,
  pp.~73--93.
\newblock Soci\'et\'e math\'ematique de France, 1987.
\newblock \url{http://www.numdam.org/item/AST_1987__154-155__73_0}.

\bibitem{Pitts1081}
J.~T. Pitts, {\em Existence and Regularity of Minimal Surfaces on Riemannian
  Manifolds. (MN-27):}.
\newblock Princeton University Press, 1981.
\newblock \url{http://www.jstor.org/stable/j.ctt7zv66w}.

\bibitem{Colding2003}
T.~H. Colding and C.~D. Lellis, ``The min--max construction of minimal
  surfaces,'' \href{http://arxiv.org/abs/arXiv:math/0303305}{{\ttfamily
  arXiv:math/0303305}}.

\bibitem{Fischetti2014}
S.~Fischetti, D.~Marolf, and A.~C. Wall, ``A paucity of bulk entangling
  surfaces: Ads wormholes with de sitter interiors,''
  \href{http://arxiv.org/abs/arXiv:1409.6754}{{\ttfamily arXiv:1409.6754}}.

\bibitem{Casini2005}
H.~Casini, C.~D. Fosco, and M.~Huerta, ``{Entanglement and alpha entropies for
  a massive Dirac field in two dimensions},''
  \href{http://dx.doi.org/10.1088/1742-5468/2005/07/P07007}{{\em J. Stat.
  Mech.} {\bfseries 0507} (2005) P07007},
\href{http://arxiv.org/abs/cond-mat/0505563}{{\ttfamily arXiv:cond-mat/0505563
  [cond-mat]}}.
%%CITATION = COND-MAT/0505563;%%.

\end{thebibliography}\endgroup
}

\bibliographystyle{utphys}

\end{document}